\documentclass{aa}  

\usepackage{graphicx}
\usepackage{txfonts}
\usepackage{xcolor}
\usepackage{caption}
\usepackage{comment}
\usepackage{longtable}
\usepackage{multirow}
\usepackage{natbib}%,twoopt}
\usepackage[urlcolor = black, citecolor = blue, colorlinks = true]{hyperref} %% to avoid \citeads line fills
\bibpunct{(}{)}{;}{a}{}{,}             %% natbib format for A&A and ApJ

\newcommand{\gaia}{\emph{Gaia}}

\newcommand{\logrhk}{$\rm log\,R^{\prime}_\mathrm{HK}$}
\newcommand{\msun}{$M_{\odot}$}

\newcommand{\vsini}{$v$\,sin\,$i_\star$}

\newcommand{\teff}{$T_{\rm eff}$}
\newcommand{\logg}{log\,{\it g$_\star$}}

\newcommand{\mearth}{$M_{\oplus}$}

\newcommand{\mjup}{M$_J$}
\newcommand{\smw}{S$_{\text{MW}}$}

\begin{document}

   \title{The GAPS programme with HARPS-N at TNG}
   
   \subtitle{LXXVII. Occurrence rates of small close-in planets in the presence of cold Jupiters}

\author{A. Ruggieri\inst{1,2}
          \and
          S. Desidera\inst{2}
          \and
          M. Pinamonti\inst{3}
          \and
          D. Barbato\inst{2}
          \and
          A. Sozzetti\inst{3}
          \and
          A. S. Bonomo\inst{3}
          \and
          L. Naponiello\inst{3}
          \and
          M. Damasso\inst{3}
          \and
          S. Benatti\inst{4}
          \and
          I. Carleo\inst{3}
          \and
          R. Gratton\inst{2}
          \and
          K. Biazzo\inst{5}
          \and
          A. F. Lanza\inst{1}
          \and
          G. Piccinini\inst{4,6}
          \and
          N. Nari\inst{7,8,9}
          \and
          G. Mantovan\inst{10,2}
          \and
          J. Maldonado\inst{4}
          \and
          D. Nardiello\inst{2,11}
          \and
          A. Fiorenzano\inst{12}
          \and
          A. Ghedina\inst{12}
          \and
          A. Bignamini\inst{13}
          }
    \authorrunning{A. Ruggieri et al.}
   \institute{INAF, Astrophysical Observatory of Catania, Via S. Sofia 78, I - 95123 Catania, Italy\\ %1
              \email{alessandro.ruggieri@inaf.it}
         \and
             INAF, Astrophysical Observatory of Padova, Vicolo dell'Osservatorio 5, I - 35122, Padova, Italy %2
         \and
            INAF, Astrophysical Observatory of Turin, Via Osservatorio 20, I - 10025, Pino Torinese (TO), Italy  %3
         \and
            INAF, Astronomical Observatory of Palermo, Piazza del Parlamento 1, I - 90134, Palermo, Italy  %4
         \and
             INAF, Astronomical Observatory of Rome, Via Frascati 33, I - 00178, Monte Porzio Catone (RM), Italy  %5
        \and
        Department of Physics and Chemistry, University of Palermo, Via Archirafi 36, I-90123, Palermo, Italy %6
         \and
         Instituto de Astrofísica de Canarias, 38205 La Laguna, Tenerife, Spain %7
         \and
         Light Bridges S.L., Observatorio del Teide, Carretera del Observatorio, s/n Guimar, 38500 Tenerife, Canarias, Spain %8
         \and
         Departamento de Astrofísica, Universidad de La Laguna, 38206 La Laguna, Tenerife, Spain %9
         \and
        Centro di Ateneo di Studi e Attività Spaziali “G. Colombo” – Università di Padova, Via Venezia 15, 35131 Padova, Italy  %10
          \and
          Department of Physics and Astronomy "Galileo Galilei", University of Padova, Vicolo dell'Osservatorio 3, I - 35122, Padova, Italy %11
          \and
          Fundación Galileo Galilei – INAF, Rambla José Ana Fernández Pérez, 7, E-38712 Breña Baja, Santa Cruz deTenerife, Spain %12
          \and
          INAF – Astronomical Observatory of Trieste, via Tiepolo 11, 34143, Trieste %13
        }

   \date{Received ; accepted }

% \abstract{}{}{}{}{} 
% 5 {} token are mandatory
 
  \abstract
   {The architecture of our Solar System, with inner small planets (ISPs) and outer giants, may or may not be common. Understanding whether a correlation exists between ISPs and outer cold giants is key to evaluating how common systems with a similar architecture to our own are.}
   {This study aims to build a large, homogeneous sample of systems hosting cold Jupiters (CJs, $a_{peri} > 1$ au, $m\sin i > 0.1$\ \mjup) detected via radial velocities (RVs), and to assess the presence of additional ISPs ($P < 400$ d, $3 \leq m\sin i \leq 31.7$ \mearth), studying the correlation between these two types of objects.}
   {We selected 137 stars known to host a CJ, including 23 that also harbour a hot Jupiter and were treated separately. Data from various instruments were compiled, including unpublished data gathered with the High Accuracy Radial velocity Planet Searcher for the Northern hemisphere (HARPS-N) within the Global Architectures of Planetary Systems (GAPS) programme, and consistently fitted using PyORBIT. We derived RV detection maps and calculated occurrence rates for ISPs, cross-validating results with two independent codes. The sample was divided into subgroups to evaluate how system parameters influence planet occurrence.}
   {We confirmed the 213 already known planets in the 137 systems and also identified six new candidates. We divided them, based on mass and period, into Neptunes ($10 < m\sin i < 31.7$\ \mearth) and super-Earths ($3 < m\sin i \leq 10$\ \mearth), and into hot ($1 \leq P \leq 10$ d), warm ($10 < P \leq 100$ d), and cool ($100 < P \leq 400$ d). We found occurrences of $\sim 5\%$, $\sim 13\%$, and $\sim 12\%$ for hot, warm, and cool Neptunes, respectively, and $\sim 11\%$ and $\sim 16\%$ for hot and warm super-Earths, respectively. Systems with dynamically stable inner regions show higher rates of small planets. These findings are consistent with previous studies showing no strong correlation between ISPs and CJs at average stellar metallicity and mass, and suggest that hot Jupiters may be more commonly associated with external giants.}
   {}


    \titlerunning{The GAPS Programme with HARPS-N at TNG TBD}
    \authorrunning{Ruggieri et al.}
   \maketitle
   
%
%-------------------------------------------------------------------

\section{Introduction}
\label{sec:intro}

The radial velocity (RV) method was the first one to be used to discover and characterize exoplanets in the 1990s. After 30 years, significant progress has been made, and this technique is still a reliable asset to enhance our knowledge of the field. We are already able to probe the m/s regime, or even the sub-m/s one in some cases, with several spectrographs, including ESPRESSO \citep[Echelle SPectrograph for Rocky Exoplanets and Stable Spectroscopic Observations,][]{espresso}, the HARPS and HARPS-N twins \citep[High Accuracy Radial velocity Planet Searcher,][]{harps, Cosentino2012}, HIRES \citep[High Resolution Echelle Spectrometer,][]{hires}, APF \citep[Automated Planet Finder,][]{apf}, and CARMENES \citep[Calar Alto high-Resolution search for M dwarfs with Exo-earths with Near-infrared and optical Echelle Spectrographs,][]{carmenes}. We have more advanced algorithms for extracting RVs, such as TERRA \citep[Template-Enhanced Radial velocity Re-analysis Application,][]{terra} and SERVAL \citep[SpEctrum Radial Velocity AnaLyser,][]{serval}, or activity indicators, such as ACTIN \citep[Activity Indices Calculator,][]{dasilva2018, daSilva2011}, from spectra. Besides, since the Earth induces a reflex motion in the Sun with a semi-amplitude of 9 cm/s, a one-order-of-magnitude improvement is needed to possibly detect an Earth twin around a Sun-like star, which is currently one of the main goals in exoplanetary science. In addition, even if we now have more advanced techniques to account for stellar activity \citep[e.g.][]{Grunblatt2015, LopezMorales2016, Perger2021, mantovan2024a}, this phenomenon is not yet fully understood and can often complicate the exoplanetary search. \\

\noindent Another issue is that exoplanetary systems display remarkable differences in their architectures, including types of planets that are absent in the Solar System (super-Earths (SEs), sub-Neptunes, and hot Jupiters (HJs)), systems of tightly packed rocky planets \citep{weiss2018}, and objects much more massive than Jupiter, which complicate the interpretation of formation and migration models. Thus, understanding whether the Solar System is common or not is not trivial. As has been delineated by many authors, it is not simple to define what a Solar System analogue is in the first place \citep{barbato_b_2018}, but, as a first and broad approximation, we can say that it is a system with inner small planets (ISPs, which we define here as a planet with $P < 400$ d and $3 \leq m\sin i \leq 31.7$ \mearth) and an outer giant planet. According to the NASA Exoplanet Archive\footnote{\url{https://exoplanetarchive.ipac.caltech.edu/}} \citep{christiansen2025}, the only small rocky planets ($m_p < 2$ \mearth) discovered with the RV method around FGK stars are HD 20781 b \citep{udry2019}, HD 215152 b and c \citep{delisle2018}, and tau Ceti g and h \citep{feng2017}, although more were discovered with the transit method and later characterized with RV follow-up \citep[e.g.][]{barros2022, bonomoetal2023, brinkman2025}. \\

\noindent Given the complexity of this topic, it becomes increasingly important to move beyond individual detections and towards a more global approach. In particular, to better understand the diversity of planetary architectures and the underlying physical mechanisms that shape them, it is crucial to adopt a statistical approach that accounts for the limitations and biases of our detection techniques. However, this requires careful consideration of observational biases and methodological differences that can significantly affect our results and their interpretation. Furthermore, it is often difficult to compare results from different works because each author uses different selection criteria and focuses on a specific type of planet, stellar spectral type, or detection method, and so on. \\

\noindent Nevertheless, several works have added one piece to this large puzzle. Around M dwarfs, the most common stellar type, ISPs are much more abundant than Sun-like stars, as found by many authors \citep[e.g.][]{bonfils2013, tuomi2014, gaidos2016, pinamonti2022}, but, at the same time, cold Jupiters (CJs, which we define as a planet with $m\sin i > 0.1$ \mjup, $a_{peri} > 1$ au, and $a < 10$ au) are rare, about a few percent, around this kind of star \citep[][and references therein]{pass2023}. For FGK stars, the situation is very different as the occurrence of CJs is around 10 - 15\% \citep[e.g.][]{cumming2008, wittenmyer2020, fulton2021}. The California Legacy Survey \citep[CLS,][]{rosenthal2021} yielded a large number of results in this regard, deriving occurrence rates for both small inner and outer giant planets, plus the conditional probability of hosting one type given the other and vice versa \citep{rosenthal2022}. In particular, they claim that outer giants may favour the formation of smaller siblings in the inner regions, except when they are located relatively close to the central star. However, from their Tables 2 and 3, we note that P(Inner) and P(Inner|Outer) are compatible at a $1\sigma$ level when considering only ISPs with masses lower than 20 \mearth. \cite{zhu2018} and \cite{bryan2019} both estimated the conditional probability of hosting a small inner planet in a system with a cold giant as $f_{\text{ISP|CJ}} \sim 100\%$, showing a strong correlation between the two types of planets. On the other hand, \cite{bonomoetal2023} did not find an excess of Jovian analogues in systems with transiting ISPs. The situation becomes even more complicated when considering the effect of stellar mass and metallicity. In particular, Jupiter-like planets seem to be more abundant around more metallic stars \citep{santos2004,fischer2005,johnson2010}, while \cite{buchhave2012} found little or no correlation between the presence of small planets and metallicity. On the contrary, \cite{adibekyan2012}, \cite{petigura2018}, and \cite{zhu2019} found a weak increase in the occurrence of small planets with increasing stellar metallicity. The dependence of the ISP-CJ relation on stellar mass and metallicity has been investigated in a few recent works. For instance, \cite{bryan2024} found a strong correlation between ISPs and CJs for FGK stars with [Fe/H] $> 0$ but not for less metallic stars. \cite{bryan2025} confirm that this correlation is only present for high metallicities but also point out that it is not present for M dwarfs, regardless of metallicity. \cite{bonomo2025} found a possible enhancement of CJs in ISP systems, but only at super-solar mass and metallicity ($1.0<M_{*}<1.2\,M_{\odot}$, [Fe/H]$>0.1$), in general agreement with \cite{bryan2025}. Additionally, based on CLS data, \cite{mulders2025} noticed that ISPs are more abundant when accompanied by Saturn-mass planets rather than super-Jupiters, trying to explain the apparent discrepancy between other studies (e.g. \citealt{bryan2019} vs \citealt{bonomoetal2023}) with a difference in the mass of the CJs considered. However, it should also be noted that \cite{mulders2025} included giant planets with $a < 1$ au in their CJ definition, and thus considered objects that experienced a certain degree of migration and, overall, a different evolution history. The same applies to the work by \cite{rosenthal2022}. Therefore, it is not straightforward to make comparisons between results from the literature when these rely on different definitions of CJs and/or ISPs.\\

\noindent This work aims to consider a large and well-studied sample of stars hosting CJs, selected using rigorous criteria to ensure that all targets are well suited for searching for small planets at low separations. In particular, our goal is to study whether a correlation exists between CJs and ISPs, understanding what physical parameters might affect the connection between the two classes of objects. \\

\noindent In Section \ref{sec:sample_selection}, we list the selection criteria used to build our sample. In Section \ref{sec:data_gathering}, we illustrate the data used in this work. In Section \ref{sec:dataanalysis4}, we describe the methods and tools used for our analysis. In Section \ref{sec:targets_removed}, we explain why we removed a few objects from our list despite them satisfying our selection criteria. In Section \ref{sec:sample_characteristics}, we describe our sample. In Section \ref{sec:statistics}, we analyse the derived occurrence rates for small and large planets in various cases. Finally, in Section \ref{sec:conclusion4}, we summarize our findings, indicating a possible direction for future work.

\section{Sample selection}
\label{sec:sample_selection}
Understanding whether ISPs are more likely to form in systems that host CJs requires a robust statistical approach. Specifically, estimating the conditional probability, $f_{\text{ISP|CJ}}$ -- that is, the likelihood of finding a small planet given the presence of an external gas giant -- requires a large and well-characterized sample of planetary systems. To this end, we constructed our target list by querying the NASA Exoplanet Archive on 24 October 2023, applying a set of rigorous physical and observational criteria to isolate systems suitable for this analysis. Our selection criteria are the following:

\begin{enumerate}

    \item Stars known to host a CJ; that is, planets with masses or minimum masses higher than 0.1 \mjup\ and periastron larger than 1 au detected with the RV method. Since our objective is to study the link between external gas giants and ISPs, we removed the systems in which the closest gas giant had $a > 10$ au, as these are likely too distant to affect the inner regions of the respective systems. This is the same approach taken by several works, such as \cite{wittenmyer2020}, \cite{rosenthal2021}, \cite{bonomoetal2023}, and \cite{bonomo2025}.
    
    \item Bright stars; that is, those with $V < 10$.

    \item Inactive stars, choosing \logrhk $< -4.8$ as a reference value. We used the \logrhk values reported in \cite{borosaikia2018} for the available stars. We picked the higher one as a conservative approach when two or more values were given. We assumed the stars not present in the catalogue to have low or negligible activity. Nevertheless, we ensured that activity did not significantly affect the RV signal during the subsequent analysis. Specifically, we checked for all targets if there were any significant peaks in the periodogram of the activity indicator time series or any correlations between said indicators and RVs.

    \item Unevolved stars, meaning the ones with \logg $> 3.5$, to avoid red giant branch and other peculiar stars. This was done because of the large jitter that this kind of star has and the probable engulfment of the closest planets, potentially altering the statistics. The \logg values were taken from the Gaia DR3 catalogue \citep{gaiacatalog}. For some objects, the \logg value was not reported in the catalogue, for several possible reasons; therefore, we further removed a select number of stars based on their position on the HR diagram. In particular, we removed the stars that departed from the main sequence, keeping dwarfs and sub-giants.
    
    \item For binaries or other multiples, we kept only those with a separation larger than 5 arcsec to avoid the additional RV errors that may arise in the case of close visual companions due to the contamination of the spectra from the light of the companion \citep[e.g.][]{fiorenzano2005, cunha2013}.

    \item All the systems with an inner ($a < 1$ au) gas giant were treated separately as they have rather different architectures, and therefore experienced different evolution histories. A similar approach was taken in other works, such as \cite{barbato_b_2018}, \cite{bonomoetal2023}, and \cite{bonomo2025}.

    \item Since, as described in Section \ref{sec:intro}, planetary systems around M dwarfs display different properties compared to FGK stars, we restricted ourselves to the latter stellar types, setting $M_* > 0.6$ \msun.

    \item After the previous steps, we manually checked the remaining objects to find any other instances that needed to be removed but somehow escaped the selection process. In particular, we further removed the following objects: bet Pic (pulsating star), NGC 2682 Sand 978 (giant star), HD 62509 (giant star), bet Umi (giant star), and HD 6860 (giant star).

\end{enumerate}

\noindent After this selection process, we were left with 144 planetary systems. However, as detailed in Sect. \ref{sec:targets_removed}, we later removed 7 more targets before proceeding with the actual statistical analysis. \\

\noindent Although the bulk of our sample was selected from archival data to ensure statistical significance, an important contribution came from the GAPS (Global Architecture of Planetary Systems) collaboration through the intensive monitoring of a small (16 systems) but carefully chosen sub-sample of northern targets. These were selected for their known external gas giants and poor sensitivity to inner low-mass planets in existing datasets. The goal was to improve detection completeness for potential ISPs using observations with the high-resolution spectrograph HARPS-N \citep[High Accuracy Radial velocity Planet Searcher for the Northern hemisphere,][]{Cosentino2012} mounted at the Telescopio Nazionale Galileo. The GAPS observational programme led to the publication of a few new planets \citep{benatti2020, ruggieri2024b, ruggieri2025}; a detailed presentation of the GAPS sample, datasets, and analyses will be given in a separate work (Ruggieri et al., in prep.). By combining this focused observational effort with the archival data of the larger sample, we aim to derive $f_{\text{ISP|CJ}}$ and explore how planetary properties influence the architecture of planetary systems. \\

We point out that the selection criteria used in this work are the same as those used to build the GAPS subsample, as well as the sample of a similar programme performed from the southern hemisphere with HARPS \citep{barbato_b_2018}. However, in the GAPS programme, the systems were selected on a case-by-case basis based on which ones, at that time, no ISPs had been detected for, and only a few and/or low-quality RV data were available, leaving room for significant improvement with new HARPS-N observations.

\section{Data gathering}
\label{sec:data_gathering}
After selecting our targets, we searched for the RV data in the literature. When available, we also added unpublished HARPS-N data gathered in the framework of the GAPS programme and presented in Ruggieri et al. (in prep.). More specifically:

\begin{itemize}
    \item We used the data presented in \cite{rosenthal2021} for the instruments HIRES, HAMILTON, and APF. In a very small number of cases, this catalogue did not contain the HIRES data so we used the ones by \cite{Butler2017} instead. 

    \item For HARPS, we downloaded the data directly from the ESO archive\footnote{\url{https://archive.eso.org/scienceportal/home}} with RVs obtained from the instrument pipeline.

    \item We used the data in \cite{wittenmyer2009} gathered at the Hobby-Eberly Telescope at MacDonald observatory with the HRS instrument. 

    \item We used the DACE\footnote{\url{https://dace.unige.ch}} online platform as a source for data taken with the ELODIE and CORALIE instruments. 

    \item We downloaded the available data collected with the SOPHIE instrument directly from the archive\footnote{\url{http://atlas.obs-hp.fr/sophie/}}. 

    \item For other instruments, we resorted to the individual discovery papers of the respective targets, such as PFS, UCLES, ESPRESSO, MIKE, TULL, MINERVA, UVES, SARG, HJST, HJS, and HDS.  
    
\end{itemize}

\begin{figure*}[htbp]
\sidecaption
%\hspace{-1.5cm}
%\centering
\includegraphics[width = 12cm]{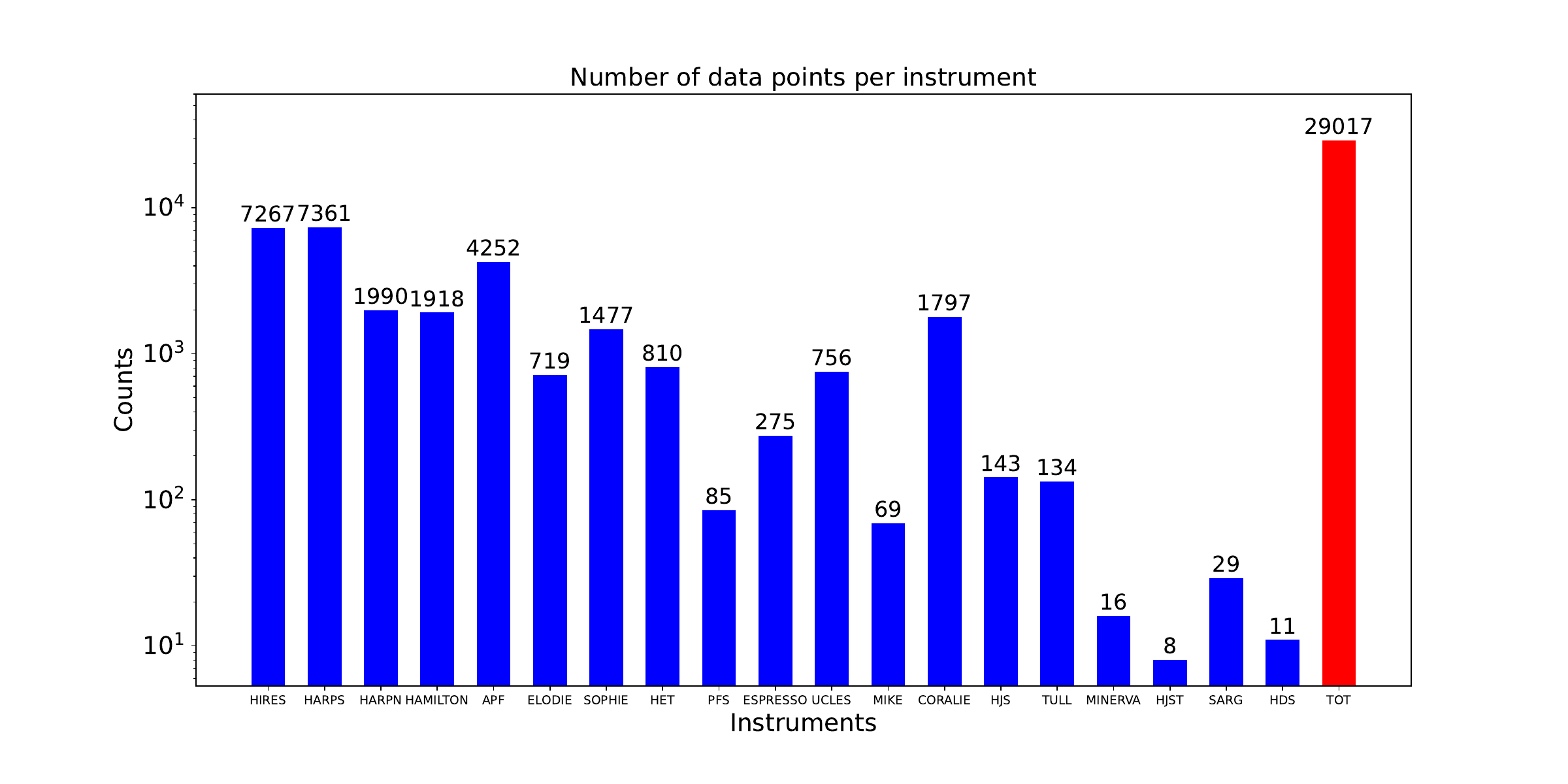}
\caption{Number of data points used in our RV analysis taken with each instrument. The last bar on the right represents the total.}
    \label{fig:histogram_data}
    \end{figure*}

\noindent We automatically removed the data points for which the uncertainty on the RV measure was larger than 10 m/s. This was done because we are interested in estimating our sensitivity to small planets; thus, an uncertainty this large is of little help in our analysis. However, there are some cases in which we kept even the data points with uncertainties larger than 10 m/s. The first one regards those objects for which the total number of data points was small, as the removal would have severely affected our results. In the second case, we kept these points when their inclusion was needed to constrain the orbit of a long-period planet better, significantly improving our phase coverage. For 125 targets (91\% of the total), the number of data points removed in this way is less than 10\%, and for 61 of those 125 targets, we removed no data at all. The median percentage of removed points is 0.5\%. The removed points mostly come from older and/or less precise spectrographs, such as ELODIE, CORALIE, or HAMILTON, which are less suitable for searching for small planets. Thus, overall, the number of removed points is low with respect to the total, and the impact on the sensitivity for low-mass planets, which is the main focus of our work, is expected to be very low. In addition, we removed the data points that appeared to be clear outliers. The total number of data points used for each instrument and target is available at the CDS via anonymous file transfer protocol (FTP) to \url{http://cdsarc.u-strasbg.fr/} (130.79.128.5) or via \url{http://cdsweb.u-strasbg.fr/cgi-bin/qcat?J/A+A/}. Figure \ref{fig:histogram_data} shows a histogram of the total number of data points used for each instrument. For each target, we calculated the median RV uncertainty and the median RV sampling; that is, the median time between two consecutive observations (considering all datasets). The former quantity ranges from 0.4 to 9.0 m/s for our sample, with a mean of 1.95 m/s and a median of 1.56 m/s. The latter varies from 0.003 to 105.27 days, with a mean of 13.17 d and a median of 6.00 d.

\section{Data analysis}
\label{sec:dataanalysis4}
Our analysis procedure was the same for all systems analysed. First of all, we used the \texttt{astropy} package to extract generalized Lomb-Scargle (GLS) periodograms of the target to start recovering the known planets. Next, we used PyORBIT \citep{Malavolta2016, Malavolta2018} to analyse our RV data with the optimization algorithm \texttt{PyDE} \citep{storn1997} and the affine invariant MCMC sampler \texttt{emcee} \citep{foreman2013}. For each planet, the code fits the data using a Keplerian curve with the following parameters: the RV semi-amplitude, $K$, the orbital period, $P$, the mean longitude, $L = \omega + M_0$ \citep[setting $\Omega=0$ as shown in Sect. 4.2 in][]{ford2006}, $\sqrt{e}\sin\omega$, and $\sqrt{e}\cos\omega$, where $\omega$ is the argument of periastron of the planet and $e$ the orbital eccentricity \citep[following the parametrization by][]{Eastman2013}. The semi-major axis and the minimum mass ($m\sin i$) were then derived using the stellar mass. In most cases, we used the default uniform priors set by PyORBIT over a finite region of the parameter space\footnote{See the appropriate documentation at \url{https://pyorbit.readthedocs.io/running_pyorbit/parameter_defaults/}}. For the targets described in the Appendix, we explicitly state when we set a different prior. We ran the algorithm for 200'000 steps, with a burn-in parameter of 50'000, and using four walkers for each fitted parameter. To verify convergence of the chains, we used the integrated autocorrelation analysis \citep{goodman2010}, the Gelman-Rubin diagnostic using a reference value of 1.01 \citep{gelman1992}, and visual inspection. In all cases, we set a different zero-point offset and jitter term for each spectrograph as free parameters, also accounting for upgrades of the instruments. Specifically, we used different offsets for HIRES \citep{Butler2017}, HARPS \citep{LoCurto2015}, SOPHIE \citep{sophie+}, and CORALIE \citep{coralie1, coralie2, coralie3}. Additionally, we studied the impact of the second HARPS upgrade during the COVID lockdown on 23 March 2020, as indicated on the ESO webpage\footnote{
\url{https://www.eso.org/sci/facilities/lasilla/instruments/harps/news.html}}, but found no significant offset between data taken before and after such date for our targets. In three cases, namely HD 3765, HD 7199, and HD 75898, we used Gaussian process (GP) regression to model stellar activity. Specifically, we used a quasi-periodic kernel, using the formulation by \cite{Grunblatt2015}. We set no prior on the GP parameters except for the amplitude, which we limited to being below 15 m/s in order to limit the activity contribution to the RV signal based on the results by \cite{lovis2011}. Long-term polynomial trends were included when already known in the literature or if clearly visible in the residuals. In the end, we extracted the GLS of the residuals to search for potential new candidates and performed an additional RV fitting when a significant signal was detected. Subsequently, we selected the best-fit model based on the Bayesian information criterion \citep[BIC,][]{schwarz1978} value and the selection criterion by \cite{kass1995}. Specifically, we used $\Delta\text{BIC} > 10$ as a threshold to determine whether a new planetary companion is likely present in the system. Finally, we used the equations derived in \cite{kipping2011} to estimate the parameters of companions causing long-term polynomial trends in RVs when applicable. All the resulting planetary parameters are made available at the CDS via anonymous FTP to \url{http://cdsarc.u-strasbg.fr/} (130.79.128.5) or via \url{http://cdsweb.u-strasbg.fr/cgi-bin/qcat?J/A+A/}. The results for stellar activity parameters and long-term polynomial terms are shown in Appendix \ref{sec:appendix_act_pol}. \\

\noindent To derive the occurrence rates of close-in small planets, we used two different codes that work differently to compare the results. The first code is an upgraded version of the one used in \cite{ruggieri2024b} for HD 75898 and HD 11506, as explained in Section 3.7.1 of that work. We injected planetary signal over a $100 \times 100$ logarithmic grid in orbital period and minimum mass, over the ranges $[1,400]$ d in period and $[3\,M_\oplus,0.1\,M_\text{J}]$ in mass. For each bin, we simulated ten Keplerian signals with a randomly generated mass and period within the bin, and a random uniform epoch of the periastron. Eccentricity was fixed as 0, since low-to-moderate eccentricity does not significantly affect signal detectability \citep[e.g.][]{pinamontietal2017} and high eccentricity is not expected for short-period low-mass planets.
Each generated planetary signal was injected into the RV residual time series after the removal of the known signals, and then the detection completeness of every period-mass bin, $C_i(\Delta_{P,M})$, where $i$ is the index corresponding to each star in the sample, was computed as the fraction of injected signals recovered with $\Delta \text{BIC}<-10$ with respect to a constant model. The global completeness of the sample can be computed simply as the average of the completeness for each target:
\begin{equation}
    C(\Delta_{P,M}) = {1 \over N} \sum_{i=0}^N C_i(\Delta_{P,M}).
\end{equation}
Given the completeness, $C$, the planetary occurrence rates, $f_\text{occ}$, can be computed from the number of detected planets, $n$, and the number of stars in the sample. This can be done in two different ways, following two definitions of planetary frequencies, i.e. either the number of planets per star or the number of stars with planets. In the first case we can derive $eta_\text{occ}$ inverting the Poisson distribution \citep[e.g.][]{pinamonti2022}:
\begin{equation}
\label{eq:poisson}
    \eta(n|s,f_\text{occ}) = {(s f_\text{occ})^n e^{-s f_\text{occ}} \over n!},
\end{equation}
where $n = n( \Delta_{P,M})$ is the number of detected planets in the chosen period-mass interval, and $s = s( \Delta_{P,M}) = S \cdot C(\Delta_{P,M})$ is the number of sensitive target, i.e. the number of stars in the sample, $S$, times the average completeness of the survey.\\
If the occurrence rate is computed as the number of stars with planets instead, it can be derived from the Binomial distribution \citep[][, and references therein]{barbato_b_2018}:
\begin{equation}
\label{eq:binom}
    F(n|s,f_\text{occ}) = {s! \over n! (s-n)!} f_\text{occ}^n (1-f_\text{occ})^{s-n},
\end{equation}
with the same notation as in Eq. \ref{eq:poisson}. All the planetary occurrence rates reported in the following sections were computed as planets per star (indicated with $\eta$), following Eq. \ref{eq:poisson}, with the exception of those discussed in Sect. \ref{sec:bonomo} in comparison with \cite{bonomoetal2023} and \cite{bonomo2025}, which were computed as the number of stars with planets (indicated with $F$). \\

The second code is an updated version of the one that was successfully applied in previous works \citep{barbato_b_2018,barbato2019,barbato2023,matthews2023,gratton2024}. First, we explored a grid of 10$\times$10 companion minimum masses and orbital periods ranging from 3 \mearth\ to 0.1 \mjup\ and from 1 to 400 days. For each realization of mass-separation, we generated 100 synthetic RV curves corresponding to the relevant companion with randomly drawn values of eccentricity, $e$, argument of periastron, $\omega$, and mean longitude, $\lambda_0$, injecting the synthetic signal in the RV post-fit residual time series. We then considered each of the $10^{4}$ injected signals as detected if the corresponding FAP value is $\leq10^{-3}$. Finally, for each mass-separation realization, we computed the detection completeness as the ratio between detectable injected signals and total injections, as done for the previous code, obtaining a detection completeness map for each target in the sample. By averaging all maps together, we produced a final full-sample detection map, from which we then computed the occurrence rate of planetary populations of interest by inverting the binomial distribution \citep[see e.g.][]{burgasser2003,sozzetti2009,faria2016,barbato_b_2018} defined in Eq.~\ref{eq:binom}. We computed the occurrence rate uncertainties corresponding to the central 68\% confidence interval of the integrated binomial probability function \citep[see e.g.][]{sozzetti2009,faria2016}. \\

A caveat that needs some discussion is the effect of orbital eccentricity on the analysis carried out in this work. As shown by \cite{pinamonti2017}, at increasing eccentricities, the GLS becomes less efficient at detecting planetary signals. Since we used the GLS to search for additional signals in our systems, this may affect our chance to detect them, leading to an underestimation of the occurrence rates. In our sample, there are 29 ISPs, and 25 of them have low ($e < 0.25$) eccentricities. The others are BD-114672 c ($e = 0.587$), HD 30669 c ($e = 0.496$), HD 39091 d ($e = 0.432$), and HD 181433 b ($e = 0.395$), indicating a tail at intermediate eccentricities for short- and medium-period planets. Nevertheless, the mean and median eccentricities of this group of planets are 0.152 and 0.092, respectively, confirming that, overall, ISPs tend to have small eccentricities \citep[as shown by e.g.][]{vaneylen2019}.

After analysing our sample with both codes, we compared the results to make sure that they agreed with each other. We find that the results obtained with the two codes agree at most within $2\sigma$ in all cases, and within $1\sigma$ in most cases; therefore, in the following, we only report the values obtained with the first tool for simplicity. Additionally, we repeated the analysis using the same mass and period ranges used in previous works to make a direct comparison. In particular, we refer to the results by \cite{rosenthal2022} for a blind RV survey, by \cite{barbato_b_2018} for a comparison with similar work on a smaller sample, and by \cite{bonomoetal2023, bonomo2025} as examples of complementary works (see Section \ref{sec:literature}). We would like to stress once again that, in this work, when we talk about occurrence rates, we are referring to the number of planets per star rather than the number of stars hosting a given type of planet. The only exception is our comparison with \cite{bonomoetal2023}, since the authors calculated the number of stars hosting planets, and so we did the same.

\section{Targets removed from the analysis}
\label{sec:targets_removed}
After downloading the available data for each target, we proceeded with our RV fitting procedure. However, we had to remove a few more objects that fulfilled our selection criteria but for which we could not carry out a complete and satisfactory analysis. These stars are:

\begin{itemize}
    \item HD 220773: The giant planet around this star has recently been dismissed by \cite{carleo2024b} as non-existent based on HARPS-N data obtained by the GAPS team.

    \item HD 43197: This star hosts two giant planets, of which one falls in our sample (planet c), while the other does not because it is located at $a \sim 0.9$ au. The external one has been announced by \cite{feng2022}, combining RV and astrometry, but we find that with RVs only (51 HARPS data), the orbit is very poorly constrained, so we removed this object.

    \item HD 56957: For this star, \cite{feng2022} announced a 6 \mjup\ object at 6.4 au combining RV and astrometry. Here, the number of data points is very low (4 from HIRES and 11 from HARPS), taken over almost 12 years. Therefore, a pure RV analysis is impossible. 

    \item HD 97037: Also for this star, a giant planet has been announced by \cite{feng2022} with astrometry + RV, with a period of $P \sim 6450 \pm 150$ d. As shown in the best-fit plot provided in the online material by \cite{feng2022}, the solution is driven by the offset between old and new HAMILTON data (Lick13 and Lick6 vs Lick8 using their nomenclature). The effect of this offset due to instrumental changes has been studied and quantified by \cite{fischer2014}, who found an average offset of about 13 m/s. On the contrary, the solution by \cite{feng2022} shows an offset of the order of 100 m/s, about one order of magnitude larger. Additionally, when performing an RV-only fit, convergence is not reached, and the resulting orbital parameters are significantly different compared to the literature result (much higher period and semi-amplitude). Thus, we conclude that retrieving HD 97037 b using only RV data (without astrometry) leads to inconclusive results and discard it from our target list.

    \item HD 113337: This is a mid-F known to host a $\sim 3$ \mjup\ planet at 1 au \citep{borgniet2014} and a 20 \mjup\ brown dwarf (BD) at 6.8 au \citep{feng2022}. However, as reported in the discovery paper of planet b, the star is only $150_{-50}^{+100}$ Myr old. Its young age is confirmed by the presence of a disk \citep{xuan2020}. Even though the star has a low chromospheric activity level (\logrhk $= -4.80$), this is not indicative of the stellar jitter for an object of this stellar type. In particular, if we try to fit the 246 SOPHIE and SOPHIE+ data points with a two-planet model, we obtain jitter terms of 39 and 51 m/s, respectively. We conclude that this star is not suitable for our purpose and thus discard it from our statistical analysis. 

    \item HD 175167: This G5 star hosts a giant planet with $m\sin i \sim 8$ \mjup\ and $P \sim 1290$ d discovered by \cite{arriagada2010} using 13 MIKE spectra taken over 5 yr. Since we have found no additional available data in the literature, we removed this object from our catalogue as the dataset is too small and sparse to provide any reasonable information regarding ISPs. 

    \item HD 217958: This G5 star has a 0.5 \mjup\ planet at 3.8 au discovered by \cite{feng2022}. This result has been obtained by combining UCLES, PFS, and astrometric data. However, the mentioned RV data have not been made publicly available anywhere, making the analysis impossible.
\end{itemize}

\noindent At this point, we are left with 219 planets around 137 stars. Of these, 161 planets around 114 stars make up our ‘regular’ sample\footnote{From now on, we refer to the ‘regular’ sample as that without inner giant planets.}, including 156 previously known planets and five new candidates found during our analysis (see Section \ref{sec:newresults}) and labeled as robust. Specifically, we have 82 single-planet systems, 24 systems with two planets, 5 systems with three planets, 1 system with four planets, 0 systems with five planets, and 2 systems with six planets. The other 23 stars host inner gas giants, including 17 two-planet systems, three have three planets, one has four, one has five, and one has six planets, for a total of 58 planets that we treated separately. Among these, there is one new candidate found during our analysis.

\begin{figure*}[htbp]
\sidecaption
%\hskip-1.4cm
   %\centering
   \includegraphics[width = 12cm]{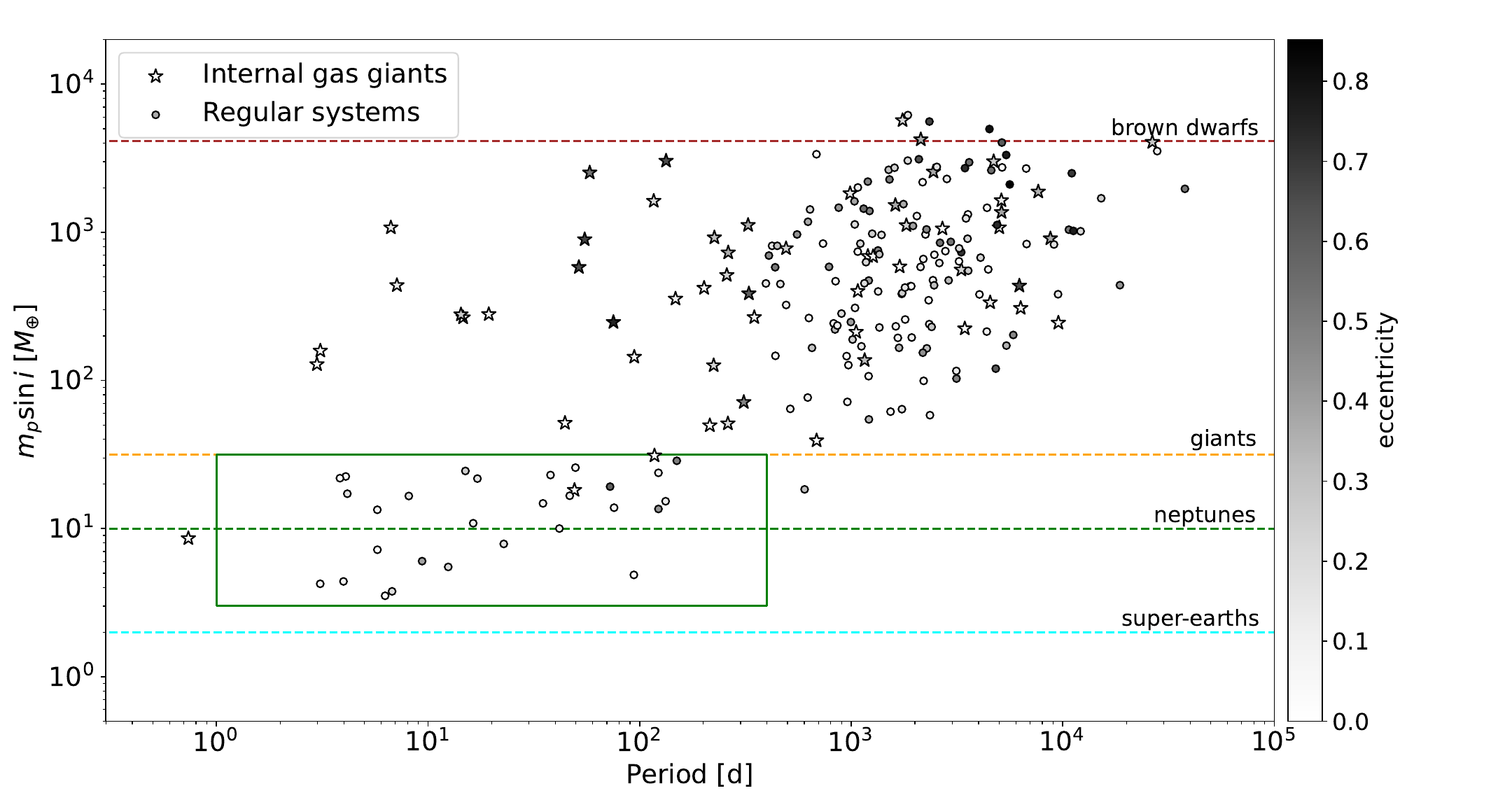}
      \caption{Minimum masses vs orbital period of all the planets in our global sample, colour-coded for eccentricity. The green box represents the region of the parameter space that we considered for our occurrence rates analysis (see Section \ref{sec:statistics}).
              }
    \label{fig:full_plot}
    \end{figure*}

\noindent A list of all our target stars, with their main physical parameters, and a list of all our planets, with their masses and orbital parameters, are made available at the CDS. Figure \ref{fig:full_plot} shows the minimum masses and orbital periods of all the planets in our sample, with the colour bar on the right representing eccentricity values. 

\section{Characteristics of our sample}
\label{sec:sample_characteristics}
\subsection{Host stars}
\label{sec:central_stars}
\begin{figure}[htbp]
\centering
\includegraphics[width=\linewidth]{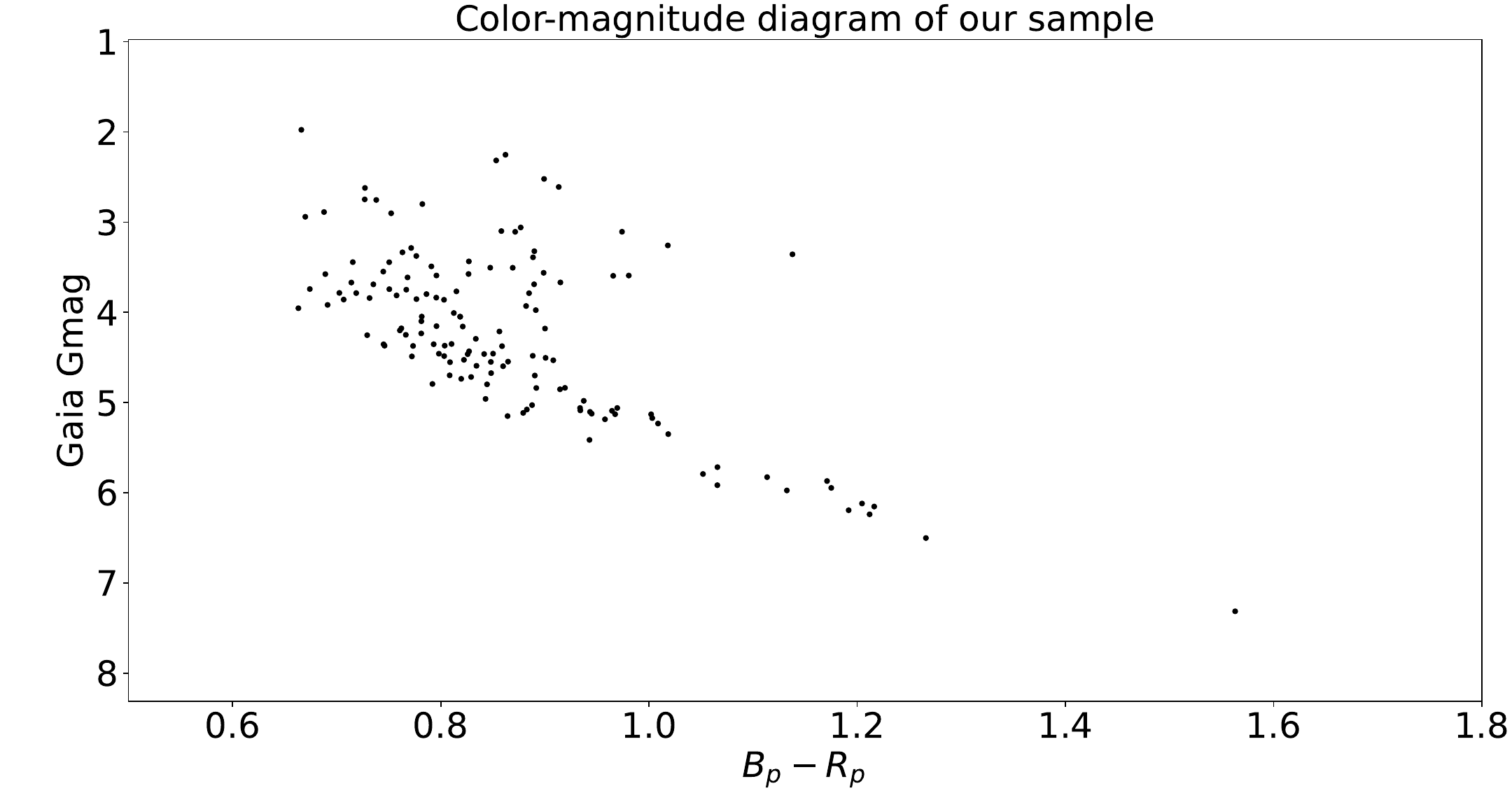}
\includegraphics[width=\linewidth]{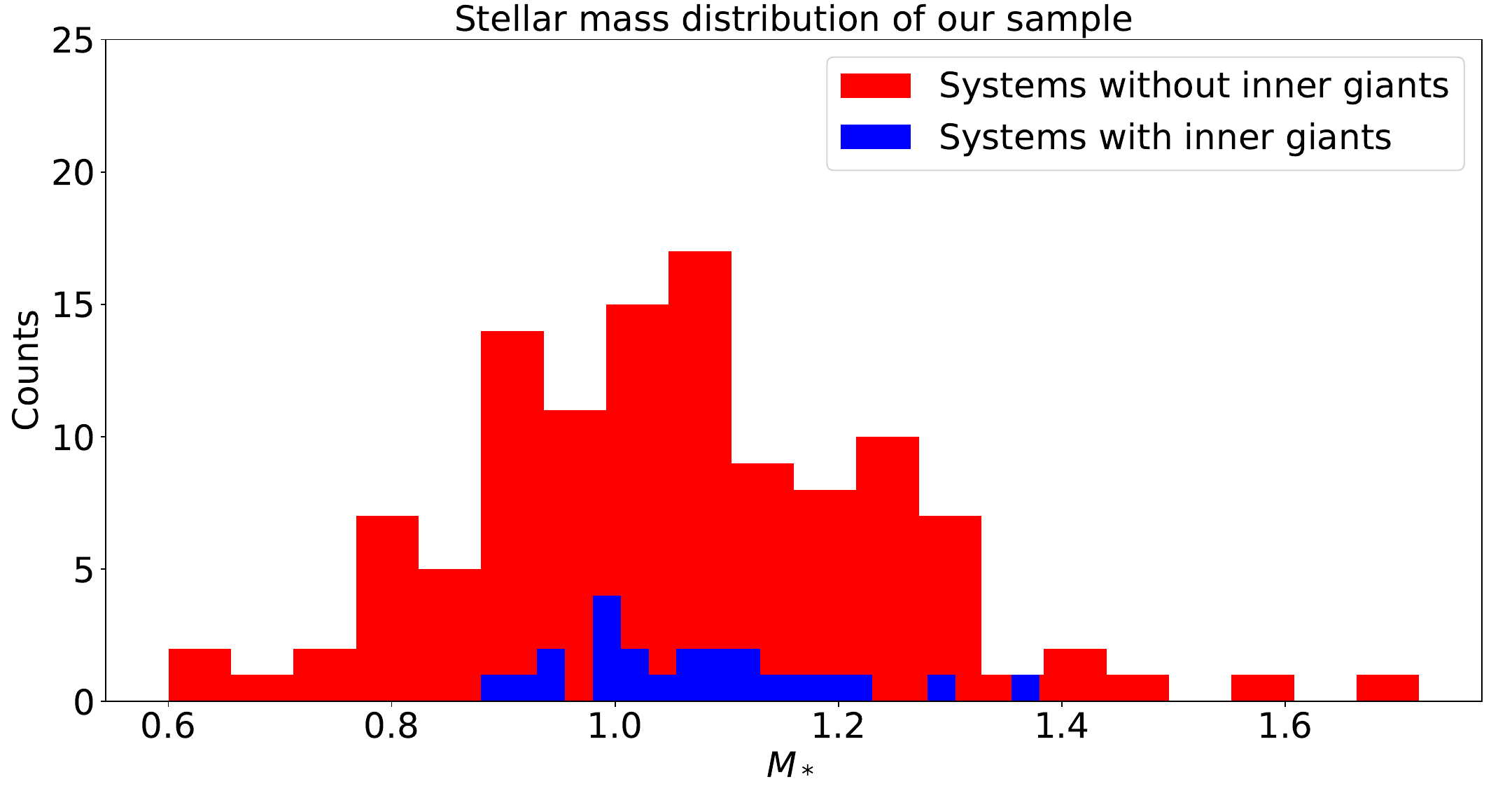}
\includegraphics[width=\linewidth]{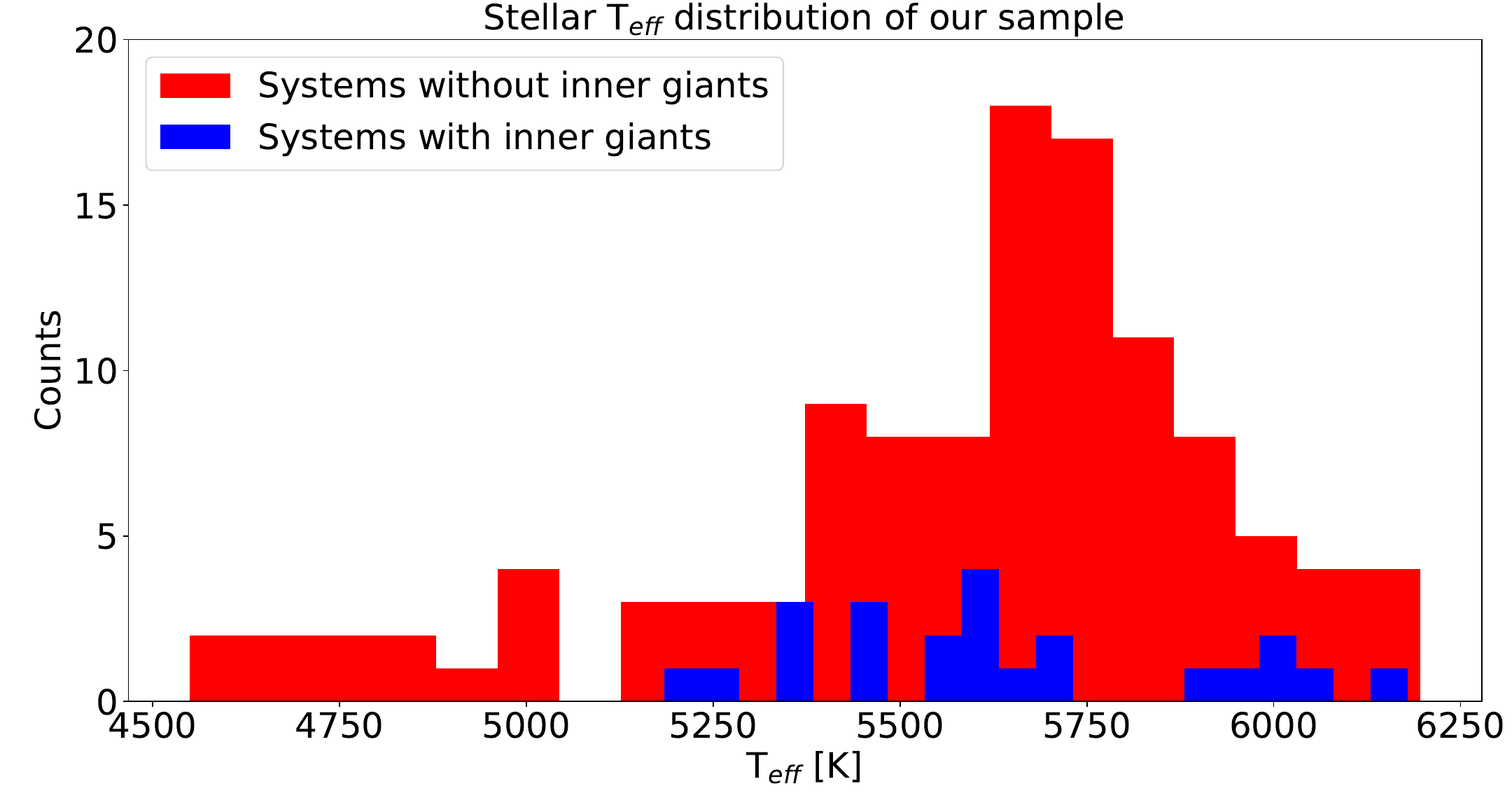}
        \caption{Colour-magnitude diagram (top), mass (centre), and \teff\ (bottom) distribution of the host stars in our sample. The $G$ and $B_p - R_p$ values have been taken from the \gaia\ DR3 catalogue \citep{gaiacatalog}.}
        \label{fig:sample_data}
\end{figure}

As mentioned in the last section, our sample includes 138 stars, of which 23 host inner gas giants, in addition to the CJ, while the other 114 do not. Figure \ref{fig:sample_data} shows the HR diagram (obtained using \gaia\ DR3 $G$ magnitude and $B_p-R_p$ colour) of our sample, the distribution of the stellar masses (taken from the most recent work for each target), and the \teff\ distribution (temperatures from the \gaia\ DR3 catalogue).
%As we can see, our sample is rather homogeneous, except for a couple of slightly evolved, more massive stars, since we removed two M dwarfs that satisfied our selection criteria as explained before. We did so to focus on a sample made up exclusively of Solar-like stars. 
Overall, our sample well represents the types of stars around which planets are typically searched with the RV method, except for surveys that target M dwarfs specifically. The mean and median masses of the regular sample are 1.07 and 1.06 \msun, respectively (indicating a quite symmetrical distribution), while for systems with inner giants, these are 1.11 and 1.08 \msun. Overall, the mean and median metallicities of our sample are -0.08 and -0.06, respectively.

\subsection{Planets}
\begin{figure*}[htbp]%\vspace*{-2.6cm}
\centering
\includegraphics[width=0.49\textwidth]{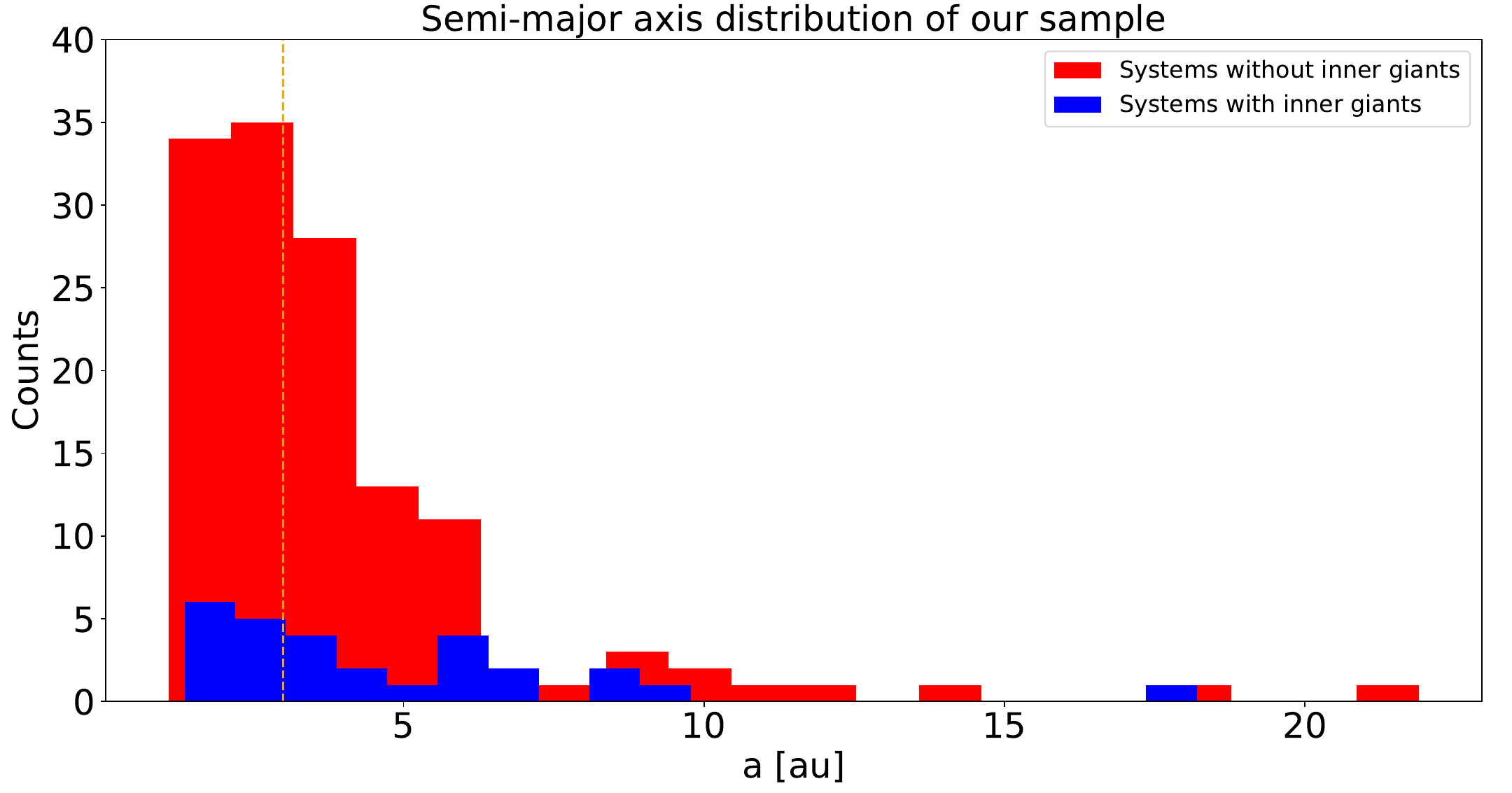}
\includegraphics[width=0.49\textwidth]{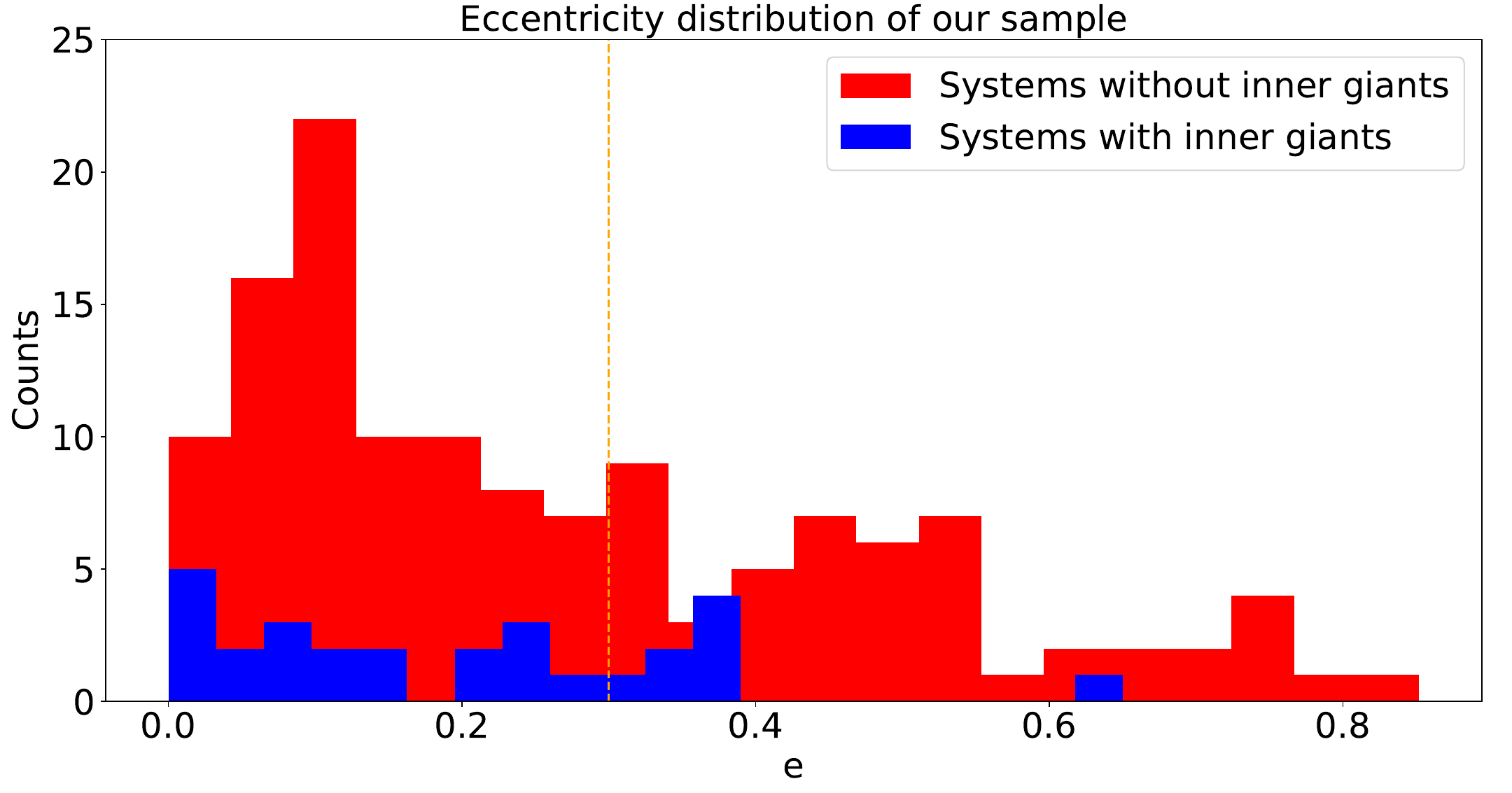}
\includegraphics[width=0.49\textwidth]{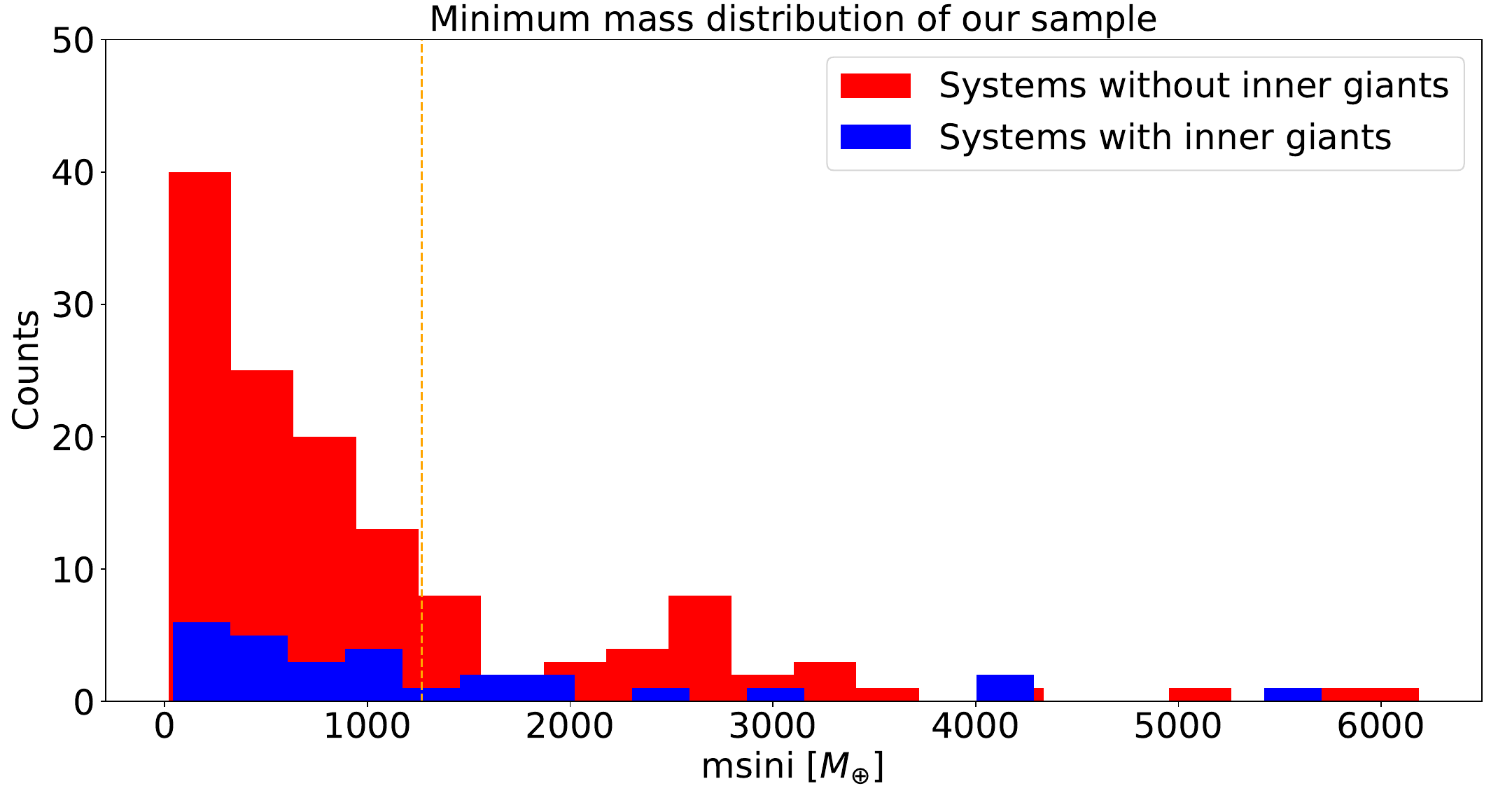}
\includegraphics[width = 0.49\textwidth]{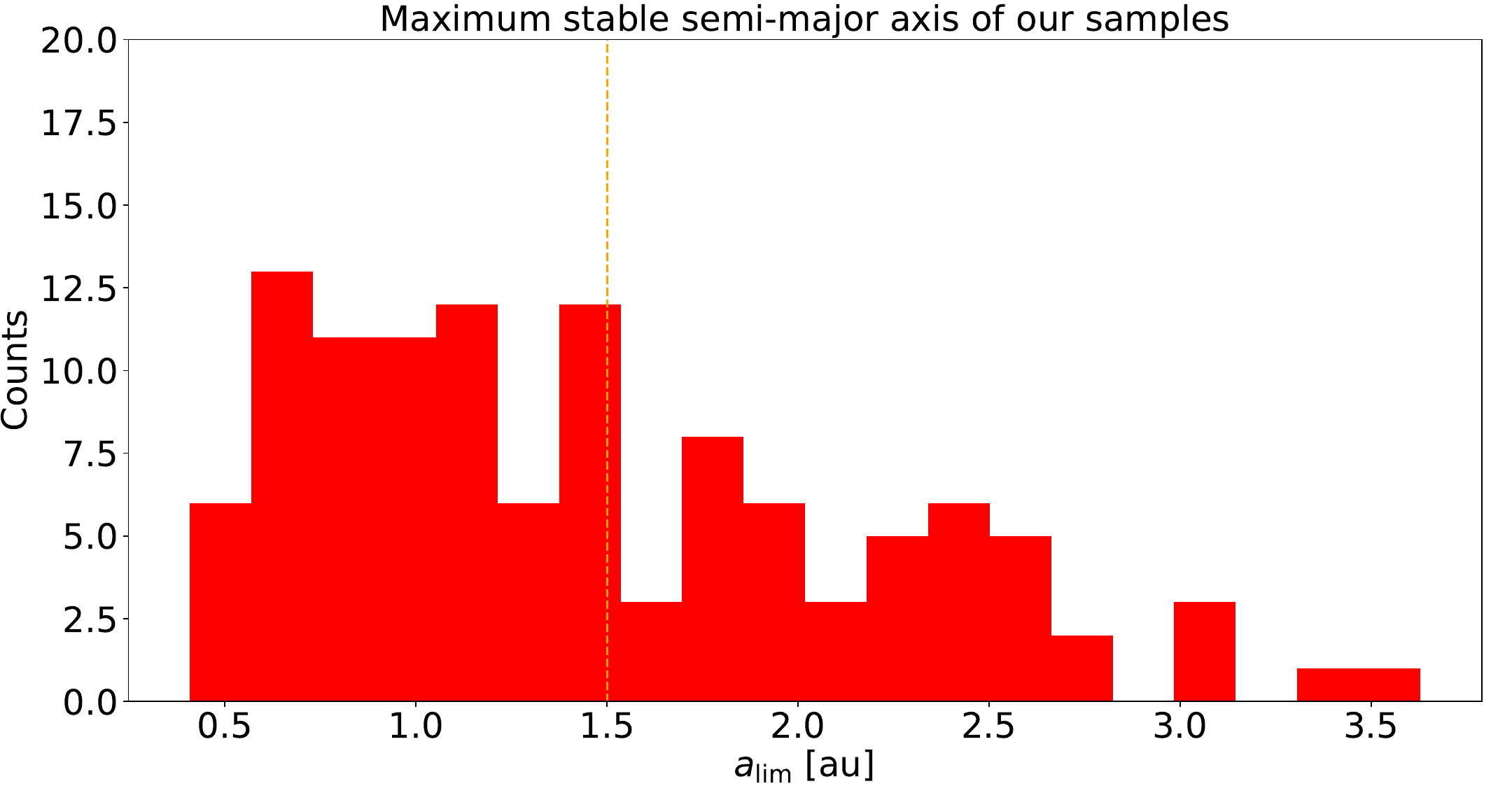}
\caption{Distributions of (from top left to bottom right) CJ semi-major axis, eccentricity, $m \sin i$, and maximum stable orbital radius, $a_{lim}$ (as defined in Sect. \ref{sec:dynamic_influence}), for our sample as obtained from our RV results. The dashed vertical lines ($a = 3$ au, $m\sin i = 4$ \mjup, $e = 0.3$, and $a_{lim} = 1.5$ au) represent the boundaries used to test the properties of different sub-samples, as detailed in Section \ref{sec:dynamic_influence}.}
\label{fig:planets_data}
\end{figure*}
After the selection process described in the previous sections, our list of planets comprises 161 objects in our regular sample (of which 5 are new candidates, as detailed in the next section), and 58 planets found in systems with inner giants (one of which is a new candidate). The last group includes all objects belonging to systems with at least one inner giant, so, for instance, 55 Cnc e is included in it because it is a small planet ($m\sin i = 8.6$ \mearth), but the system does have other inner giants. Figure \ref{fig:planets_data} shows the distributions of semi-major axis, eccentricity, and minimum mass for our CJs, distinguishing between the regular systems and those with inner giants. As expected for RV‑selected CJs, the semi‑major axis distributions for both the regular systems and those with inner giants peak towards smaller separations and decrease at larger values of $a$, with only 7 planets with $a > 10$ au over 137 systems, about 5\%. On the one hand, this is because these are typically hard to study with RVs alone, except for systems that have been regularly monitored since the 1990s (e.g. with HIRES). Our sample selection further accentuates the paucity of objects on very wide orbits since, by construction, we excluded systems in which the closest CJ lies at $a > 10$ au (Section \ref{sec:sample_selection}), so the few objects we list with $a > 10$ au are additional giants in multi‑giant systems rather than representatives of the field‑star distribution at those separations. Even so, this is consistent with both the demographics of RV-systems \citep[e.g. ][]{cumming2008,wittenmyer2016,fernandes2019} and constraints from direct imaging \citep[e.g. ][]{bowler2018,nielsen2019}. Once again, we emphasize that the $\sim 5\%$ fraction of $a > 10$ au planets in our list refers to a CJ‑host sample and should not be interpreted as a field‑star occurrence rate. In terms of eccentricity, we see a spike for the regular systems around 0.1, but there is a long tail up to eccentricities as high as 0.8, and the systems with inner giants show similar behaviour. As far as the minimum masses are concerned, we can see the peak at low masses ($m\sin i \lesssim 1$ \mjup), indicating that, in addition to the small inner companions, there are also a lot of objects with masses lower than that of Jupiter.

\subsection{New results}
\label{sec:newresults}
For most of the objects analysed, the RV fitting procedure did not yield additional results beyond what was found in the literature, confirming the known planets. However, in some cases, we have evidence for long-term trends, potential short-period candidates, activity cycles, or doubts about some literature planets. A detailed analysis of these systems is found in the appendix. We implicitly confirm the literature results for all the targets we do not mention. To summarize, we found evidence for new candidates in the following systems: HD 3765, HD 204941, HD 30669, HD 170469, HD 103891, HD 13908, HD 10697, and HD 136925. Of these, HD 170469 and HD 204941 are the most dubious cases, and thus we did not consider them in our statistical analysis. In addition, we added polynomial trends for HD 24040, HD 163607, HD 73267, HD 191806, and HD 106270, indicating the presence of additional objects with very long periods. We also derived the presence of an activity cycle affecting RVs for HD 7199. For HD 156098 and HD 47186, we found orbital solutions that are quite discrepant compared to the literature. Additionally, we confirm the new planet HD 23079 c announced very recently by \cite{delisle2025}, but we found no convincing evidence for the new planet HD 196067 c announced in the same work (possibly due to the use of a partly different dataset). Finally, we found a slight disagreement on the orbital parameters of HD 204941 b with respect to \cite{Dumusque2011}. Section \ref{sec:appendix_candidates} in the appendix contains a detailed analysis of our new candidates, for which we summarize the main findings in the following: \\

\noindent \paragraph{HD 3765}: We found evidence for a new cool Neptune with $P = 132.71 \pm 0.19$ d, $e = 0.114_{-0.080}^{+0.120}$, $K = 2.16 \pm 0.28$ m/s, $\omega = -72_{-68}^{+193}$ degrees, and $m \sin i = 15.3 \pm 2.0$ \mearth. \\

\noindent \paragraph{HD 204941}: Even though the evidence in favour of this new candidate is not particularly convincing, we signal that a new planet might be present with $P = 28.729 \pm 0.009$ d, $e = 0.133_{-0.092}^{+0.13}$, $K = 1.66 \pm 0.28$ m/s, $\omega = 10_{-88}^{+74}$ degrees, and $m\sin i = 5.4 \pm 1.1$ \mearth. Additionally, we found values in slight disagreement with \cite{Dumusque2011} for planet b; that is, $K = 4.6 \pm 0.3$ m/s, $P = 1529 \pm 13$ d, $e = 0.160_{-0.064}^{+0.069}$, $\omega = -38_{-29}^{+21}$ degrees, and $m\sin i = 56.1 \pm 6.5$ \mearth. \\

\noindent \paragraph{HD 30669}: We found evidence for an eccentric cool Neptune with $K = 4.2 \pm 0.5$ m/s, $P = 149.9 \pm 0.2$ d, $e = 0.496_{-0.097}^{+0.088}$, $\omega = -62 \pm 19$ degrees, and $m \sin i = 28.7 \pm 3.5$ \mearth. \\

\noindent \paragraph{HD 170469}: As for HD 204941, this new candidate does not have sufficient evidence to be labeled as robust, but we report that its likely parameters are $P = 8.2676 \pm 0.0015$ d, $K = 4.2 \pm 1.1$ m/s, $e = 0.19_{-0.13}^{+0.26}$, $\omega = 30_{-138}^{+65}$ degrees, and $m \sin i = 13.8 \pm 3.2$ \mearth. \\

\noindent \paragraph{HD 103891}: We found a new cool giant planet in a 2:1 mean motion resonance with planet b, whose parameters are $P = 951.5 \pm 5.1$ d, $K = 8.07 \pm 0.60$ m/s, and $m \sin i = 0.460 \pm 0.034$ \mjup. Additionally, we have evidence that planet b is on a circular orbit. \\

\noindent \paragraph{HD 13908}: We found evidence for a long-period and moderately eccentric super-Jupiter with $K = 55.1_{-6.1}^{+6.9}$ m/s, $P = 7661_{-1481}^{+1507}$ d, $e = 0.347_{-0.12}^{+0.087}$, $\omega = 78_{-14}^{+12}$ degrees, and $m\sin i = 5.92_{-0.76}^{+0.84}$ \mjup. \\

\noindent \paragraph{HD 10697}: We found evidence for a new hot Neptune with $K = 5.03 \pm 0.78$ m/s, $P = 8.1127 \pm 0.0006$ d, $e = 0.18_{-0.12}^{+0.15}$, $\omega = 109_{-263}^{+52}$ degrees, and $m\sin i = 16.6 \pm 2.6$ \mearth. \\

\noindent \paragraph{HD 136925}: We found evidence for a cool eccentric Saturn-like planet with $K = 6.94_{-0.85}^{+0.96}$ m/s, $P = 310.76_{-0.48}^{+0.61}$ d, $e = 0.48_{-0.15}^{+0.13}$, $\omega = 64_{-72}^{+49}$, and $m\sin i = 71.4_{-8.7}^{+9.2}$ \mearth. \\

\section{Statistical analysis of our sample}
\label{sec:statistics}

\subsection{Detection maps}
\label{sec:general_results}
\begin{figure}[htbp]%\vspace*{-3cm}
\centering
\includegraphics[width=\linewidth]{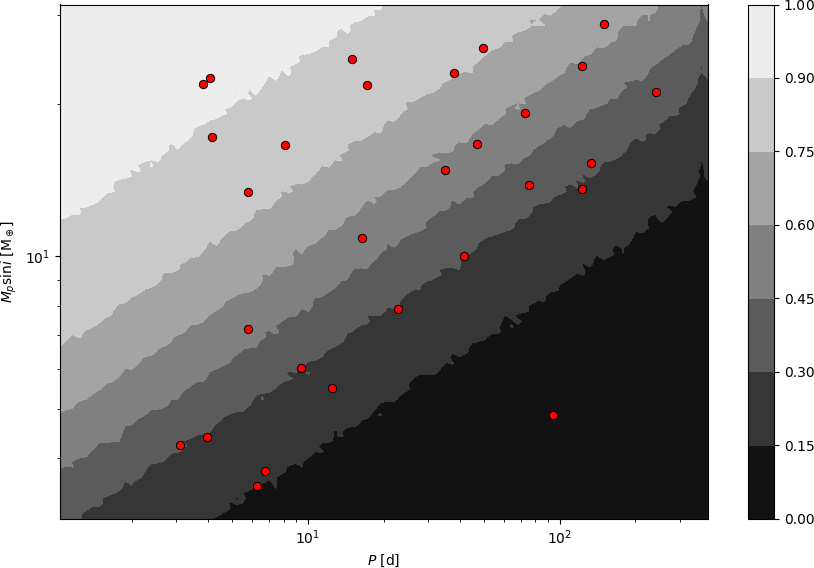}
\caption{Detection map of our sample (excluding system with inner giant planets). Red dots represent the known planets, including the new candidates discussed in Section \ref{sec:newresults}.}
        \label{fig:fullsample_rvdetection}
\end{figure}
After fitting the RV data for all the systems in our sample, we proceeded with our statistical analysis as described in Section \ref{sec:dataanalysis4}. As we said, we treated the 23 systems with at least one internal giant planet (i.e. $m\sin i > 0.1$ \mjup\ and $a < 1$ au) separately, focusing at first on the 114 systems without inner giants, but we included them for further analysis later on (see Section \ref{sec:jups}). More precisely, since this work focuses on the presence of ISPs, we consider the following region of the parameter space. In terms of mass, we consider objects from 3 \mearth\ to 0.1 \mjup, since the latter was our boundary when selecting the giant planets of the sample. We use 3 \mearth\ as a lower limit because our global sensitivity at lower masses is practically zero. In terms of period, we limited ourselves to the range from 1 to 400 d as the latter corresponds to a bit more than 1 au for Sun-like stars, and this was our inner boundary to define ‘external’ planets. Figure \ref{fig:fullsample_rvdetection} shows the completeness map obtained for our sample. In the figure, each point of the explored parameter space is colour-coded according to the sample detection completeness computed as described in Sect.~\ref{sec:dataanalysis4}, ranging from 0\% (no signal detectable) to 100\% (all investigated signals are detectable).

\noindent We only point out that the first code, for which we report the results, gives on average higher completenesses at low masses and vice versa compared to the second code, leading to minor differences in the results. The two always agree at less than $2\sigma$, but our explanation is that the second code calculates the FAPs for the detections analytically, leading to an overestimation of the potential detections \citep[e.g.][]{pinamonti2017, VanderPlas2018}. There can be large differences between one target and another depending on their architectures, the amount of available data, and their sampling. In any case, we see that, overall, our sample displays a good degree of completeness ($\gtrsim 45\%$) at least for Neptune-mass ($\sim 17$ \mearth) planets up until $P \sim 100$ d and up to 20-30 d for 10-\mearth\ objects.
\begin{table}[htbp]
\centering
\caption{Planetary occurrence rates of our sample.}
\label{tab:occ_t1}
\begin{tabular}{c c c c}
\hline
\hline
 \\
  &  & Period [d] &  \\
\\
$m_p \sin i$ [\mearth] & [1, 10] & ]10, 100] & ]100, 400] \\
\\
\hline
\\\relax
 ]10, 31.7] & $4.9_{-1.3}^{+3.4}$ & $13.7_{-3.1}^{+5.8}$ & $10.1_{-2.8}^{+8.0}$ \\
   & & & \\
  & $n = 5$ & $n = 10$  &  $n = 5$ \\
  & & & \\
  & $C = 88.7\%$ & $C = 64.2\%$  &  $C = 34.7\%$ \\
\\
\hline
\\\relax
 [3, 10] & $11.5_{-2.9}^{+6.9}$ & $^*16.5_{-4.9}^{+16.0}$ & - \\
   & & & \\
 & $n = 6$ &  $n = 3$ & $n = 0$ \\
   & & & \\
 & $C = 45.7\%$ &  $C = 15.9\%$ & $C = 4.80\%$  \\
\\
\hline
\end{tabular}
\tablefoot{Number of planets per star ($\eta$) expressed as percentages for the different types of planets in our sample. The systems with internal gas giants are excluded. In all cases, $n$ represents the number of detections, and $C$ is the average completeness in the given region of the parameter space. All uncertainties correspond to the $1\sigma$ confidence intervals.\\
*Due to low completeness, this number should be taken with caution.}
\end{table}

\noindent To better compare our occurrence rates, we divided the explored region of the parameter space into six parts. We defined ‘hot’ planets as ones with $1 \leq P < 10$ d, ‘warm’ those with $10 \leq P < 100$ d, and ‘cool’ those with $100 \leq P \leq 400$ d. Then, we separated SEs and Neptunes using $m\sin i = 10$ \mearth\ as a boundary between the two, as usually done in the literature \citep[e.g.][]{barbato_b_2018}. All the occurrence rates, detections, and average completeness for each part of the parameter space are shown in Table \ref{tab:occ_t1}. We point out that we have no detections of cool SEs and extremely low completeness in this region. Thus, we did not calculate the expected occurrence rate for these objects, as the result would be unreliable. \\

\noindent As we can see, we have 20 Neptunians but only 9 SEs, as these are harder to detect. In particular, these SEs are found in the following systems: HD 23079 (94 spectra over 25 years but concentrated in a few seasons), HD 39091 (782 spectra over 23 yr), HD 86226 (275 spectra over 24 yr), HD 164922 (1033 spectra over 17.6 yr), HD 181433 (210 spectra over 19.3 yr), and HD 219134 (1268 spectra over 27 yr), the latter hosting four SEs as well as a warm Neptunian. In practice, the objects around which these SEs have been found are among the best-studied exoplanetary systems in our sample and in general, and their very intensive monitoring may be due to case-by-case choices (e.g. linked to the presence of additional planets), possibly introducing a bias in the results. Unfortunately, these are not representative of the whole sample, in which there are also systems with only 20-30 sparse data points for which it is impossible to see warm SEs even if they were present. This is exemplified well by the fact that the completeness in this part of the parameter space is very low, despite a few systems that have been greatly monitored for one reason or another. In practice, these few systems alter and bias the number of SEs that we can expect from our sample; therefore, the values reported in Table \ref{tab:occ_t1} should be taken cautiously. To quantify the effect of the poorly sampled systems on our sample, we repeated the analysis, removing all the systems with fewer than 50 RV measurements. As expected, the results are compatible with those obtained from the full sample at $0.1\sigma$ or less, since stars with low completeness are appropriately weighted when calculating the mean detection map. \\

\noindent For the sake of completeness, we recalculated the occurrence rates excluding our new candidate planets described in Section \ref{sec:newresults}. The only differences are found for hot and cool Neptunes, but the results agree with those including the candidates at $0.3\sigma$ and $0.7\sigma$, respectively. Therefore, from now on, we always include the new candidates in the next calculations.

\subsection{Dynamic influence of external giants}
\label{sec:dynamic_influence}
As described in the previous sections, this analysis aims to see whether Solar System analogues are common; that is, if ISPs are common around stars hosting an external gas giant. In particular, in the Solar System, Jupiter has a mass more than twice that of all other planets combined, meaning that it is by far the dominant object after the Sun. Today, Jupiter's influence in shaping the Solar System during its evolution is very well known \citep[see e.g.][and references therein]{feng2024}, and it is clear that, together with Saturn, it causes stability in the inner regions of our Solar System. Therefore, we decided to investigate the dynamic influence that the gas giants of our sample have on the respective inner regions. To this aim, as shown by the dashed lines in Figure \ref{fig:planets_data}, we decided to split the sample (without inner giants) into two using different thresholds for different parameters. For eccentricity, we chose 0.3 as a reasonable threshold between low and moderate-to-high eccentricities and because there seems to be some sort of minimum in the distribution, indicating that there might indeed be two classes of systems with different properties. As far as the minimum mass is concerned, the distribution strongly drops for $m\sin i \gtrsim 4$ \mjup, which is a value typically used to distinguish between Jupiter-like objects and the so-called ‘super-Jupiters’ \citep{stevens2013, perryman2018}. For this reason, we also used this value as a division to test the properties of two different sub-samples. Similarly, we divided our sample based on the semi-major axis of the CJ using $a = 3$ au as a limit, since at higher values the distribution decreases (this is, at least partly, due to observational biases). We found these sub-samples to display different properties, but only for warm planets. In particular, we find that 8/10 warm Neptunes and 3/3 warm SEs are found in systems where the outer giant has a low ($< 0.3$) eccentricity. Similarly, 9/10 warm Neptunes and 3/3 SEs have an outer companion with $m\sin i < 4$ \mjup. Therefore, we decided to make a further division following the same approach taken by \cite{barbato_b_2018}: we calculated the Hill radius of the external giant planet for each system using

\begin{equation}
R_H = a_1 (1-e_1) \sqrt[3]{\frac{m_1}{3M_s}}
,\end{equation}

\noindent where $a_1$ is the semi-major axis, $e_1$ the eccentricity, $m_1$ the $m\sin i$, and $M_s$ the stellar mass. For the systems with more than one outer gas giant, we used the one with a smaller orbital radius. We are aware that the interactions between the outer giants may cause additional zones of dynamical instability in the inner regions of the system but a case-by-case investigation would be beyond the scope of the work. From the Hill radius, we calculated the maximum orbital distance at which an inner companion can remain stable as

\begin{equation}
a_{lim} = a_1(1-e_1) - 2\sqrt{3}R_H,
\end{equation}

\noindent as shown in, for example, \cite{murray1999}. We found this to be a good parameter to investigate the dynamic influence of outer giants because it contains information on all the key parameters that affect dynamic interactions; that is, the semi-major axis, eccentricity, and mass. After calculating this quantity for every system, we separated the systems with $a_{lim} < 1.5$ au (68 stars) from those with $a_{lim} \geq 1.5$ au (47 stars) and recalculated the occurrence rates. We used the value of 1.5 au because our selection criterion for outer giants was periastron $> 1$ au, so, accounting for some eccentricity, this roughly gives us $a > 1.5$ au. In practice, the systems for which $a_{lim} < 1.5$ au have less stable inner regions compared to the other sub-sample due to perturbations caused by their CJ, either because this is very massive, relatively close, or has an eccentric orbit (or even a combination of the three). The bottom right panel of Figure \ref{fig:planets_data} shows the distribution of $a_{lim}$ for our sample. 
    
\noindent Once again, for hot planets, whether Neptunes or SEs, we find no significant difference in the two cases, likely because the objects found here are very close to the central star and are therefore safer from their outer companions. However, we found that all 13 warm planets are located in systems with $a_{lim} \geq 1.5$ au, despite this being the smallest subsample of the two. Quantitatively, for warm SEs, we find an occurrence rate of $13.3_{-3.9}^{+13.0}\%$ when $a_{lim} \geq 1.5$ au and $< 7.6\%$ in the opposite case, with a significance of only $1.6\sigma$ due to the large uncertainty (caused by a low completeness). However, for warm Neptunes we find an occurrence rate of $12.3_{-2.7}^{+5.3}\%$ when $a_{lim} \geq 1.5$ au and $< 1.8\%$ in the opposite case, the significance being $3.8\sigma$. This result is not completely unexpected because, as we said, the selection based on the value of $a_{lim}$ accounts for the dynamic stability of the inner regions, but being able to robustly quantify this effect is very important. This is in agreement with the results by \cite{rosenthal2022}, who found that, at a $1\sigma$ level, the presence of an outer giant favours the formation of inner smaller planets, unless the outer giant is located between 0.3 and 3 au. This is qualitatively similar to our analysis based on $a_{lim}$, whose value can indeed be low if the outer companion is relatively close. Additionally, several works found that eccentric cold giants can inhibit the formation of close-in rocky planets, in agreement with our observational results. \cite{agnew2017} examined the dynamical stability of Earth-sized planets in the habitable zone \citep{kopparapu2013,kopparapu2014}, finding that a gas giant on a wide or low-eccentric orbit permits the existence of such inner companions. We also point out that some of their most stable systems (labeled ‘blue resonant’ systems by the authors) are in our sample. \cite{bitsch2020} simulated different formation scenarios, finding that those that best represent the observed eccentricity distribution of giant planets also experience more scattering events, leading to a low number of SEs at orbital separations lower than about 1 au. However, they found that less massive CJs on nearly circular orbits have a less disruptive effect. A similar result, based on the Generation 3 Bern model, is found by \cite{schlecker2021}, who determined that, even though inner SEs and CJs are correlated, dynamically hot external giants inhibit the formation of warm ($a \lesssim 0.3$ au) SEs. They also found an architecture-composition link, with high-density SEs more likely to be found in systems with a CJ.

\subsection{Comparison with literature values}
\label{sec:literature}
As previously stated, we repeated our previous analysis using the same regions of the parameter space of literature works to make more straightforward comparisons. All our results are shown in Table \ref{tab:occ_lit}, where the first row indicates which work we are referring to below. In the following sections, we describe these results, comparing them with the ones obtained by the mentioned authors. 
\begin{table*}[htbp]
\centering
\caption{Comparison of ISP planetary occurrence rates in our sample and literature works.}
\begin{tabular}{c c c c}
\hline
\hline
 \\
 & \cite{barbato_b_2018} & \cite{bonomo2025} & \cite{rosenthal2022} \\ 
 \\
%  & P [d] & & P [d] & & P [d] & & P [d] \\
%\\
%$m_p \sin i$ [\mearth] & P < 150 d & $m_p \sin i$ [\mearth] & P < 100 d & $m_p \sin i$ [\mearth] & P < 50 d & $m_p \sin i$ [\mearth] & [1.2, 365.25] d \\
 & $m\sin i \in [10, 30]$ \mearth, P < 150 d & $m\sin i \in [3, 20]$ \mearth, P < 100 d & $m\sin i \in [3, 30]$ \mearth, P $\in $ [1.2, 365.25] d \\
\\
\hline
\\\relax
This work & $\eta = 22.9_{-4.1}^{+6.5}\%$ & $F = 24.1_{-5.1}^{+7.3}\%$ &  $\eta =55.6_{-8.6}^{+12.5}\%$ \\
\\
 & $n = 19$, $C = 73\%$ & $n = 11$, $C = 45\%$ & $n = 29$, $C = 44\%$ \\
\\
% & $C = 73\%$ & $C = 45\%$ & $C = 44\%$ \\
%\\
Literature & $< 9.84\%$ & $32 \pm 11\%$ & $69 \pm 19\%$ \\
\\
\hline
\end{tabular}
\tablefoot{Number of planets per star ($\eta$), except for the central column, where the fraction of stars with at least one planet ($F$) is reported, expressed as percentages obtained in this work, adapted to the parameter space regions explored by previous works, as highlighted on top. The systems with internal gas giants are excluded. In all cases, $n$ represents the number of detected planets and $C$ is the average completeness in the given region of the parameter space. The last row reports the literature value against which we compare. All uncertainties correspond to the $1\sigma$ confidence intervals.}
\label{tab:occ_lit}
\end{table*}

\subsubsection{Results from HIRES survey}
We now compare our results with those obtained by \cite{rosenthal2022}, based on CLS data. Their sample size is 719 stars (including some M dwarfs), of which 27 ($\sim20\%$ of the total) belong to our regular sample. They considered planets between 0.02 and 1 au, so, considering that both our and their samples are made up of FGK stars, we translated that to the period range [1.2, 365.25] d\footnote{We used the period instead of semi-major axis for consistency with our results discussed in previous sections}. The mass range goes from 3 to 30 \mearth. We point out that \cite{rosenthal2022} defined CJs in a partly different manner compared to us. The mass regime is the same, but they include giant planets between 0.23 and 1 au. However, as found by many authors \citep[e.g.][]{petigura2018,wittenmyer2020,fulton2021}, the semi-major axis distribution of gas giants shows a notable gap roughly between 0.2 and 1 au (often called ‘period valley’). Thus, the parameter space included by \cite{rosenthal2022} is relatively poorly populated, and this difference in the definition of CJs should not considerably affect the comparison between their and our results in a significant way. The authors report their absolute and conditional probabilities for ISPs and outer giants in Table 2. We consider the last column of this table, as it shows the average number of planets per star, which is the same quantity we derived. Furthermore, their results refer to a blind survey, while we selected our sample to have at least a CJ and then calculated how many ISPs are present. This means that we have to refer to their conditional probability P(Inner|Outer); that is, the probability of having a low-mass planet at low separations when an external Jupiter-like is present. They found $\text{P(Inner|Outer)} = 0.69 \pm 0.19$. As we can see from Table \ref{tab:occ_lit}, we have a total of 29 confirmed planets in this part of the parameter space, leading to a value of $\eta\text{(Inner|Outer)} = 55.6_{-8.6}^{+12.5}\%$. Therefore, our results agree within $1\sigma$ with those by \cite{rosenthal2022}.

\subsubsection{Results from similar work}

We compared our results with those of \cite{barbato_b_2018} as their sample was selected with the same criteria. We point out that all their targets (20 systems) are also present in our list, and, similarly to their results, we found no additional Neptunian planet with $10 \leq m\sin i \leq 30$ \mearth\ and $P < 150$ d among the common systems. Therefore, we have 19 objects in this parameter space region with a completeness of 73\%. In their analysis, they report an occurrence rate of $< 9.84\%$ for these objects. As shown in Table \ref{tab:occ_lit}, our value in the same mass-period region is $\eta = 22.9_{-4.1}^{+6.5}\%$, which is roughly twice as much as their upper limit with $2.1\sigma$ significance. Thus, the discrepancy is overall not too large and is probably due to their smaller sample size. In particular, our results indicate that there are roughly 0.2 planets of this kind per star, so, since their sample is small, it was drawn from the larger set of stars not hosting inner Neptunians. %\textbf{ha senso? altre idee?}

\subsubsection{Results from complementary work}
\label{sec:bonomo}
We also compared our results with the recent and complementary works by both \cite{bonomoetal2023} and \cite{bonomo2025}. The former work uses a sample of 38 systems with no overlap with ours, while the latter is based on a sample of 213 stars, of which 13 ($\sim11\%$) are also in our list. For ISPs, \cite{bonomoetal2023} considered planets ranging from 1 to 20 \mearth\ and P < 100 d. As for CJs, they selected objects with $a > 1$ au, while we used $a_{peri} > 1$ au. Thus, to make a more robust comparison, we recalculated the occurrence rates for their sample, but discarding the systems that do not satisfy our periastron criterion. Additionally, they considered planets with $m\sin i > 0.3$ \mjup, so we restricted our sample with such a criterion. As shown in Table \ref{tab:occ_lit}, we have 11 planets in this part of the parameter space, remembering that, in this case, we are calculating the fraction of stars with at least one planet rather than the number of planets per star, with an average completeness of 45\%. However, they followed a path opposite to ours: they started from systems with transiting ISPs and then searched for additional CJ companions. In practice, they derived $\text{F}_{\text{CJ|ISP}}$; that is, the conditional probability of having an outer giant planet in systems hosting small planets at low periods. On the other hand, we calculated $\text{F}_{\text{ISP|CJ}}$, and so we needed to convert our result using the equation

\begin{equation}
\text{F}_{\text{ISP}} \times \text{F}_{\text{CJ|ISP}} = \text{F}_{\text{CJ}} \times \text{F}_{\text{ISP|CJ}},
\end{equation}

\noindent which is the Bayes' theorem. Here, $\text{F}_{\text{ISP|CJ}}$ is our value reported in Table \ref{tab:occ_lit}, $\text{F}_{\text{ISP}} = 28.1_{-5.1}^{+6.6}\%$ using the result from \cite{rosenthal2022}, and $\text{F}_{\text{CJ}} = 10 \pm 1\%$ as very recently found in \cite{bonomo2025}. Inserting these numbers in the previous equation, we derived $\text{F}_{\text{CJ|ISP}} = 9.1 \pm 3.6\%$, a result perfectly compatible with the $6.4_{-2.1}^{+7.4}\%$ (or $9.6_{-3.0}^{+7.9}\%$ after including one of their targets with an RV trend compatible with a CJ, that is, K2-12) value found in \cite{bonomoetal2023} at $0.3\sigma$ (or $0.1\sigma$ including K2-12) and with the value of $11.6_{-1.9}^{+2.6}\%$ found in \cite{bonomo2025} at a $0.6\sigma$ level. We note that the definitions of ISPs are slightly different between \cite{rosenthal2022} and \cite{bonomoetal2023}; that is, $m\sin i \in [2,20]$ \mearth\ and $a < 0.5$ au versus $m\sin i \in [1,20]$ \mearth\ and $a < 0.4$ au. However, the difference is actually very small, and we do not expect this to have a significant impact on the final calculation. In any case, our result is in contrast with the much higher values found in other works such as \cite{bryan2019} and \cite{zhu2018} that show an overabundance of CJs in systems hosting ISPs. Additionally, \cite{bonomo2025} derived $\text{F}_{\text{ISP|CJ}} = 32 \pm 11\%$, which is again fully compatible at less than $1\sigma$ with our calculated value in Table \ref{tab:occ_lit}.

\subsection{Systems with inner giant planets}
\label{sec:jups}

\subsubsection{Hot Jupiters}
Now we consider our 137-star full sample, including the systems with internal ($a < 1$ au) gas giants that were removed for the analysis presented in the previous sections. As was previously done for low-mass planets, we considered two slightly different parameter space regions to make a proper comparison with previous works about HJs. In particular, we considered $0.3 < m\sin i < 3$ \mjup\ and $P < 12 $ d as done in \cite{howard2010}, and then $0.3 < m\sin i < 13$ \mjup\ and $1 < P < 10$ d as done in \cite{wittenmyer2020}. \cite{howard2010} used a sample of 166 GK stars, of which 11 ($\sim8\%$ of the total) are also in our sample. The analysis by \cite{wittenmyer2020} is based on a target list of 203 stars, of which 16 ($\sim 12\%$ of the total) are in common with our analysis. Our results are shown in Table \ref{tab:occ_HJ_WJ}.

\begin{table*}[htbp]
\centering
\caption{Comparison of HJ planetary occurrence rates in our sample and literature works.}
\begin{tabular}{c c c c}
\hline
\hline
 \\
 & \cite{howard2010} & \cite{wittenmyer2020} & \cite{su2024} \\ 
 \\
%  & P [d] & & P [d] & & P [d] & & P [d] \\
%\\
%$m_p \sin i$ [\mearth] & P < 150 d & $m_p \sin i$ [\mearth] & P < 100 d & $m_p \sin i$ [\mearth] & P < 50 d & $m_p \sin i$ [\mearth] & [1.2, 365.25] d \\
  & $m\sin i \in [0.3, 3]$ \mjup, P < 12 d & $m\sin i \in [0.3, 13]$ \mjup, P $\in $ [1, 10] d & $m\sin i \in [0.1, 13]$ \mjup, P $\in $ [10, 100] d \\
\\
\hline
\\\relax
This work & $2.4_{-0.7}^{+2.3}\%$  & $3.0_{-0.8}^{+2.5}\%$ & $9.5_{-2.2}^{+4.3}\%$ \\
\\
 & $n = 3$, $C = 92.6\%$ & $n = 4$, $C = 95.8\%$ & $n = 9$, $C = 69.5\%$ \\
\\
% & $C = 92.6\%$ & & $C = 95.8\%$ & & $C = 69.5\%$ \\
%\\
Literature & $1.2 \pm 0.2\%$ & $0.84_{-0.20}^{+0.70}\%$ & $\sim 10\%$ \\
\\
\hline
\end{tabular}
\label{tab:occ_HJ_WJ}
\tablefoot{Number of planets per star, expressed as percentages, obtained in this work for hot and warm Jupiters, adapted to the parameter space regions explored by previous works, as highlighted on top. We found no new candidates in this part of the parameter space. In all cases, $n$ represents the number of detected planets and $C$ is the average completeness in the given region of the parameter space. The last row reports the literature values against which we compare our results.}
\end{table*}

\noindent The first thing we can notice from the table is that the occurrence rate of HJs is formally larger than reported in the literature. Specifically, \cite{howard2010} found for these objects a value of $1.2 \pm 0.2\%$ in a sample of stars similar to ours in size (166 stars vs 138) and with similar selection criteria. Using results from the AAT survey, \cite{wittenmyer2020} found a similar result of $0.84_{-0.20}^{+0.70}\%$. It is also known that HJs are even less abundant around M dwarfs, likely around $\sim 0.30\%$ \citep[e.g.][and references therein]{gan2023}. Considering other works \citep[e.g.][]{zhou2019, beleznay2022, zhu2022} also, it appears that the occurrence rate of HJs is found to be around 1\% or less for FGK stars.
Our result is $1.7\sigma$ higher than that of \cite{howard2010} and $2.0\sigma$ higher than that of \cite{wittenmyer2020}, hinting at the fact that there might be an overabundance of HJs in systems with CJs. Since we only have three and four detections in the parameter space regions considered, we did not split the sample for further testing, as the low numbers would likely yield non-significant results.

\subsubsection{Warm Jupiters}
As summarized by many recent works \citep[e.g.][]{carleo2024b,mantovan2024b,bieryla2024}, WJs are a poorly understood class of objects with a peculiar eccentricity distribution, and this might be linked to different formation and migration processes. They are not close enough to their stars to have experienced tidal circularization, but, on the other hand, they must have experienced at least a certain degree of inward migration. As we can see in Table \ref{tab:occ_HJ_WJ}, these are more abundant than HJs in our sample. The subsamples obtained by splitting based on the values of $a_{lim}$ do not show different properties, as the occurrence rates of WJs are not significantly different from the value obtained for the full sample and are always compatible with the respective counterparts. In general, this part of the parameter space has, so far, been poorly explored, as most works focus mainly on hot or cool Jupiters, neglecting the middle ground. We derived occurrence rates in the region from 0.1 to 13 \mjup\ and from 10 to 100 d. \cite{su2024}, based on the RV survey presented in \cite{Mayor2011}, defined WJs based on the position with respect to the snow line, that is, for $0.1 < a_p/a_{snow} \leq 1.0$, rather than the orbital period, and analysed how the frequency of WJs varies with the star's effective temperature. In particular, they derived an occurrence rate ranging from $\sim 4\%$ to $\sim 12\%$, strongly increasing with \teff. Considering that our sample is made up mostly of Sun-like stars, from their Figure 3, we can guess an average value of around 10\%. This is in perfect agreement with our nominal value reported in Table \ref{tab:occ_HJ_WJ}.

\section{Conclusion}
\label{sec:conclusion4}
In this work, we selected a large sample of 137 stars using clear and rigorous selection criteria to search for Solar System analogues. These systems all have a CJ planet in their outer regions and are potential targets in the search for ISPs. We conducted a systematic search for all available RV data in the literature for each system and then analysed the data consistently, confirming, in most cases, the literature results. For some targets, we added unpublished high-quality data gathered with HARPS-N by the GAPS team that enhance the sensitivity of ISPs for those systems. After studying more in-depth a few cases for which we had evidence of new candidates, activity cycles, and long-term trends, we proceeded with our statistical analysis. \\

\noindent We used two already tested codes to estimate the occurrence rates of our sample, both for ISPs, warm Jupiters, and HJs. First of all, dividing our target list into various subsamples, we found a significant difference when we split based on the value of $a_{lim}$; that is, the maximum orbital distance at which inner planets can remain unperturbed from the external gas giant. This is not qualitatively unexpected, but the really interesting part was quantifying this result. Furthermore, we compared our results for ISPs with those from previous works. In the first case, we found agreement with the conditional probability, derived by \cite{rosenthal2022}, of having an ISP in systems with an external giant companion. In the second case, we noticed a higher abundance of Neptunian planets compared to \cite{barbato_b_2018}, but, since the two samples have the same criteria, this is realistically just a statistical effect related to the small sample size of \cite{barbato_b_2018}. Finally, we found a very good agreement with the results by \cite{bonomoetal2023} and \cite{bonomo2025}. In particular, our results suggest no considerable overabundance of CJs in systems with ISPs at the average stellar metallicity and mass of our sample, in contrast with findings of some previous works \citep[e.g.][]{zhu2018,bryan2019}. We derived this result by applying the Bayes' theorem to our value of $\text{F}_{\text{ISP|CJ}}$ in combination with precise determinations from the literature of $\text{F}_{\text{ISP}}$ and $\text{F}_{\text{CJ}}$. \\

In addition, we derived an occurrence rate for HJs that is formally higher than that found in the literature with a $\sim 2\sigma$ significance compared to \cite{howard2010} and \cite{wittenmyer2020}. We also derived occurrence rates compatible with the results by \cite{su2024} for WJs, finding no significant differences when the sample is split based on the external companion's parameters.

\section{Data availability}
Tables containing the total number of data points used for each instrument and target, all the resulting planetary parameters, and a list of all our target stars, with their main physical parameters are made available at the CDS via anonymous FTP to \url{http://cdsarc.u-strasbg.fr/} (130.79.128.5) or via \url{http://cdsweb.u-strasbg.fr/cgi-bin/qcat?J/A+A/}

\begin{acknowledgements}
This paper is supported by the Fondazione ICSC, Spoke 3 Astrophysics and Cosmos Observations. National Recovery and Resilience Plan (Piano Nazionale di Ripresa e Resilienza, PNRR) Project ID CN\_00000013 "Italian Research Center on    High-Performance Computing, Big Data and Quantum Computing"  funded by MUR Missione 4 Componente 2 Investimento 1.4: Potenziamento strutture di ricerca e creazione di "campioni nazionali di R\&S (M4C2-19 )" - Next Generation EU (NGEU).
We acknowledge support from the European Union –
NextGenerationEU (PRIN MUR 2022 20229R43BH) and the “Programma di
Ricerca Fondamentale INAF 2023”. We acknowledge financial contribution from
the INAF Large Grant 2023 “EXODEMO”. 
We acknowledge the Italian center for Astronomical Archives (IA2, \url{https://www.ia2.inaf.it}), part of the Italian National Institute for Astrophysics (INAF), for providing technical assistance, services and supporting activities of the GAPS collaboration.\\
M.P. acknowledges support from ASI-INAF agreement no. 2025-10-HH.0 "Partecipazione Italiana al Gaia DPAC – Supporto alle attività di responsabilità del team scientifico”.
N.N. acknowledges funding from Light Bridges for the Doctoral Thesis “Habitable Earth-like planets with ESPRESSO and NIRPS”, in cooperation with the Instituto de Astrofísica de Canarias, and the use of Indefeasible Computer Rights (ICR) being commissioned at the ASTRO POC project in the Island of Tenerife, Canary Islands (Spain). The ICR-ASTRONOMY used for his research was provided by Light Bridges in cooperation with Hewlett Packard Enterprise (HPE).
\end{acknowledgements}

\bibliography{references} % your references Yourfile.bib
\bibliographystyle{aa.bst} % style aa.bst

\newpage

\begin{appendix}

\twocolumn

\newpage

\section{New results}
\label{sec:appendix_candidates}
\subsection{New candidates}
Here, we describe the systems for which we have convincing evidence that a new candidate, previously undetected, may be present. Table \ref{tab:candidates_rv_data} lists the number of spectra used for each target taken with different instruments. \\

\begin{figure}[htbp]
   \centering
   \includegraphics[width = 0.5\textwidth]{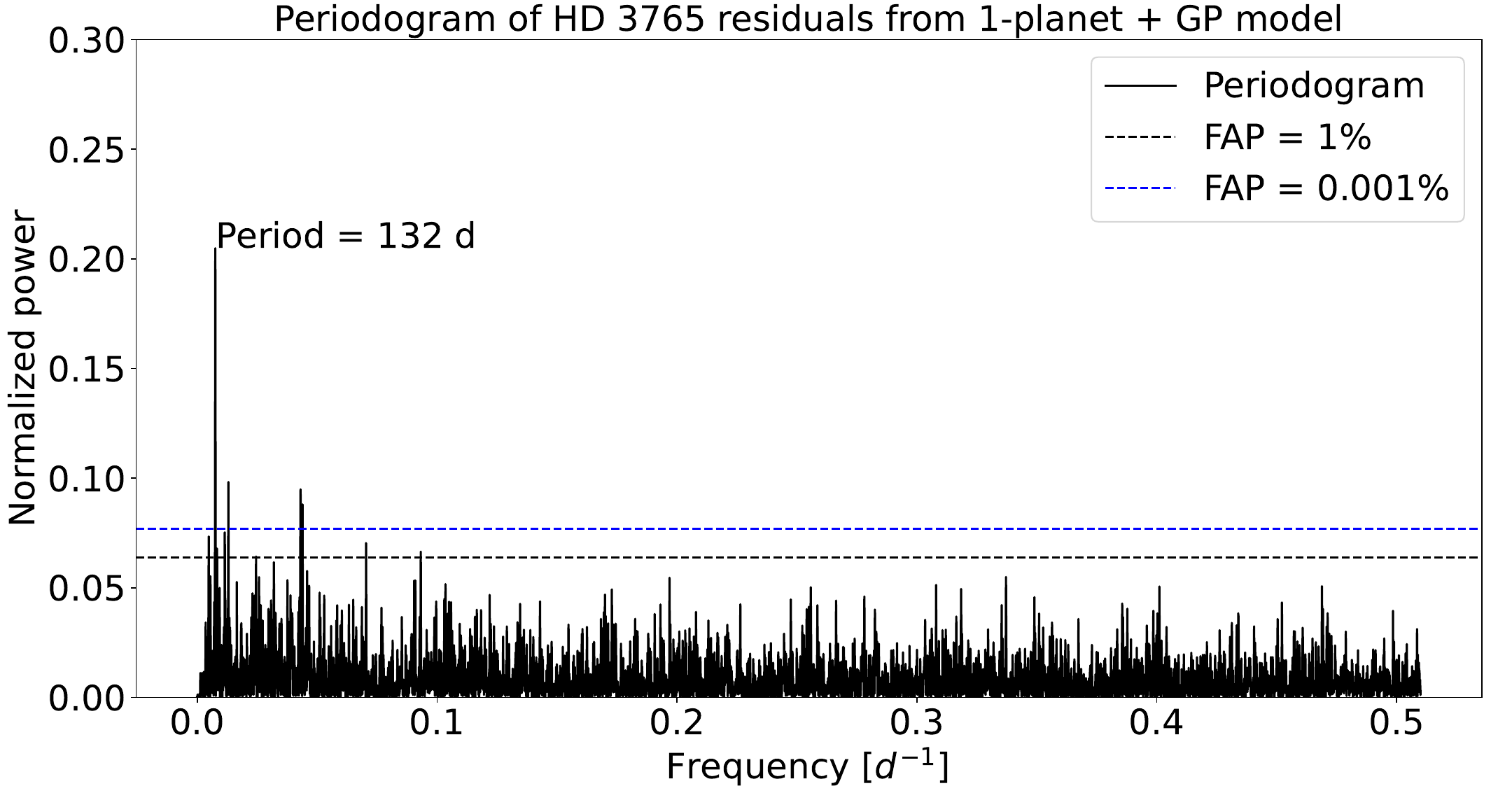}
      \caption{GLS of the RV residuals for HD 3765 after removing the signals corresponding to planet b and the activity cycle modeled with a quasi-periodic GP. 
              }
    \label{fig:hd3765_gls_residuals}
    \end{figure}

\begin{figure}[htbp]
   \centering
   \includegraphics[width = 0.5\textwidth]{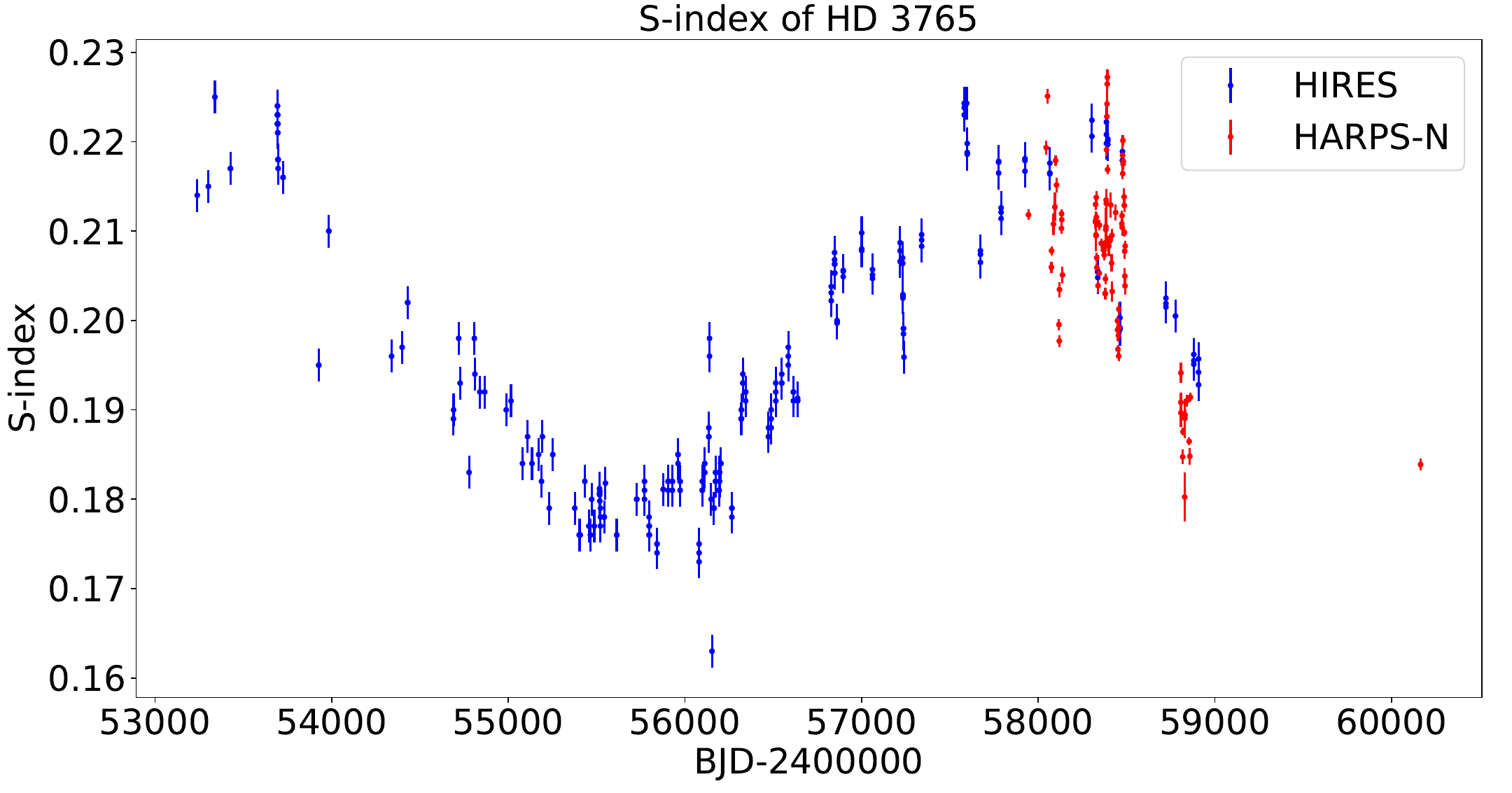}
      \caption{S-index time series of HD 3765.}
    \label{fig:hd3765_smw}
    \end{figure}

\begin{figure}[htbp]
   \centering
   \includegraphics[width = 0.5\textwidth]{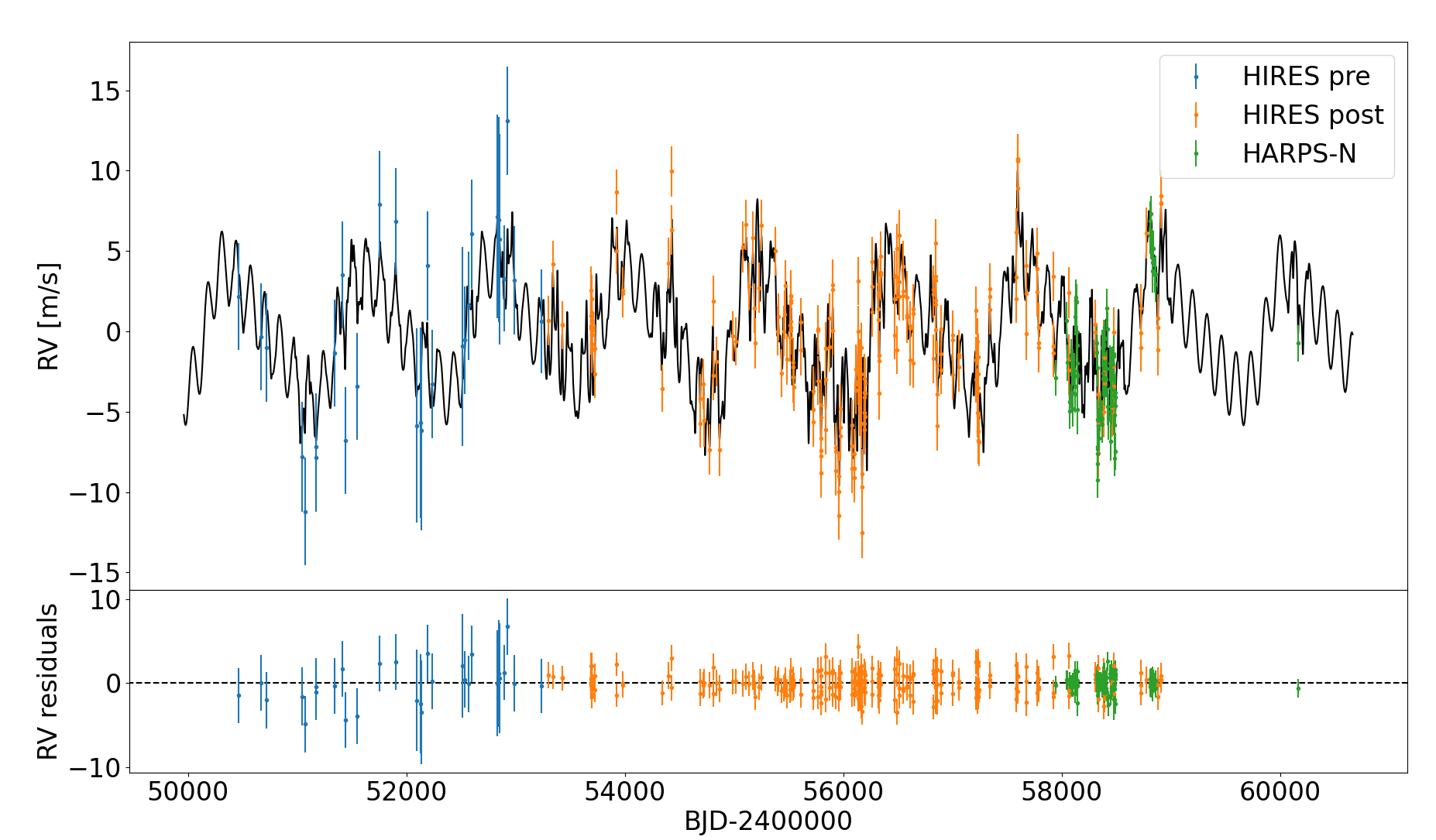}
\includegraphics[width = 0.5\textwidth]{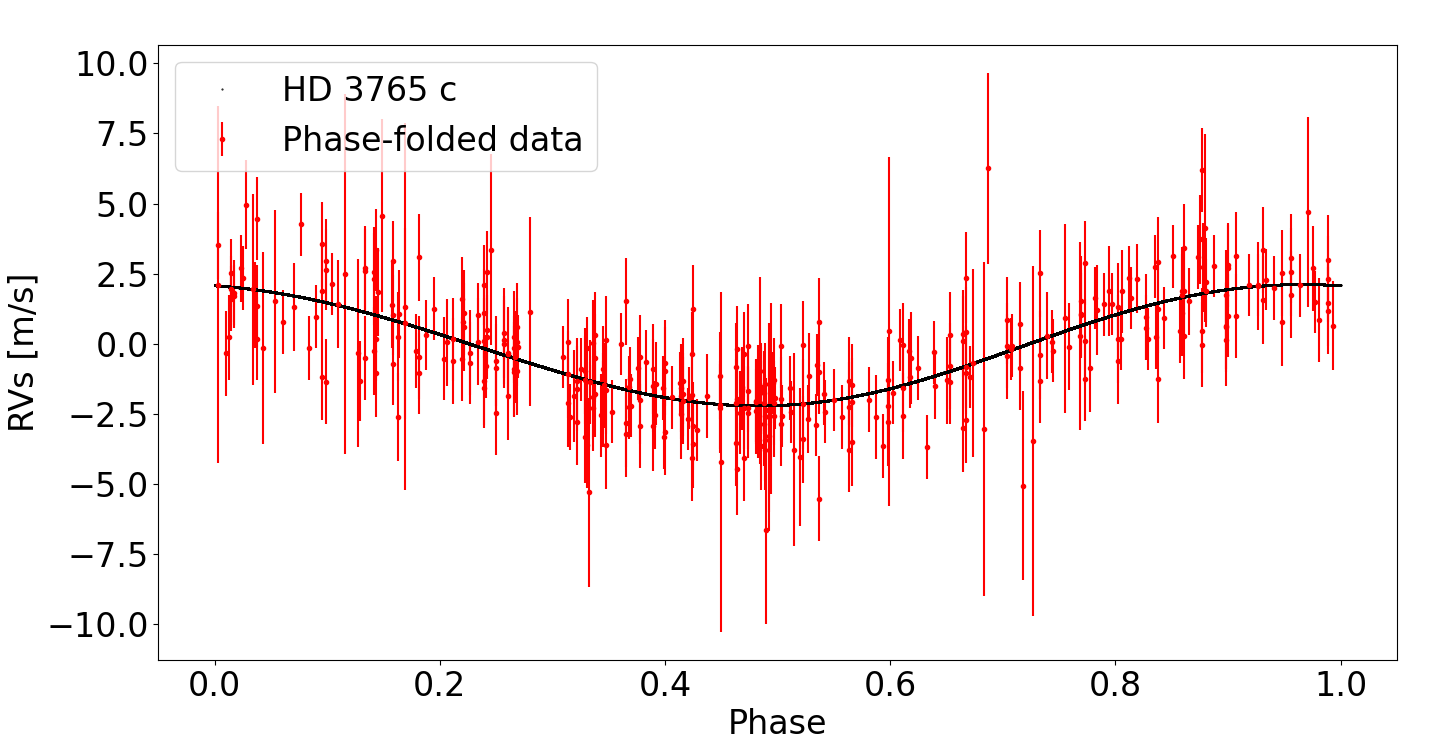}
      \caption{\textit{Top panel:} best-fit model of HD 3765. \textit{Bottom panel:} phase-folded data and Keplerian model for the new candidate HD 3765.01. 
              }
\label{fig:hd3765_phase}
\end{figure}

\paragraph{HD 3765}: this K-class star hosts a planet with a minimum mass of 0.17 \mjup\ at about 2 au ($P \sim 1200$ d) on a moderately eccentric (e = 0.3) orbit, as announced by \cite{rosenthal2021} using HIRES data. This star was used as a standard star for the observations in the GIARPS configuration \citep{claudi2017} by the GAPS team \citep{baratella2020}. Therefore, we considered the 95 available HARPS-N spectra in our analysis. For these observations, the focus of the telescope was optimized for the use of GIANO-B \citep{oliva2006}, but this does not represent a significant limitation for our analysis.  Additional data were gathered with CORALIE, ELODIE, and SOPHIE, but we did not use them due to their low precision or sparse sampling. In addition to planet b, \cite{rosenthal2021} mention an activity cycle visible in the chromospheric activity index time series and model it with a squared exponential GP, simultaneously with the RV data. Here we confirm that the GLS of the S-index time series (HIRES + HARPS-N) gives a very significant peak at $\sim 4160$ d. Therefore, we modeled the RVs with a Keplerian term, corresponding to planet b, and a GP with a quasi-periodic kernel, for which we set a uniform prior centered on the period found in the periodogram ($= 4000 \pm 400$ d). We obtain $P_{cycle} = 4125_{-273}^{+534}$ d and $K_{cycle} = 2.52_{-0.75}^{+1.20}$ m/s, compatible with \cite{rosenthal2021}. The parameters obtained for the known planet are also compatible. We searched the residuals for other potential signals and found a significant peak at 130 d with FAP $< 0.001\%$ (Figure \ref{fig:hd3765_gls_residuals}). For this reason, we repeated the analysis, adding a second Keplerian term to our model. We obtained $\Delta \text{BIC} = 76$ in favour of the model with the second Keplerian, meaning that there is convincing evidence for a second planet in the system (according to the criterion by \citealt{kass1995}). From now on, we will label this planetary candidate HD 3765.01. Then, we extracted the GLS of the residuals and found a significant peak at 22 d. At the same time, we tried to model the S-index with a sinusoidal term ($P = 4342 \pm 55$ d), since it has a perfectly sinusoidal shape (Figure \ref{fig:hd3765_smw}) and also found a few short-period peaks in the GLS of the residuals, one being at almost 22 d. This gave us a first indication that there could be a rotational signal in the RV data. To gather more information regarding the potential rotation period of the star, we considered the co-added spectrum analysed by us in \cite{baratella2020} to derive stellar parameters and estimated the projected rotational velocity according to the method in \cite{biazzoetal2022}. We therefore obtained an upper limit of \vsini$<$1.5\,km/s, which is compatible with the value of \vsini$<$1.7\,km/s obtained applying the calibration of the FWHM of the CCF into \vsini\, developed by \cite{rainer2023}. We conclude that we can only set a lower limit on the actual rotation period of the star, which could be around 46 days. In our first tests, we obtained a rotation period around 44 d (twice the 22 d peak). Since the period of the putative candidate is three times this value, special precaution is required. For this reason, we ran a set of additional models to fit the data including the known planet, the new candidate, rotational activity, and a long-term magnetic cycle. Specifically, in the fitting procedure, we included either one or two planets, and either no activity, rotational activity (GP with quasi-periodic kernel), a magnetic cycle (GP with quasi-periodic kernel), or both activity signals (double GP with one quasi-periodic kernel for rotation and a squared exponential kernel for the magnetic cycle). In all cases, the orbital parameters of planet b are mostly in agreement with each other and with the literature. We noticed that its eccentricity, and also that of the new candidate when included, tends to be higher when activity is not modeled, indicating that there is additional power being incorporated into the Keplerian term. In the 1 planet + rotation case, the resulting posterior distribution of the rotation period is strongly asymmetrical with a peak at 44 d and then extending up to the 130 d range. However, when the second planet is added, we obtained $P_{rot} = 44.7_{-1.0}^{+1.7}$ d. Regardless of the activity modeling, all models with two Keplerian terms have lower BICs than their counterparts including only planet b. Specifically, the model with the lowest BIC is that with two planets and rotational activity, with $\Delta\text{BIC} = 12$ compared to the second lowest BIC (which is the two planets + activity cycle case). Considering all these results, we conclude that the most likely explanation is that, in addition to planet b, there is a new candidate, HD 3765.01, and that the presumed rotation period of the star is $P_{rot} = 44.7_{-1.0}^{+1.7}$ d. Additionally, we think that the presence of the activity cycle found by \cite{rosenthal2021} is dubious, for the following reasons. Firstly, although it is clearly visible in the S-index time series, a peak around 4000 d is not seen in the RV residuals of none of the cases we tested without an activity cycle in the modeling. Secondly, it does not significantly reduce the BIC when included. Lastly, when both GPs are present (rotation + cycle), the RV amplitude of the cycle has a flat posterior distribution across the explored parameter space region. We are not stating that this cycle does not exist, but its contribution to the total RV signal is at least disputable. In any case, the details of magnetic cycle analysis are beyond the scope of this paper. To conclude, we confirm our findings about the new planetary candidate HD 3765.01, for which we derive the following parameters: $P = 132.70 \pm 0.19$ d, $e = 0.124_{-0.087}^{+0.150}$, $K = 2.18 \pm 0.32$ m/s, $\omega = -64_{-74}^{+202}$ degrees, and $m \sin i = 15.1 \pm 2.2$ \mearth. Figure \ref{fig:hd3765_phase} shows the best-fit model and the phase-folded Keplerian of the new candidate.
\\

\begin{figure}[htbp]
%\vspace{-2cm}
   \centering
   \includegraphics[width = 0.5\textwidth]{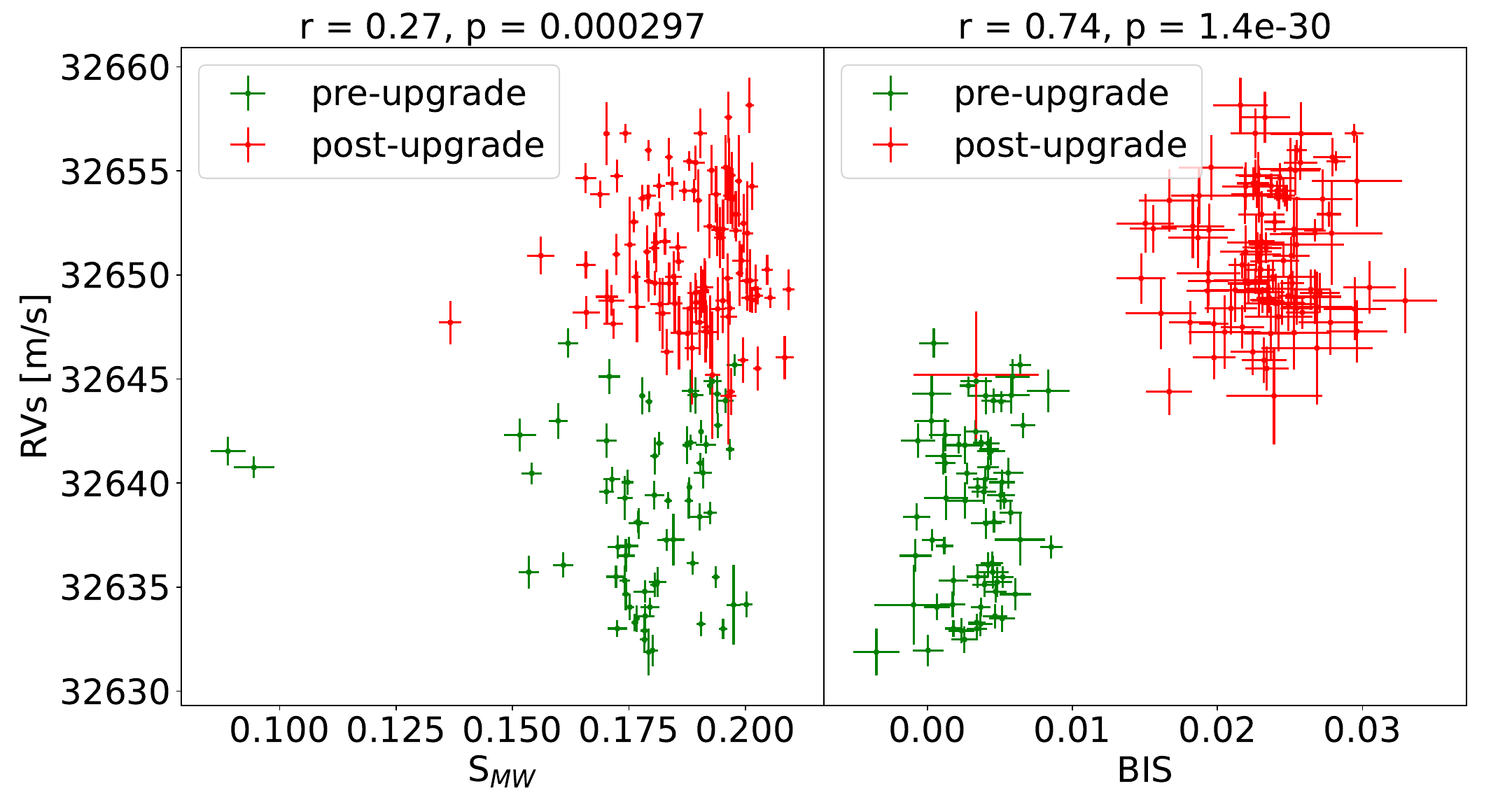}
      \caption{Correlation between RVs and two activity indicators for HD 204941, divided by pre- and post-upgrade of the instrument.}
    \label{fig:hd204941_rv_activity}
\end{figure}

\begin{figure}[htbp]
%\vspace{-2cm}
   \centering
   \includegraphics[width = 0.5\textwidth]{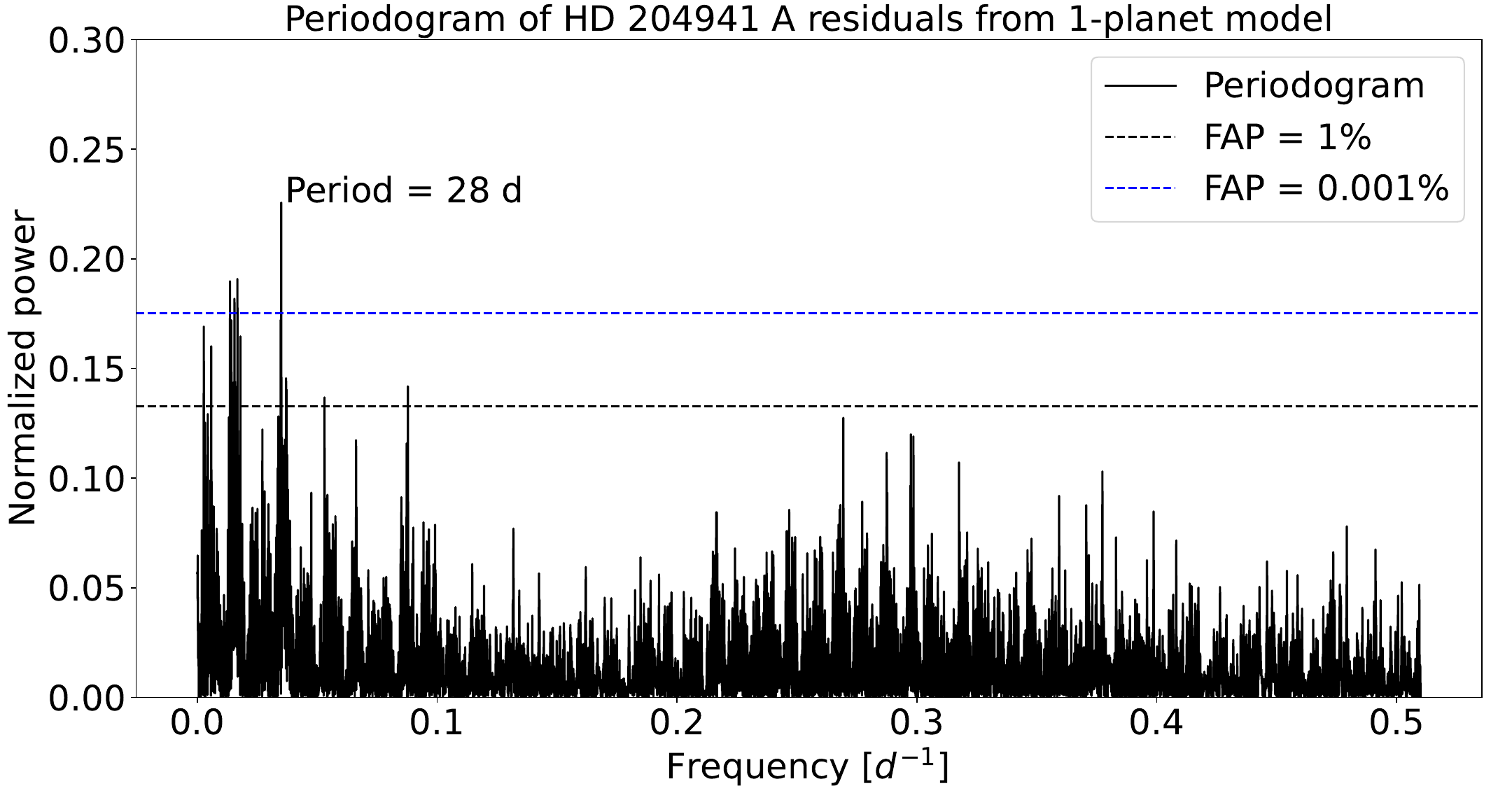}
      \caption{GLS of the RV residuals for HD 204941 after removing the signal corresponding to planet b. }
    \label{fig:hd204941_gls_residuals}
\end{figure}

\begin{figure}[htbp]
%\vspace{-2cm}
   \centering
   \includegraphics[width = 0.5\textwidth]{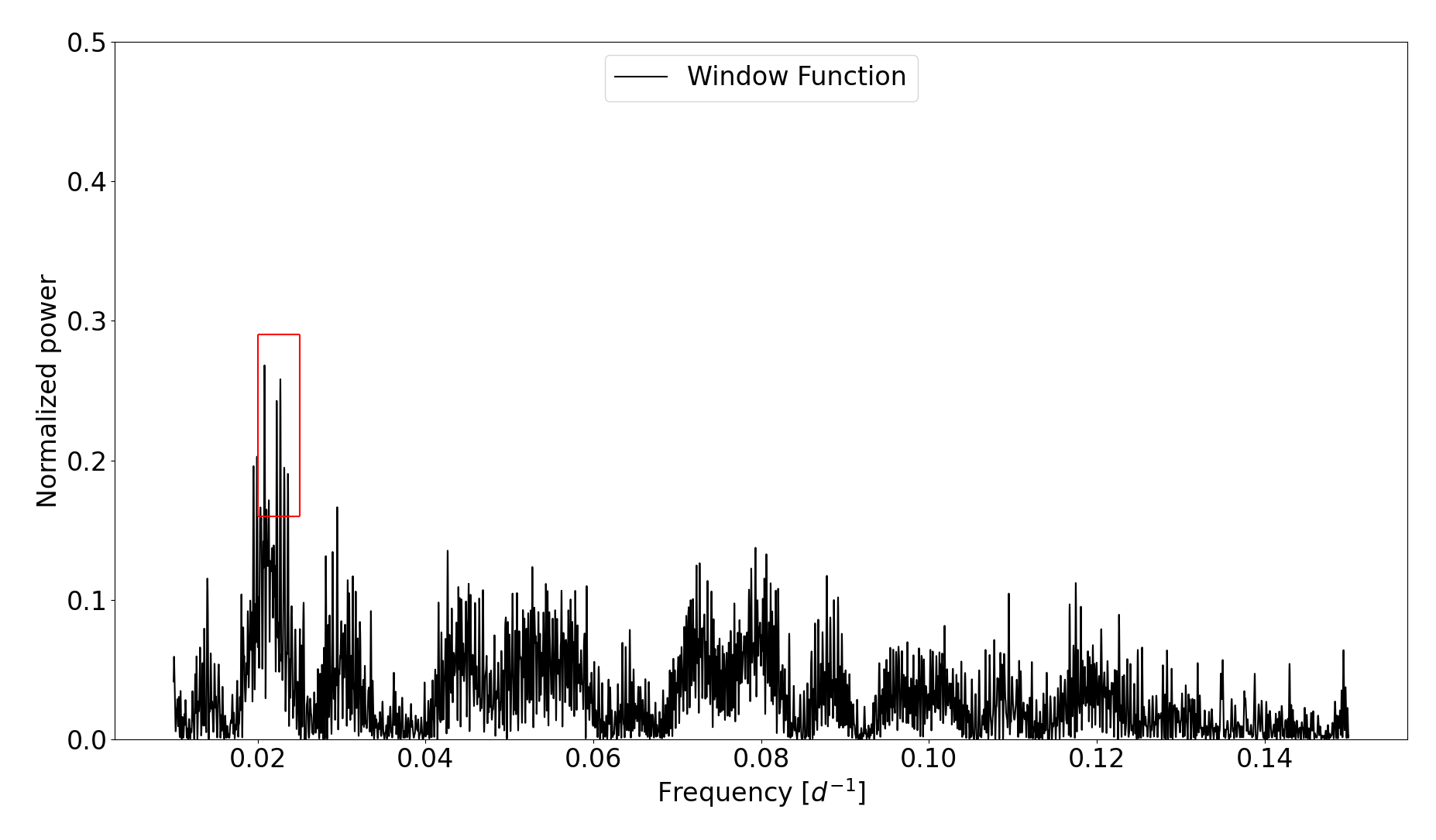}
      \caption{Window function of HD 204941.}
    \label{fig:window_hd204941}
\end{figure}

\paragraph{HD 204941}: the primary component of this binary K star hosts a 0.27 \mjup\ planet at 2.5 au ($P \sim 1700$ d) on a moderately eccentric ($e = 0.37$) orbit discovered by \cite{Dumusque2011} using 35 HARPS spectra. The authors also find evidence for an activity cycle in the \logrhk\ data with a minimum period of at least 5-6 years. Now we have 171 HARPS spectra gathered over $\sim 18$ yr and are thus able to get new insights about this system and its activity. First of all, we searched for correlations between raw RVs and activity indicators. In particular, we derived a Spearman's rank coefficient between RVs and the \smw\ of $r = 0.27$, while for RVs and BIS, we found $r = 0.74$, the latter being very significant with a p-value of $1.4 \times 10^{-30}$. However, as shown in Figure \ref{fig:hd204941_rv_activity}, this strong correlation is driven by the fact that the data points are divided into two clusters. As we can see from the plot, this correlation is clearly of instrumental origin, with the data points being divided between pre- and post-upgrade of HARPS \citep{LoCurto2015}. After accounting for the different instrumental offsets (by subtracting the median from each of the two sets, pre- and post-upgrade, independently), we also searched for signals in the GLS of the activity indices and found a significant peak in the BIS time series at $\sim 3300$ d. The same signal (3225 d) is visible in the \smw\ data, although with a higher FAP of 5\%. After fitting the RV data with one Keplerian term, corresponding to planet b, we searched for other signals in the periodogram of the residuals, finding a peak with low FAP at 28 d (Figure \ref{fig:hd204941_gls_residuals}). We can see other smaller peaks at lower frequencies, corresponding to periods roughly between 50 and 360 d, but no signal at the periods found for the activity indicators. Moreover, we find no correlation between RV residuals and activity indicators, so we conclude that, despite the periodic variations seen in the activity indices time series, the activity contribution to the RV signal is negligible. 
Regarding the signal at 28 d, in principle, it could be caused by the star's rotation rather than by a planet. To check on this, we considered the S-index data and estimated the rotation period using the equations derived by \cite{noyes1984} and \cite{mamajek2008}. Averaging over all the values, we find $P_{\rm rot,Noyes} = 44.94 \pm 5.46$ d and $P_{rot,Mamajek} = 46.96 \pm 4.49$ d. As shown in Figure \ref{fig:hd204941_gls_residuals}, there are also a few secondary, but formally significant peaks, in the frequency range between 0.0170 and 0.0137 d$^{-1}$. To inspect these additional peaks, we checked the window function of our data, shown in Figure \ref{fig:window_hd204941}. As highlighted in the red box, there is a peak at $f_{WF} = 0.0208$ d$^{-1}$ with additional peaks at slightly higher frequencies. Considering the peak at $f_0 = 0.0350$ d$^{-1}$ found in the GLS of the residuals, then the convolution with the window function gives two potential peaks at $f_0 \pm f_{WF}$, that is at 0.0558 and 0.0142 d$^{-1}$. The latter is indeed very close to the secondary peaks observed in the GLS of the residuals, thus we conclude that the formally significant peaks seen in Figure \ref{fig:hd204941_gls_residuals} are likely observational aliases of the main peak at 28 d.
We further inspected both the All Sky Automated Survey (ASAS; \citealt{Pojmanski2002}) and TESS photometry, finding no clear modulation in the timeseries. TESS photometry, however, is split into three sectors distant in time (01, 28, 92), so any periodicity longer than 25 days would hardly be measurable. With this in mind, given the possibility that there could be another planet in the system with a period around 28 d, we fit the data by adding a second Keplerian. This second model is favoured compared to the one with only the known planet with $\Delta\text{BIC} = 19.4$, representing strong statistical evidence for the presence of a second lanetary candidate, which we name HD 204941.01. However, further analysis showed that the existence of this new candidate may be at least dubious. In particular, we found that this signal at 28 d is present only in the last season of data, so that, if we remove these data, the signal disappears. We point out that this portion of the dataset has a higher temporal sampling, so this can give rise to previously unseen short-period signals, but, on the other hand, these observations have much lower exposure times compared to the previous ones (300 s vs 900 s), resulting in larger error bars. Overall, we point out that a new short-period planet could be present around HD 204941, but more investigations are needed to confirm it. In particular, more observations with high exposure times could shed further light on the nature of this signal. For this reason, we did not include this potential planet in our list for the statistical analysis described in the paper. In any case, our best-fit parameters for the potential candidate are the following: $P = 28.729 \pm 0.009$ d, $e = 0.133_{-0.092}^{+0.13}$, $K = 1.66 \pm 0.28$ m/s, $\omega = 10_{-88}^{+74}$ degrees, and $m\sin i = 5.4 \pm 1.1$ \mearth. We also note that, for planet b, we find parameters that are in slight disagreement with the ones by \cite{Dumusque2011}, that is $K = 4.6 \pm 0.3$ m/s, $P = 1529 \pm 13$ d, $e = 0.160_{-0.064}^{+0.069}$, $\omega = -38_{-29}^{+21}$ degrees, and $m\sin i = 56.1 \pm 6.5$ \mearth. These differences are not too large overall and are likely due to the larger dataset and the much longer baseline.

\begin{figure}[htbp]
%\vspace{-2cm}
   \centering
   \includegraphics[width = 0.5\textwidth]{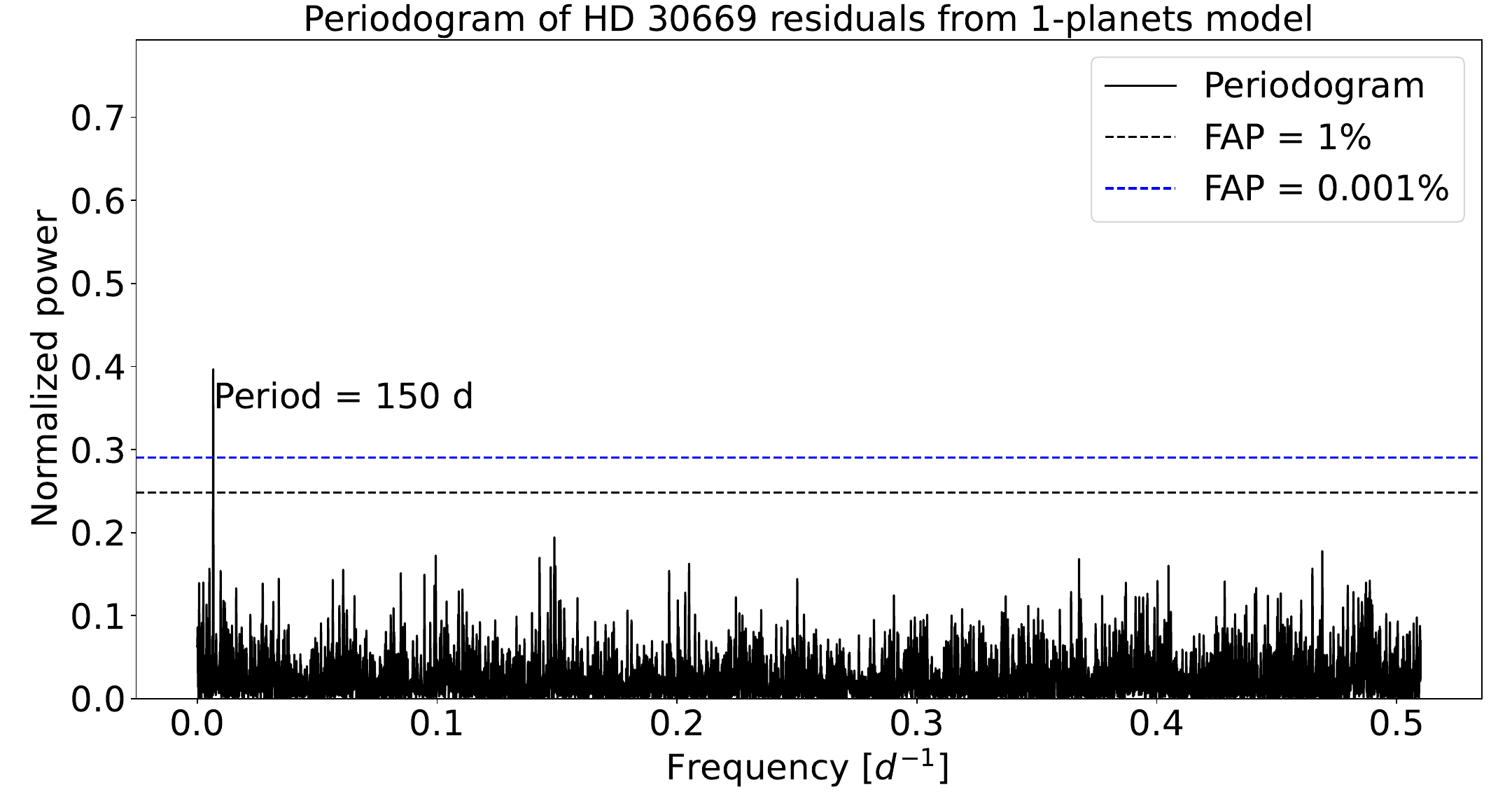}
      \caption{GLS of the RV residuals for HD 30669 after removing the signal corresponding to planet b.}
    \label{fig:hd30669_gls_residuals}
\end{figure}

\begin{figure}[htbp]
   \centering
   \includegraphics[width = 0.5\textwidth]{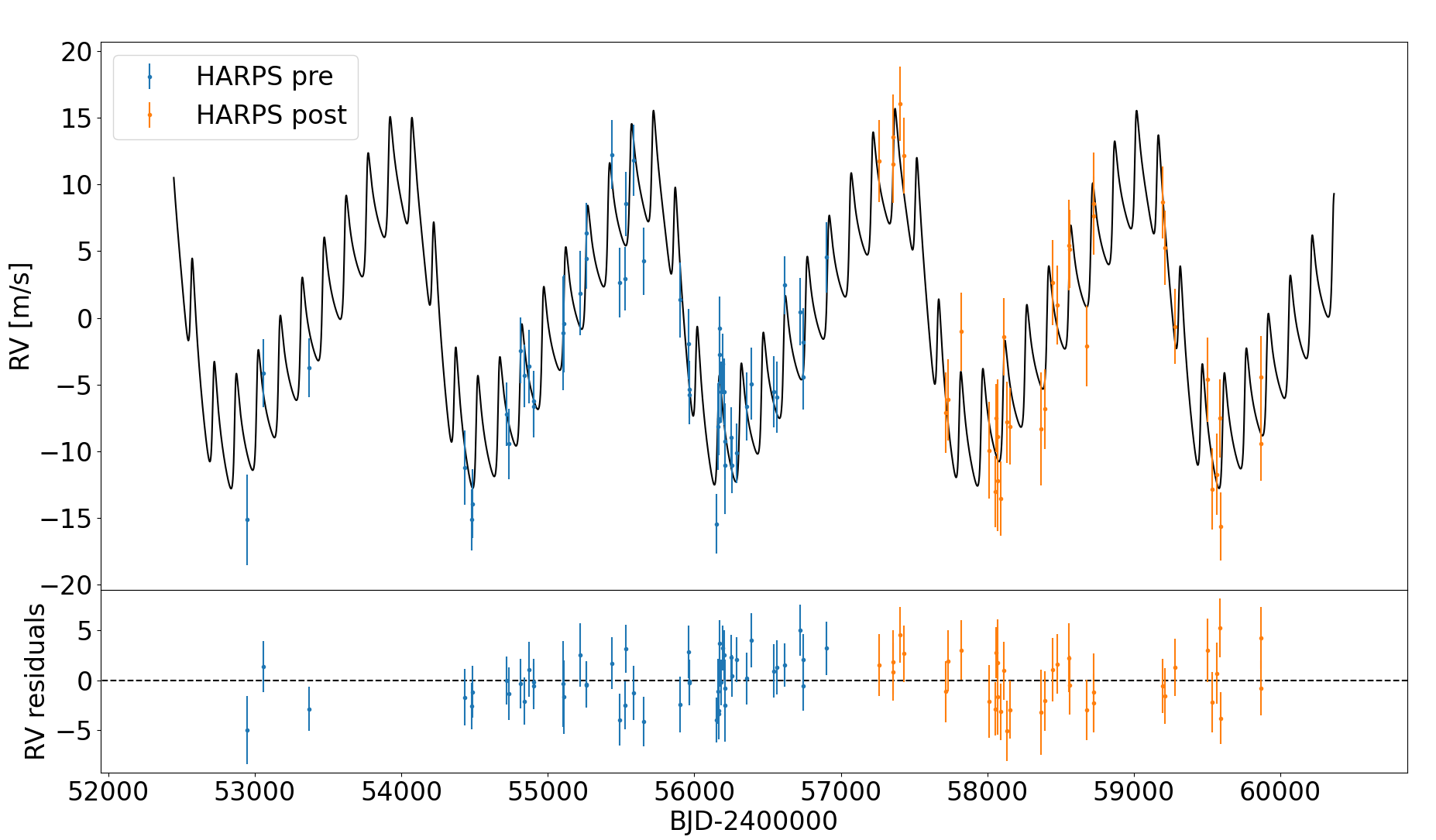}
\includegraphics[width = 0.5\textwidth]{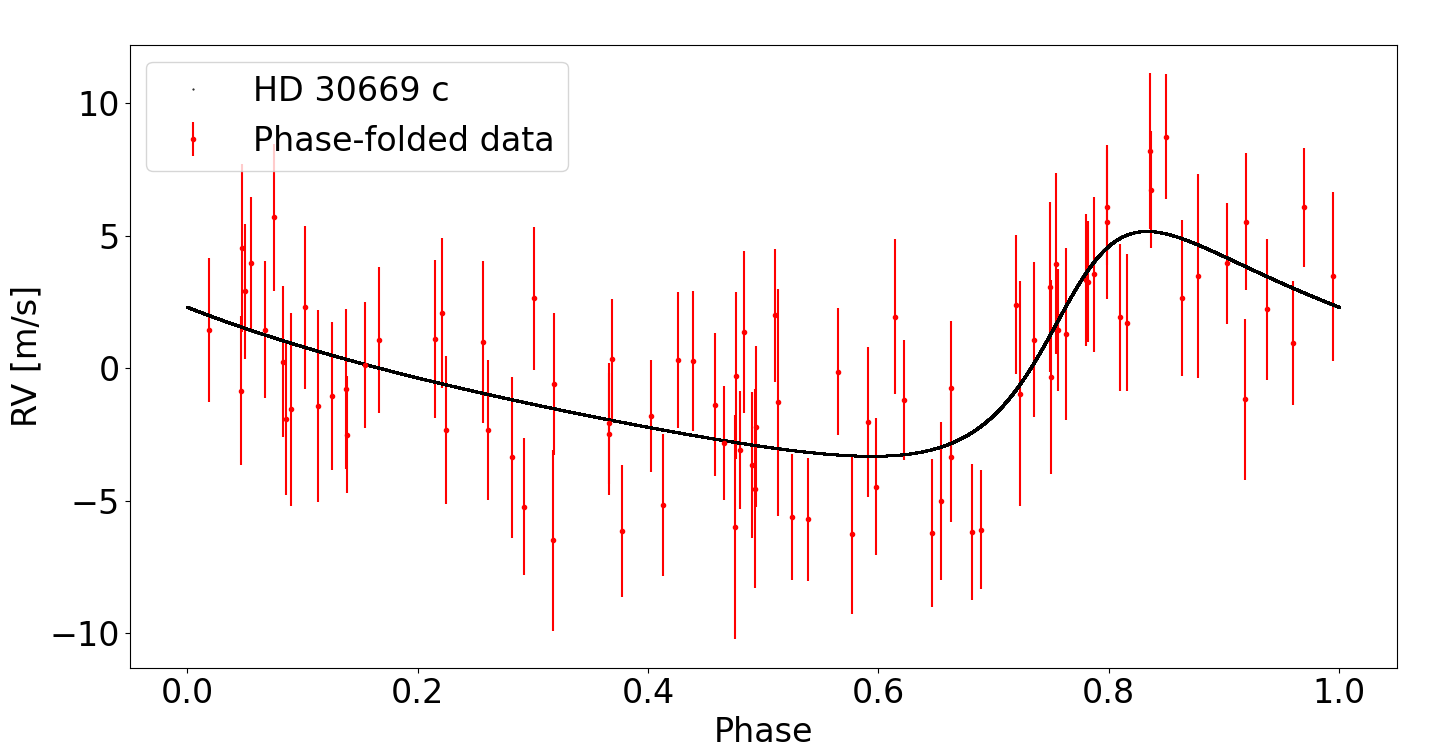}
      \caption{\textit{Top panel:} best-fit model of HD 30669. \textit{Bottom panel:} phase-folded data and Keplerian model for the new candidate HD 30669.01. 
              }
\label{fig:hd30669_phase}
\end{figure}

\paragraph{HD 30669}: this late G-type star hosts a planet with about half the mass of Jupiter and a period of 4.6 yr, announced for the first time by \cite{moutou2015} using 46 HARPS spectra, while now there are 87 available data points. The authors point out that the star is not active and notice a weak signal at 150 d in the GLS of the residuals. We confirm both the known planet and additional signal in the residuals that is now very significant, probably thanks to the expanded dataset, as shown in Figure \ref{fig:hd30669_gls_residuals}. In their work, \cite{moutou2015} state that this could be an artifact related to stellar activity. Thus, we checked if there are any periodicities in the time series of the FWHM of the CCF, BIS, \smw, H$\alpha$, and Na I., finding no signal with FAP lower than a few percent. This confirms that this signal is likely of planetary origin and so we fit our data with a second Keplerian term. This second model is favoured over the first one with $\Delta\text{BIC} = 45.5$, indicating strong evidence for the presence of a second planet in the system. We found no other significant peaks in the GLS of the residuals. For the new candidate, to be named HD 30669.01, we find the following parameters: $K = 4.2 \pm 0.5$ m/s, $P = 149.9 \pm 0.2$ d, $e = 0.496_{-0.097}^{+0.088}$, $\omega = -62 \pm 19$ degrees, and $m \sin i = 28.7 \pm 3.5$ \mearth. An interesting result is the rather high eccentricity of this object. Furthermore, we also find a higher eccentricity for planet b ($e = 0.33 \pm 0.06$) compared to \cite{moutou2015}. Since the two planets are not very far from each other ($a_b = 2.7$ au, $a_c = 0.5$ au), there might be dynamic interactions that shaped the system and led to this peculiar situation in which the periastron of the outer planet is relatively close to the apoastron of the internal one (1.8 au vs 0.8 au). To check on this stability issue, we used the python code \textit{ARDENT} \citep{stalport2025}, which allows us to derive not only the RV detection limits, but also dynamical stability limits. Our results indicate no signs of dynamical instability for orbital periods $\lesssim 200$ d and that, although the new candidate might experience periodic eccentricity variations, the semi-major axis remains stable at the current putative value. Figure \ref{fig:hd30669_phase} shows the best-fit model and the phase-folded Keplerian of the new candidate.
\\

\begin{figure}[htbp]
%\vspace{-2cm}
   \centering
   \includegraphics[width = 0.5\textwidth]{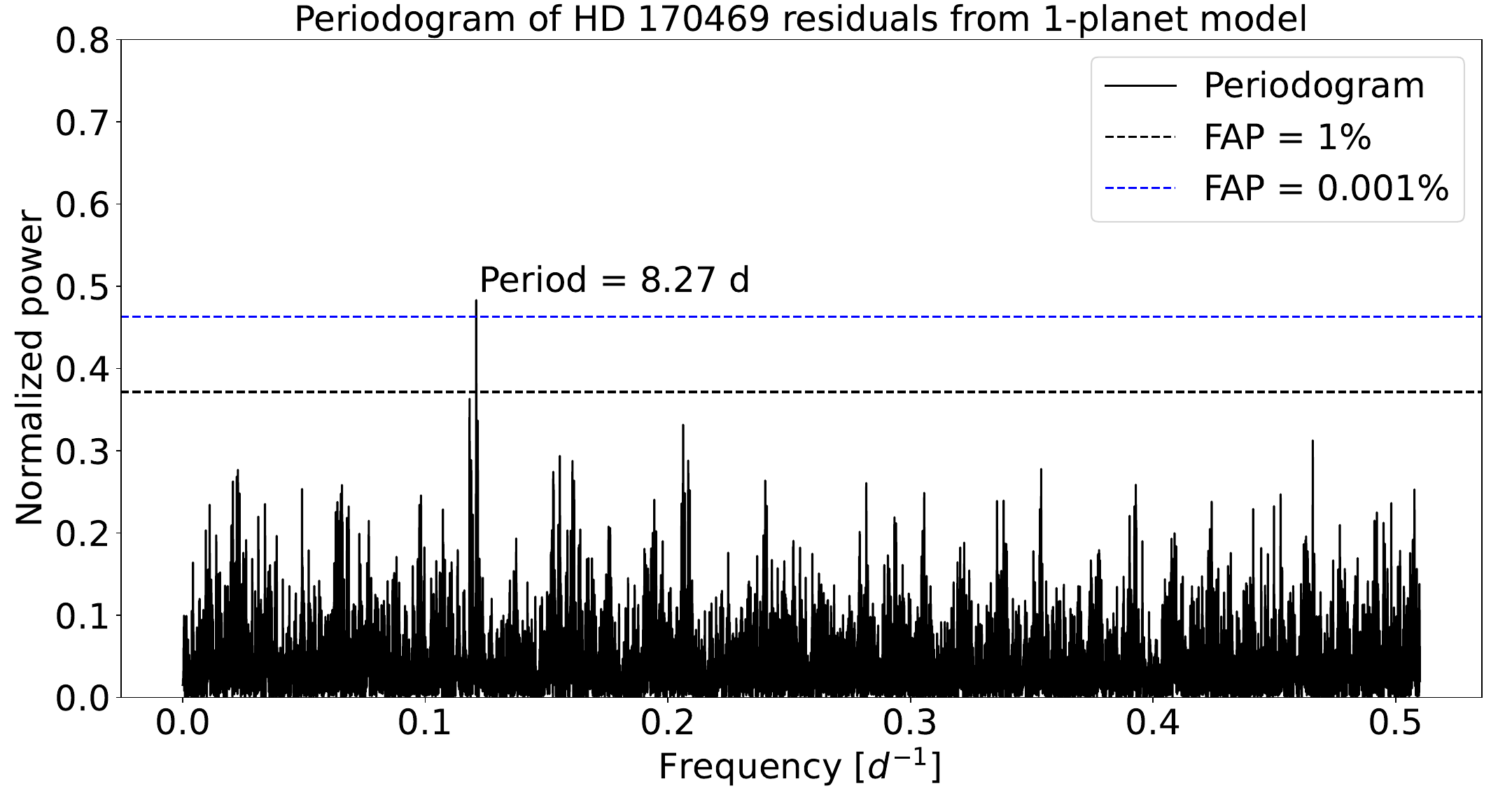}
      \caption{GLS of the RV residuals for HD 170469 A after removing the signal corresponding to planet b.}
    \label{fig:hd170469_gls_residuals}
\end{figure}

\paragraph{HD 170469}: this G5 subgiant is the primary component of a binary system and hosts a planet with about half the mass of Jupiter and a period of around 1100 d discovered by \cite{Fischer2007} and later confirmed by \cite{rosenthal2021} using HIRES data. Our dataset is the same as the previous work. After fitting the data with a single-planet model, we extracted the GLS of the residuals, finding a significant peak at $\sim 8$ d, as shown in Figure \ref{fig:hd170469_gls_residuals}. Therefore, we added a second Keplerian to our model and repeated the analysis. In both cases, the parameters obtained for planet b are perfectly consistent with the literature values. However, the model with two planets is favoured with $\Delta\text{BIC} = 11.7$, indicating strong evidence for this new planetary candidate. In this case, performing an extended analysis of stellar activity is impossible because the data published by \cite{rosenthal2021} only contain 21 \smw\ measurements over several years. Using these data, as done for HD 204941, we estimated the rotation period of the star to be $P_{\rm rot,Noyes} = 30.46 \pm 0.16$ d and $P_{rot,Mamajek} =33.14 \pm 0.24$ d, though an inspection of the available ASAS photometry did not reveal any particular periodicity (while no reduced TESS sector is available for this star). Since these values are very distant from the 8 d signal seen in the GLS, we can safely say that this signal is not linked to stellar activity. However, we admit that there can be some doubts about the presence of this new object because we only have 45 data points over $\sim 19$ yr (an average of 1 observation every 155 days), meaning that the temporal sampling is not optimal to detect such a low-period planet. Indeed, we derived the detection map of this target and found that the putative new candidate lies in a region with completeness $\lesssim 10\%$. For this reason, we did not include this object in our list for the statistical analysis described in the paper. In any case, for the potential candidate HD 170469.01, we find the following parameters: $P = 8.2676 \pm 0.0015$ d, $K = 4.2 \pm 1.1$ m/s, $e = 0.19_{-0.13}^{+0.26}$, $\omega = 30_{-138}^{+65}$ degrees, and $m \sin i = 13.8 \pm 3.2$ \mearth. 
%Figure \ref{fig:hd170469_phase} shows the phase-folded best-fit model. 

\begin{figure}[htbp]
%%\vspace{-2cm}
  \centering
   \includegraphics[width = 0.5\textwidth]{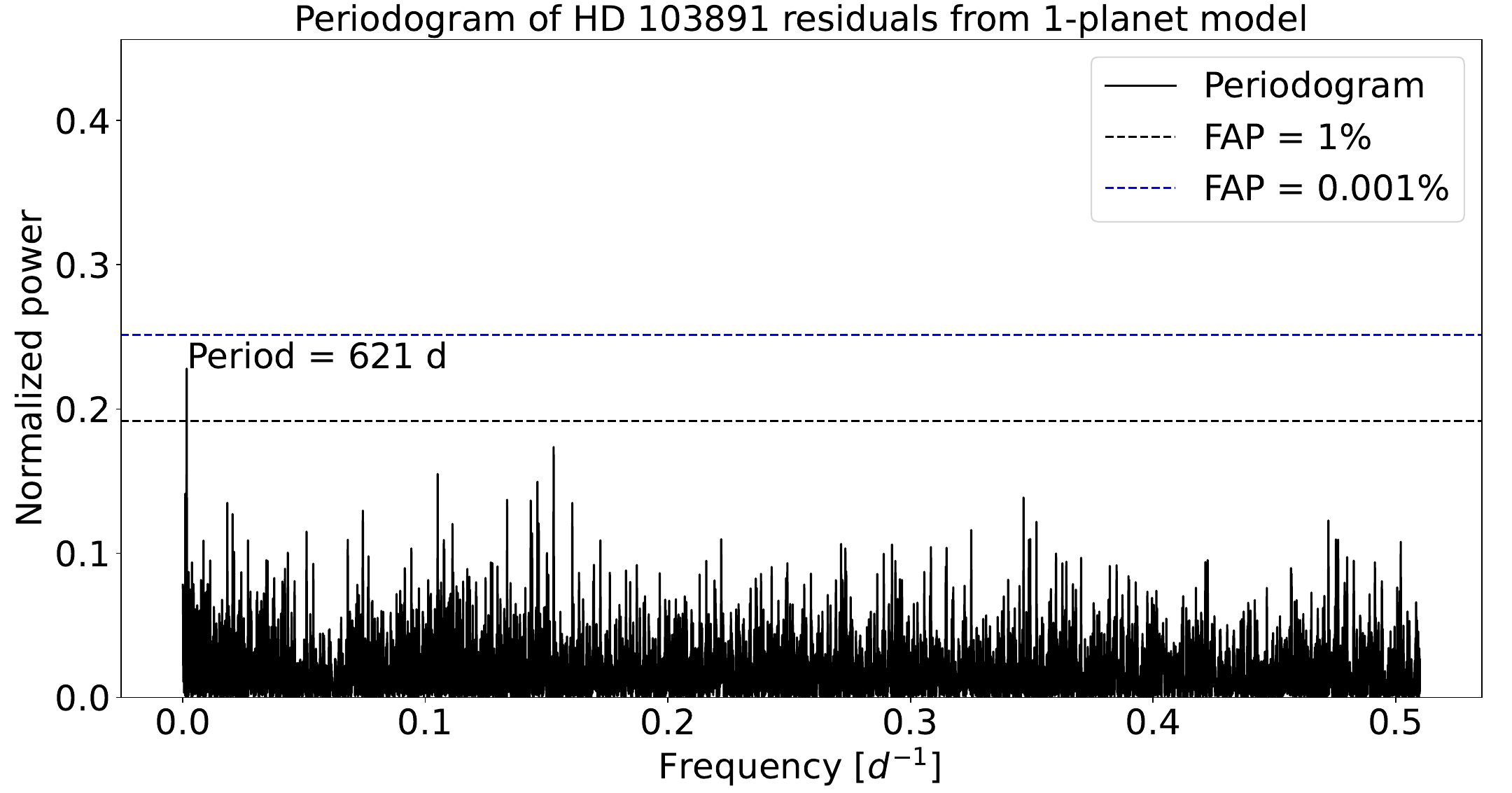}
      \caption{GLS of the RV residuals for HD 103891 after removing the signal corresponding to planet b.}
    \label{fig:hd103891_gls_residuals}
\end{figure}

\begin{figure}[htbp]
   \centering
   \includegraphics[width = 0.5\textwidth]{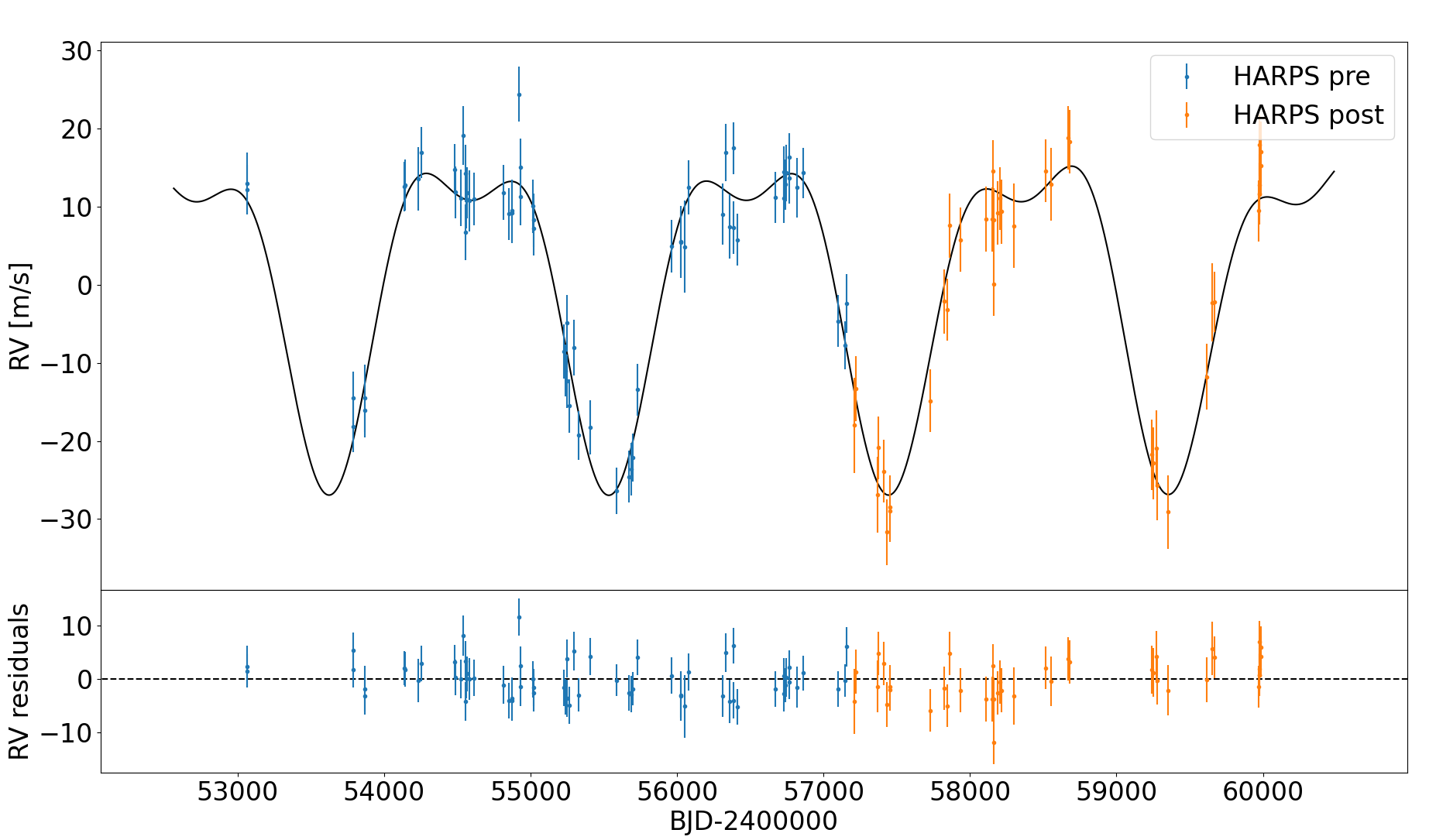}
\includegraphics[width = 0.5\textwidth]{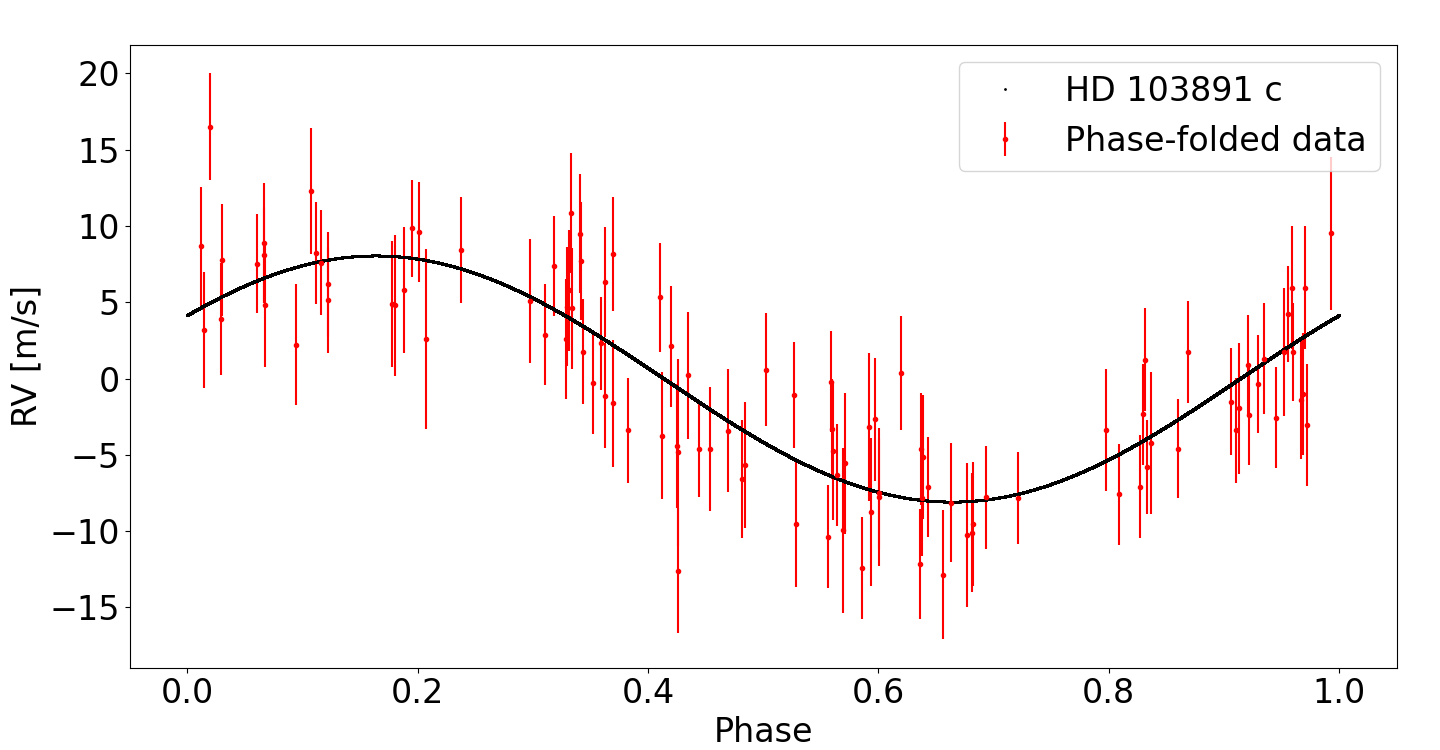}
      \caption{\textit{Top panel:} best-fit model of HD 103891. \textit{Bottom panel:} phase-folded data and Keplerian model for the new candidate HD 103891.01. 
              }
\label{fig:hd103891_phase}
\end{figure}

\begin{figure}[htbp]
   \centering
\includegraphics[width = 0.5\textwidth]{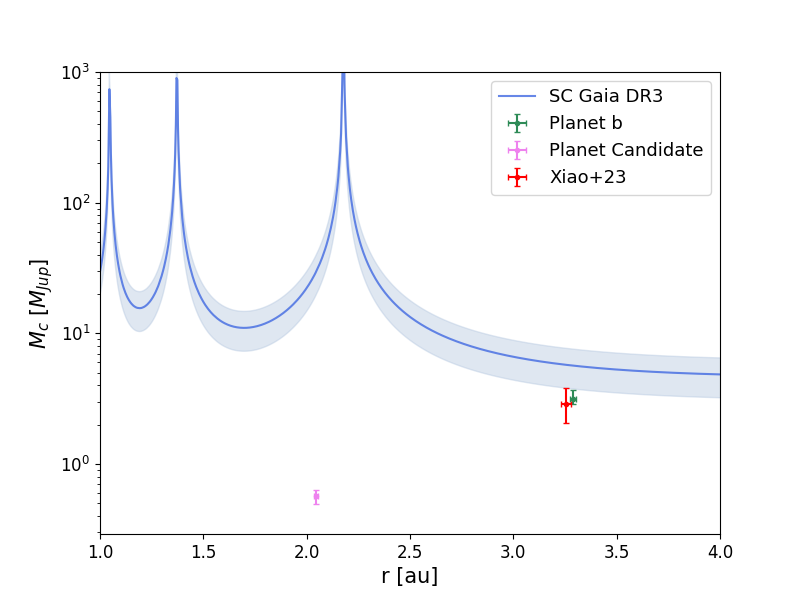}
\caption{PMa sensitivity curve calculated following \cite{Kervella2022} for HD 103891. The pink dot represents our new candidate, the green dot is HD 103891 b, and the red dot is the result by \cite{xiao2023}.}
\label{fig:hd103891_pma}
\end{figure}

\paragraph{HD 103891}: this late F-type star possesses a Jupiter-like planet on a 5.2-year orbit and a minimum mass of 1.4 \mjup\ discovered by \cite{sreevinas2022} using 91 HARPS spectra. Currently, the number of available spectra has risen to 107. In addition, \cite{xiao2023} combined RVs and \gaia\ astrometry to constrain the inclination and true mass of the object, finding $i = 27.4_{-7.1}^{+13.0}$ degrees and, therefore, $m = 2.89_{-0.84}^{+0.94}$ \mjup. In their work, \cite{sreevinas2022} also analysed activity indicators but found no periodic variations that reflect or are linked to the RVs and we confirm this result. They also point out that there are no significant signals in the residuals after removing the orbital solution for planet b but in the bottom right panel of their Figure 1, we can see a peak with FAP $< 0.1$ around 600 d. We also find this signal in our residuals, as shown in Figure \ref{fig:hd103891_gls_residuals}. As mentioned, there are no similar periodicities in the periodograms of the activity indicators, suggesting that this should be due to a second planet in the system. However, after a few tests, we found peculiar and contrasting results. When fitting the system with two Keplerians, the period of the second one is $P_c = 625_{-11}^{+330}$ d with very high eccentricity (the apoastron of this object would coincide with the periastron of planet b). If we use a Keplerian and a circular orbit, the period is then $P_c = 946_{-330}^{+11}$ d. Our interpretation is that this is a textbook case of an ‘eccentric impostor’, as described in, for example, \cite{escude2010}. Shortly, this occurs when there are two planets in circular orbits and in 2:1 Mean Motion Resonance (MMR), causing the longer-period one to incorporate the signal of the companion as a harmonic and, thus, appearing to be on an eccentric orbit. Therefore, we fitted the data again with two circular terms. This model is favoured with $\Delta\text{BIC} = 33$ over the single-Keplerian case, indicating that this is the most likely representation of the data. The parameters of planet b are consistent with literature values, except for the eccentricity. For the new candidate, to be named HD 103891.01, we find the following parameters: $P = 951.5 \pm 5.1$ d, $K = 8.07 \pm 0.60$ m/s, and $m \sin i = 0.460 \pm 0.034$ \mjup. We found no other significant peak in the GLS of the residuals. Figure \ref{fig:hd103891_phase} shows the best-fit model and the phase-folded Keplerian of the new candidate. As a sanity check, we fitted the RV with the Gaia-Hipparcos Proper Motion anomaly (PMa) data by \cite{Kervella2022} to test why \cite{xiao2023} did not recover the two-planets solution combining RVs with astrometry. We used the code published in \cite{piccinini2026}, running the MCMC for 500'000 steps, with a burn-in of 10\% and 96 walkers. If we fit only the known planet, we obtain a solution in perfect agreement with \cite{xiao2023}. However, when we fit two planets to the RVs and planet b alone to the astrometric data, we obtain a solution that is compatible with our RV-only results, with eccentricities of $e_b = 0.099_{-0.065}^{+0.081}$ and $e_c = 0.057_{-0.041}^{+0.075}$, and an orbital inclination of $i_b = 26.207_{-7.030}^{+14.762}$ degrees for planet b, compatible with the literature one. Additionally, as shown in Figure \ref{fig:hd103891_pma}, the new candidate would need a true mass around 20 \mjup\ to provide a significant astrometric signal and, in such a case, it would be very easily detectable in the RV data. As a conclusion, we obtain $\Delta \text{BIC} = 10.94$ in favour of the 2-planets model compared to the 1-planet case.
\\
\\

\begin{figure}[htbp]
%\vspace{-2cm}
   \centering
   \includegraphics[width = 0.5\textwidth]{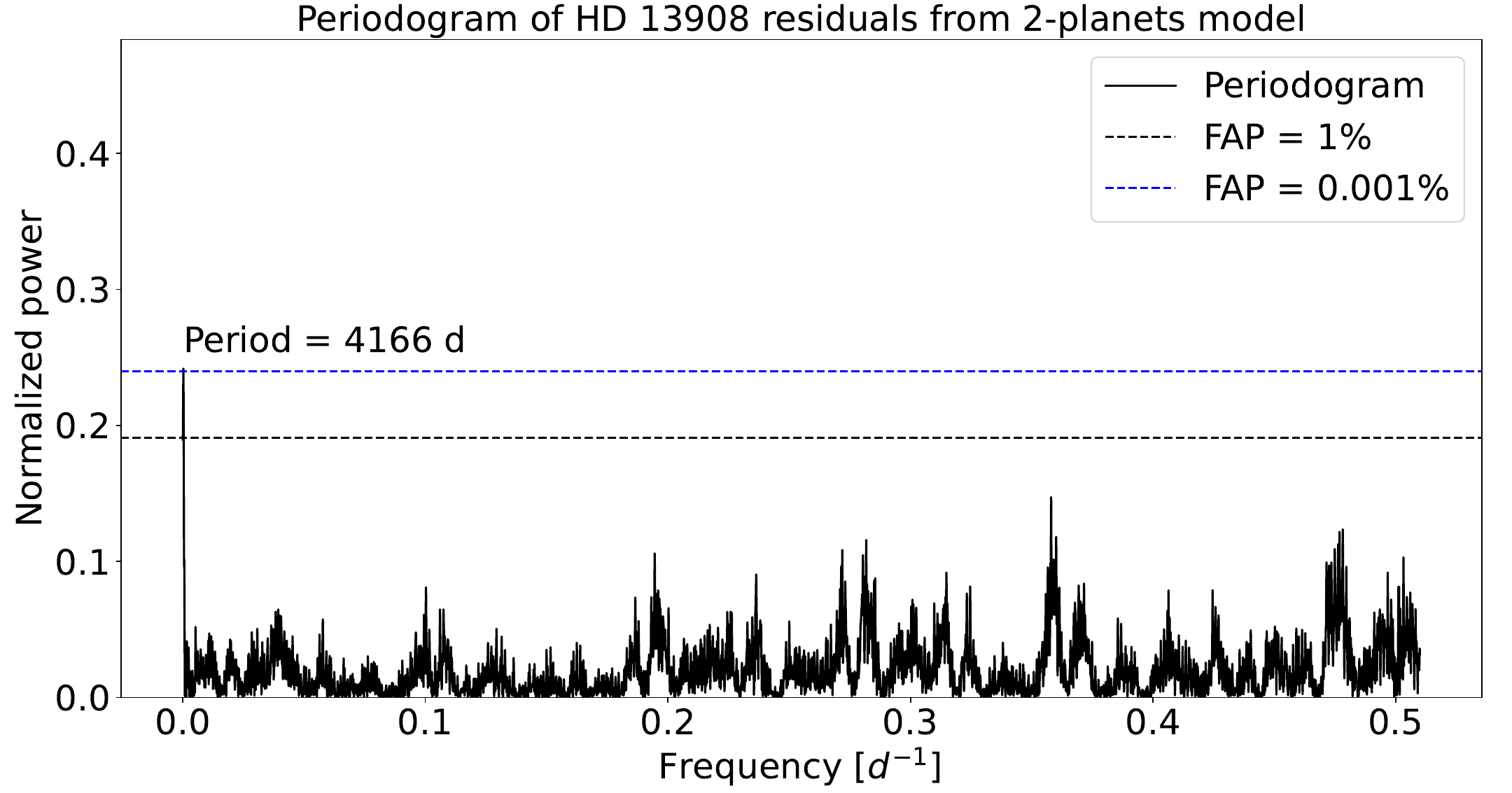}
      \caption{GLS of the RV residuals for HD 13908 after removing the signal corresponding to planets b and c.}
    \label{fig:hd13908_gls_residuals}
\end{figure}

\begin{figure}[htbp]
   \centering
   \includegraphics[width = 0.5\textwidth]{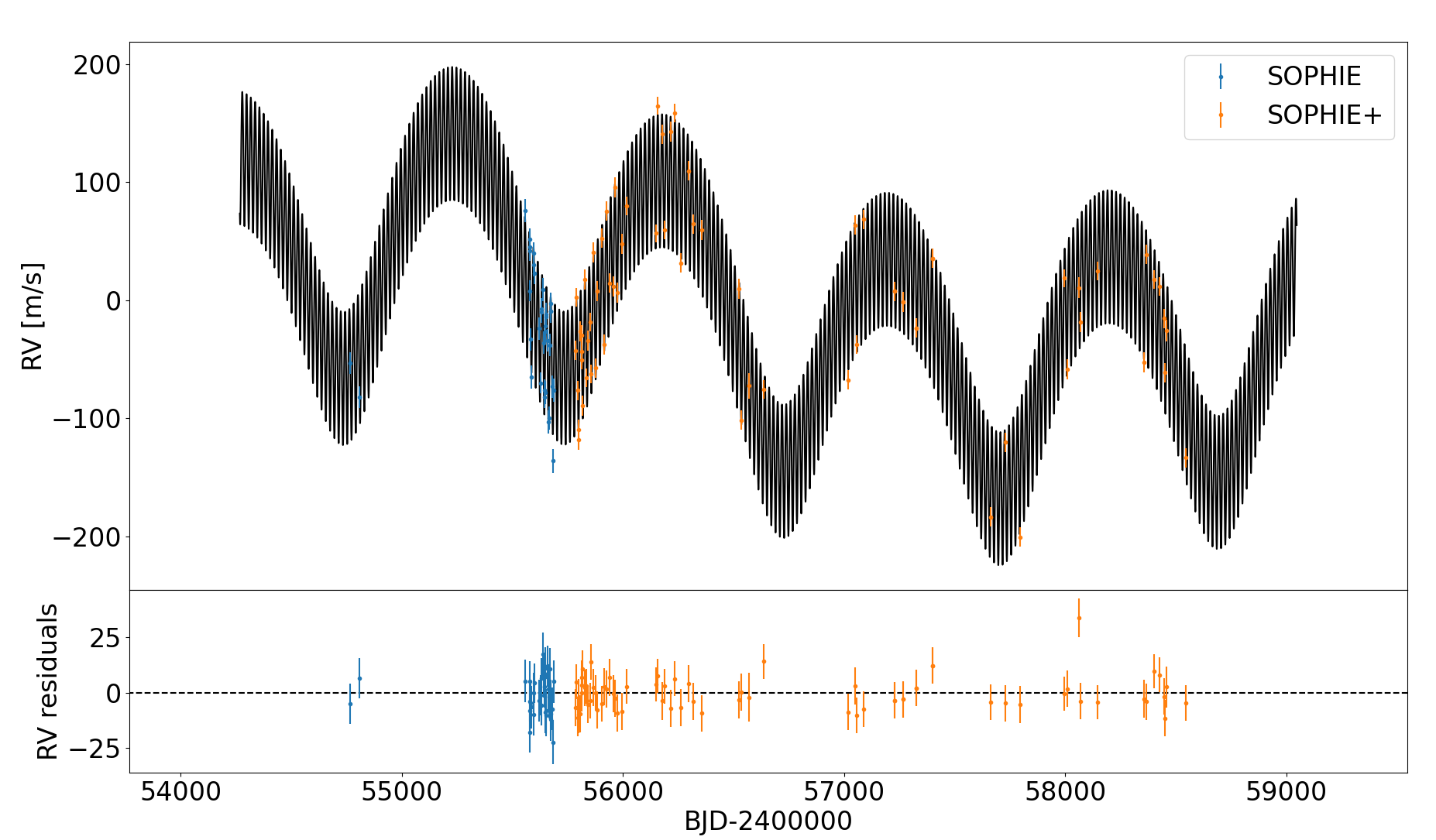}
\includegraphics[width = 0.5\textwidth]{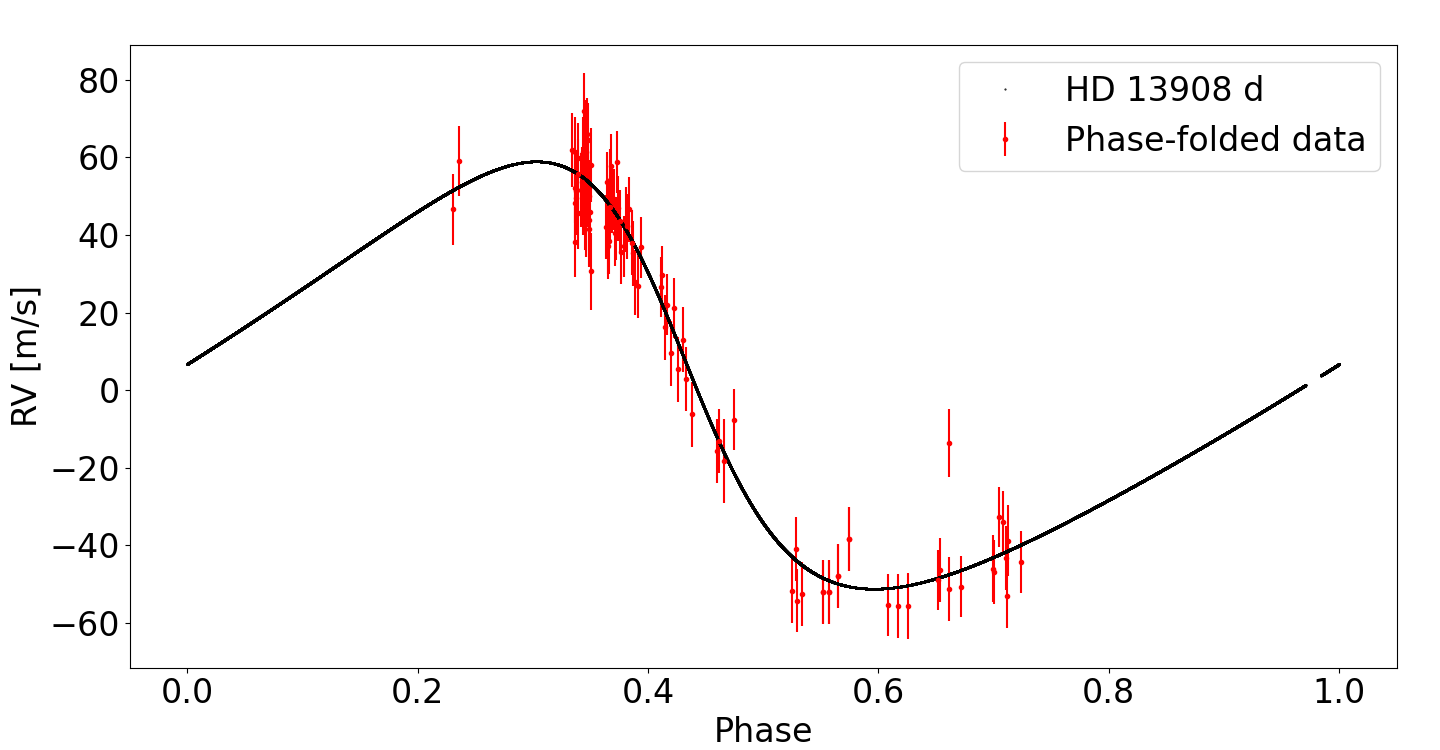}
      \caption{\textit{Top panel:} best-fit model of HD 13908. \textit{Bottom panel:} phase-folded data and Keplerian model for the new candidate HD 13908.01. 
              }
\label{fig:hd13908_phase}
\end{figure}

\paragraph{HD 13908}: this F8 star hosts a Jovian-mass planet at 0.15 au and a 5 \mjup\ planet at 2 au ($P\sim 19$ and 1000 d, respectively), both discovered by \cite{moutou2014} using 77 SOPHIE spectra. Now, there are 26 more data points available gathered over an additional $\sim 5.5$ yr. We first fitted the data with a two-planet model and searched for additional signals in the periodogram of the residuals. As shown in Figure \ref{fig:hd13908_gls_residuals}, there is a very clear signal at a bit more than 4000 d. In addition, in the two-planets model, the eccentricities of the two known planets are significantly higher compared to what was found by \cite{moutou2014} and the jitter term corresponding to the SOPHIE+ part of the set is 33.2 m/s, which is very high. Thus, we repeated the analysis with three Keplerian terms, finding that the eccentricities of planets b and c are now lower and comparable to the literature values, the SOPHIE+ jitter is now 7.4 m/s, and the model is favoured over the previous one with $\Delta\text{BIC} = 168$. At these periods, there is a possibility that the RV signal is caused by a stellar magnetic cycle. Thus, to check this hypothesis, we considered the activity indicators available on the SOPHIE archive. These are the BIS and the contrast and FWHM of the CCF, although the associated uncertainties are not provided. We extracted the GLS for these time series, but the only somewhat meaningful peak is found for the CCF contrast at $P\sim 200$ d with FAP = 0.6\%. Additionally, we searched for correlations between the three activity indicators and either raw RVs or the residuals from the 2-planets model. In all cases, we obtained a Spearman's rank coefficient $|r| \lesssim 0.2$, indicating no correlation. From these results, we conclude that, with the present dataset, there is no evidence for an activity cycle affecting HD 13908 and, thus, the planetary origin remains the most likely explanation for the long-term RV signal. For this new candidate, to be named HD 13908.01, we find the following parameters: $K = 55.1_{-6.1}^{+6.9}$ m/s, $P = 7661_{-1481}^{+1507}$ d, $e = 0.347_{-0.12}^{+0.087}$, $\omega = 78_{-14}^{+12}$ degrees, and $m\sin i = 5.92_{-0.76}^{+0.84}$ \mjup. With these new results, this system would have two planetary companions with minimum masses of $\sim 6$ \mjup\ at 2.1 and 8.3 au, respectively. So, since the \gaia-Hipparcos PMa for HD 13908 is very significant \citep[SNR $= 6.26$,][]{Kervella2022}, this is a promising target for a combined RV+astrometry analysis. Figure \ref{fig:hd13908_phase} shows the best-fit model and the phase-folded Keplerian of the new candidate.
 \\
\\

\begin{figure}[htbp]
%\vspace{-2cm}
   \centering
   \includegraphics[width = 0.5\textwidth]{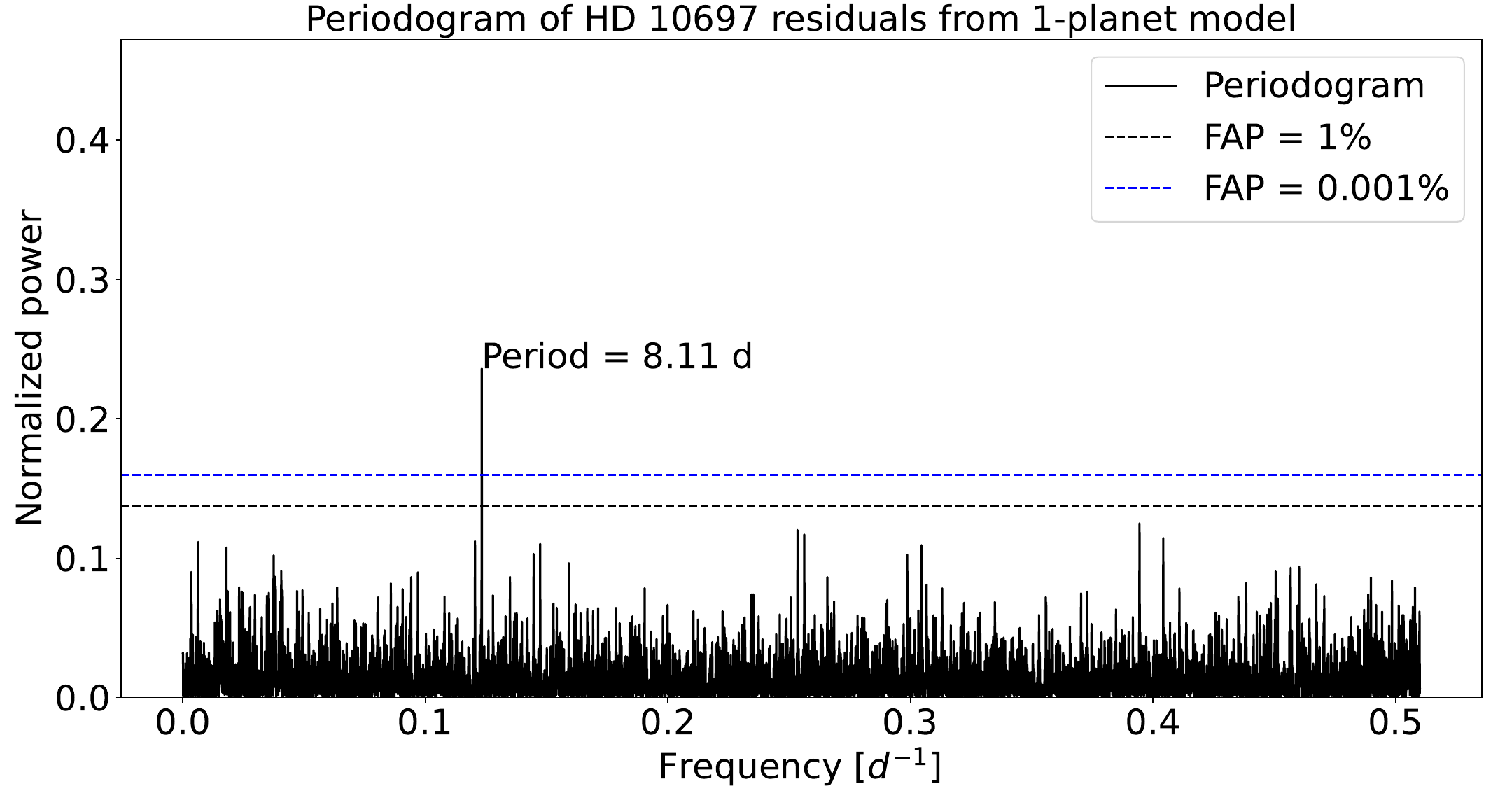}
      \caption{GLS of the RV residuals for HD 10697 after removing the signal corresponding to planet b.}
    \label{fig:hd10697_gls_residuals}
\end{figure}

\begin{figure}[htbp]
   \centering
   \includegraphics[width = 0.5\textwidth]{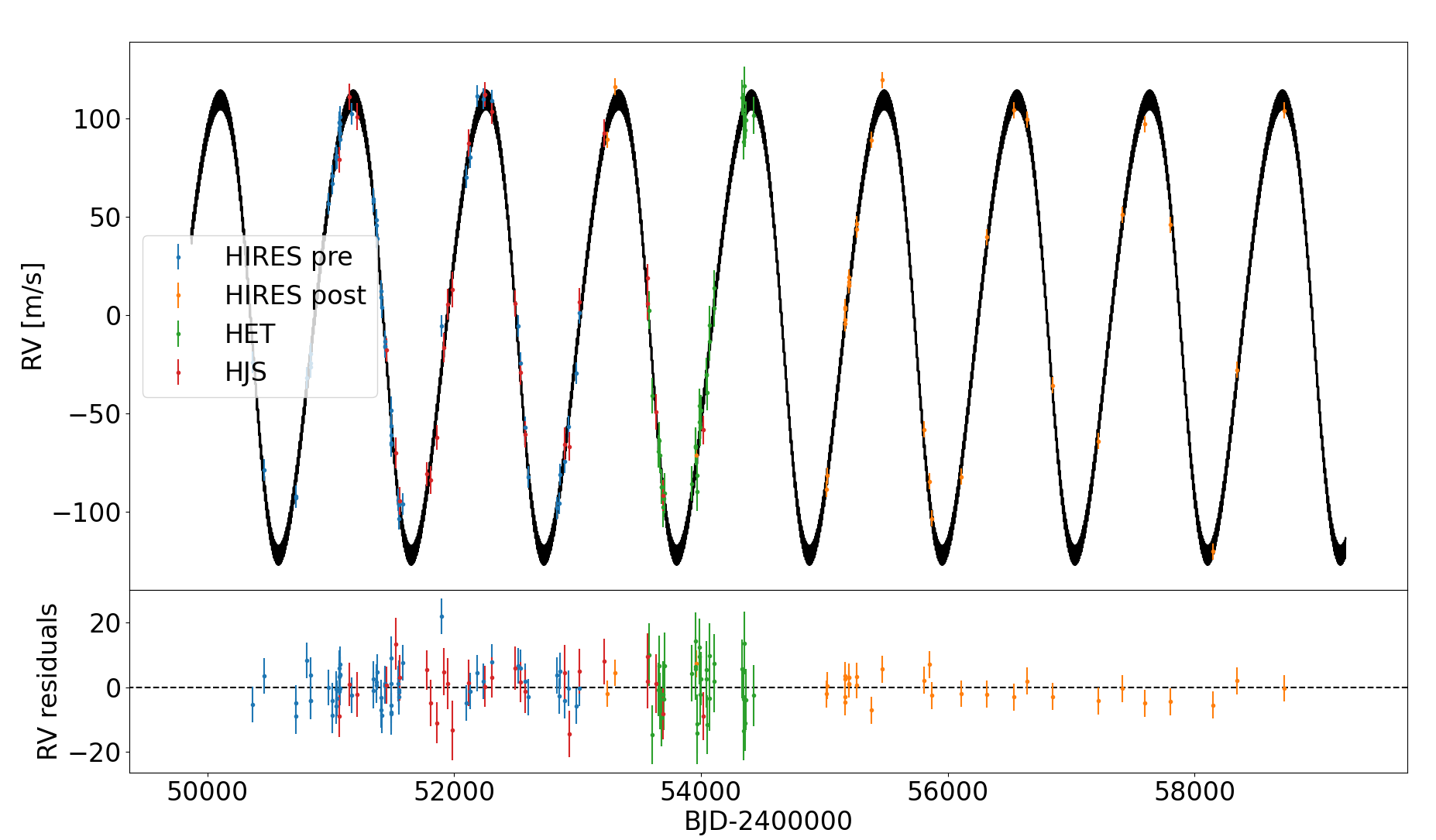}
\includegraphics[width = 0.5\textwidth]{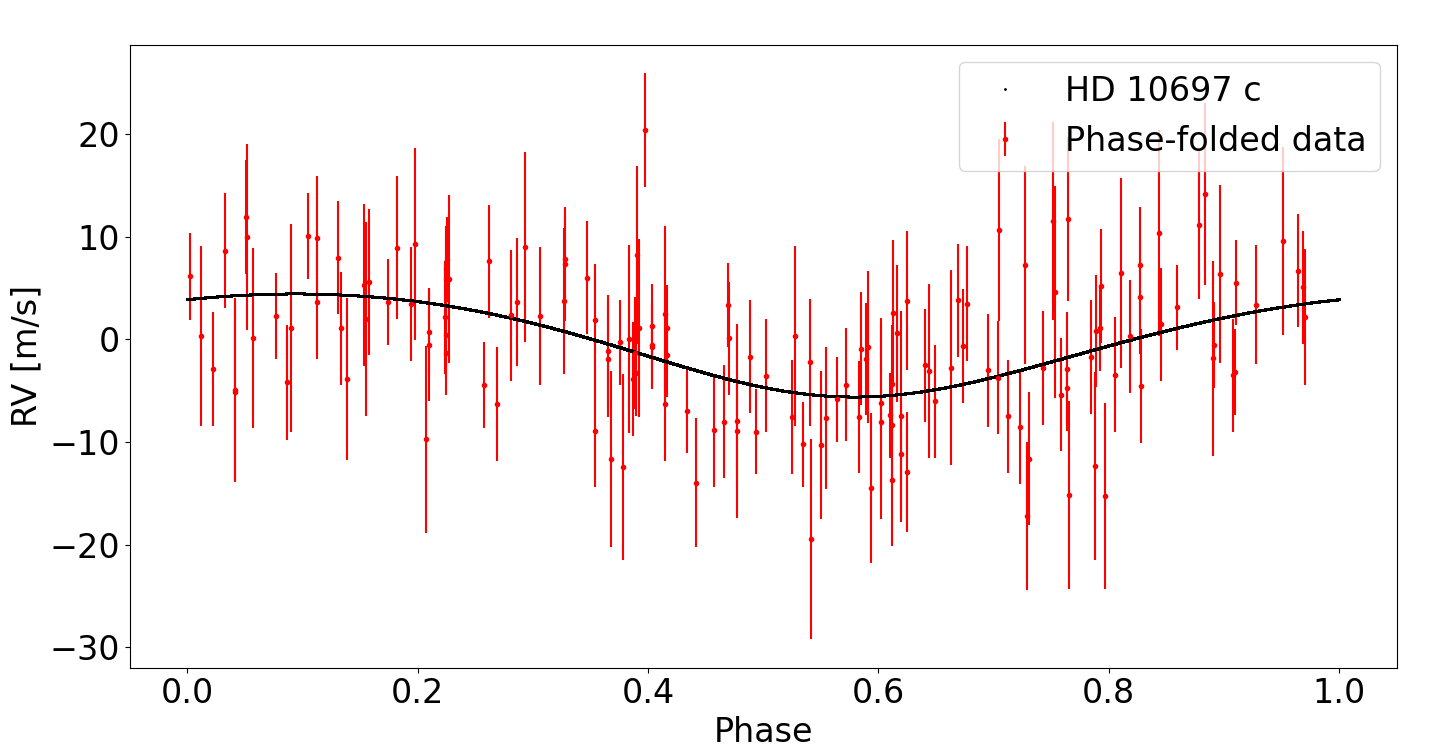}
      \caption{\textit{Top panel:} best-fit model of HD 10697. \textit{Bottom panel:} phase-folded data and Keplerian model for the new candidate HD 10697.01. 
              }
\label{fig:hd10697_phase}
\end{figure}

\paragraph{HD 10697}: this early G star hosts a 6.3 \mjup\ planet at 2 au ($P = 1075$ d) discovered by \cite{vogt2000} using 35 HIRES data. Several works have confirmed the object, the most recent being \cite{luhn2019}, \cite{rosenthal2021}, \cite{feng2022}, and \cite{xiao2023}. In addition to HIRES spectra, there are also available data taken with the Hobby-Eberly Telescope and the 2.7 m Harlan J. Smith Telescope at McDonald Observatory. We first fit the data with a one-planet model, obtaining orbital parameters in complete agreement with the literature. Then, we searched for periodicities in the GLS of the residuals, finding a very prominent peak at 8 d (Figure \ref{fig:hd10697_gls_residuals}). We checked the HIRES \smw\ data but found no dominant periodicity. In addition, we used these chromospheric activity indicators to estimate the stellar rotation period, finding $P_{\rm rot,Noyes} = 35.28 \pm 0.76$ d and $P_{rot,Mamajek} = 38.19 \pm 1.11$ d, indicating that the signal found in the GLS of the residuals is unlikely to be linked to the rotation of the star. TESS has observed HD 10697 in many sectors (17, 42, 43, 57, 70, 71, 84), so we checked both for photometric modulation and any possible transit signal. We find no coherent periodic modulation nor transits; in fact, one transit-like event happened at $\mathrm{BJD}\sim2460592$, but it is due to the passage of an object in the background. Therefore, we fit our data again with a two-planet model, finding that this is favoured over the first one with $\Delta\text{BIC} = 25$, indicating strong evidence for the presence of a second planet in the system. For the new candidate, HD 10697.01, we find the following parameters: $K = 5.03 \pm 0.78$ m/s, $P = 8.1127 \pm 0.0006$ d, $e = 0.18_{-0.12}^{+0.15}$, $\omega = 109_{-263}^{+52}$ degrees, and $m\sin i = 16.6 \pm 2.6$ \mearth. Figure \ref{fig:hd10697_phase} shows the best-fit model and the phase-folded Keplerian of the new candidate.
\\

\begin{figure}[htbp]
%\vspace{-2cm}
   \centering
   \includegraphics[width = 0.5\textwidth]{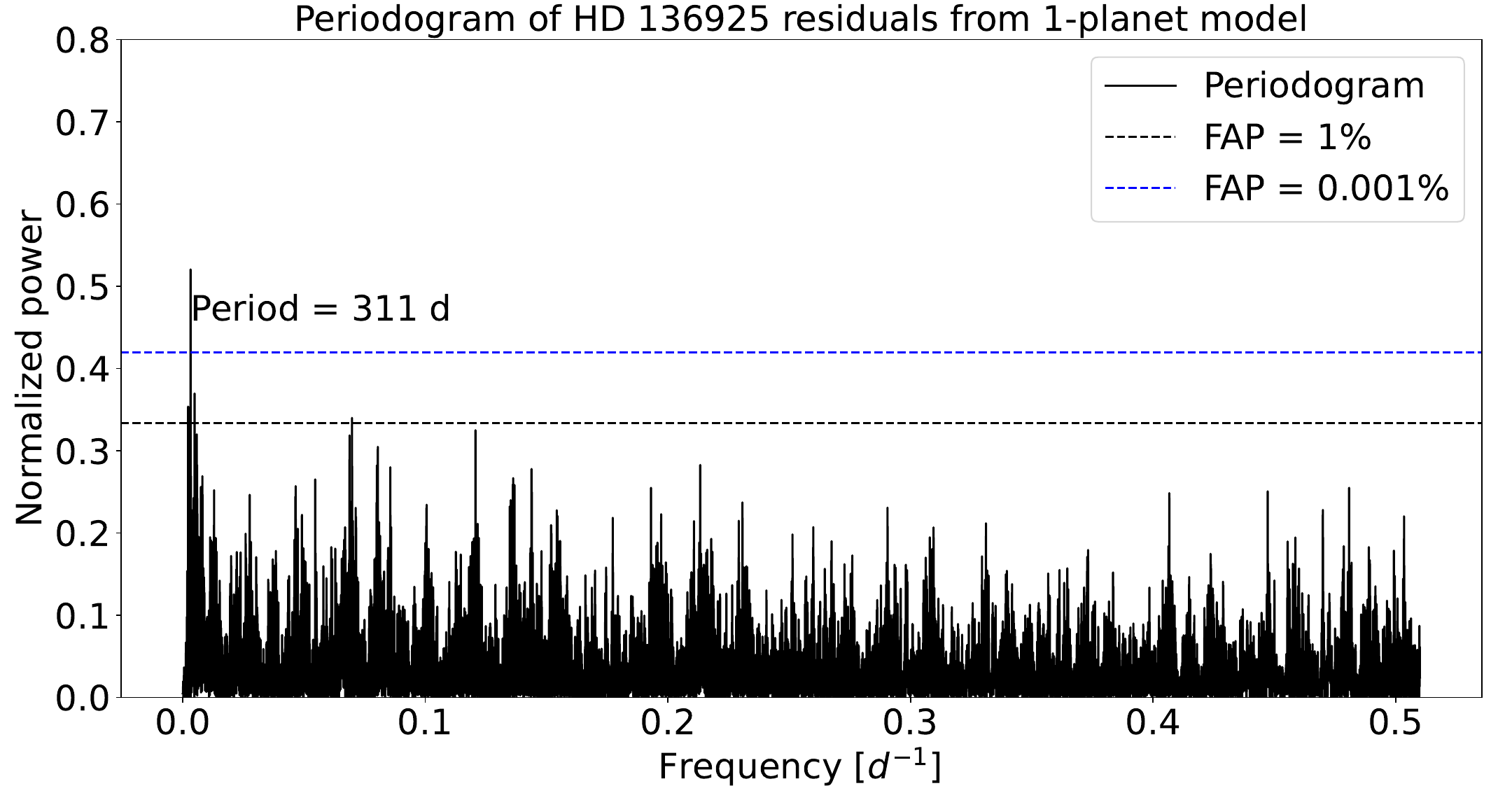}
      \caption{GLS of the RV residuals for HD 136925 after removing the signal corresponding to planet b.}
    \label{fig:hd1369257_gls_residuals}
\end{figure}

\paragraph{HD 136925}: this G0 star hosts a Jupiter-like planet with $P \sim 12$ yr, $m\sin i \sim 0.8$ \mjup, and low eccentricity discovered by \cite{rosenthal2021} with HIRES data. The authors report a peak at roughly 300 d in the GLS of the RV residuals, which we confirm (Figure \ref{fig:hd1369257_gls_residuals}). They also find a similar peak in the time series of the S-index and, therefore, label the signal as a possible false positive planet. We found a similar peak at $\sim 250$ d with FAP $= 0.2\%$. To understand whether the signal seen in the RVs is related to the one seen in stellar activity, we fit the data with different models: 
\begin{enumerate}
    \item Only planet b (Model 1, M1).

    \item Two planets terms to see if the second signal has Keplerian origin (Model 2, M2).

    \item One Keplerian and a GP with quasi-periodic kernel to see if stellar activity has a different shape and, therefore, is better explained with a GP rather than a Keplerian curve (Model 3, M3).

    \item Two planets and the GP with quasi-periodic kernel to check whether there are both a planet and an activity cycle with a similar period, as it is for the Sun and Jupiter (Model 4, M4).
\end{enumerate} 

\begin{figure}[htbp]
   \centering
   \includegraphics[width = 0.5\textwidth]{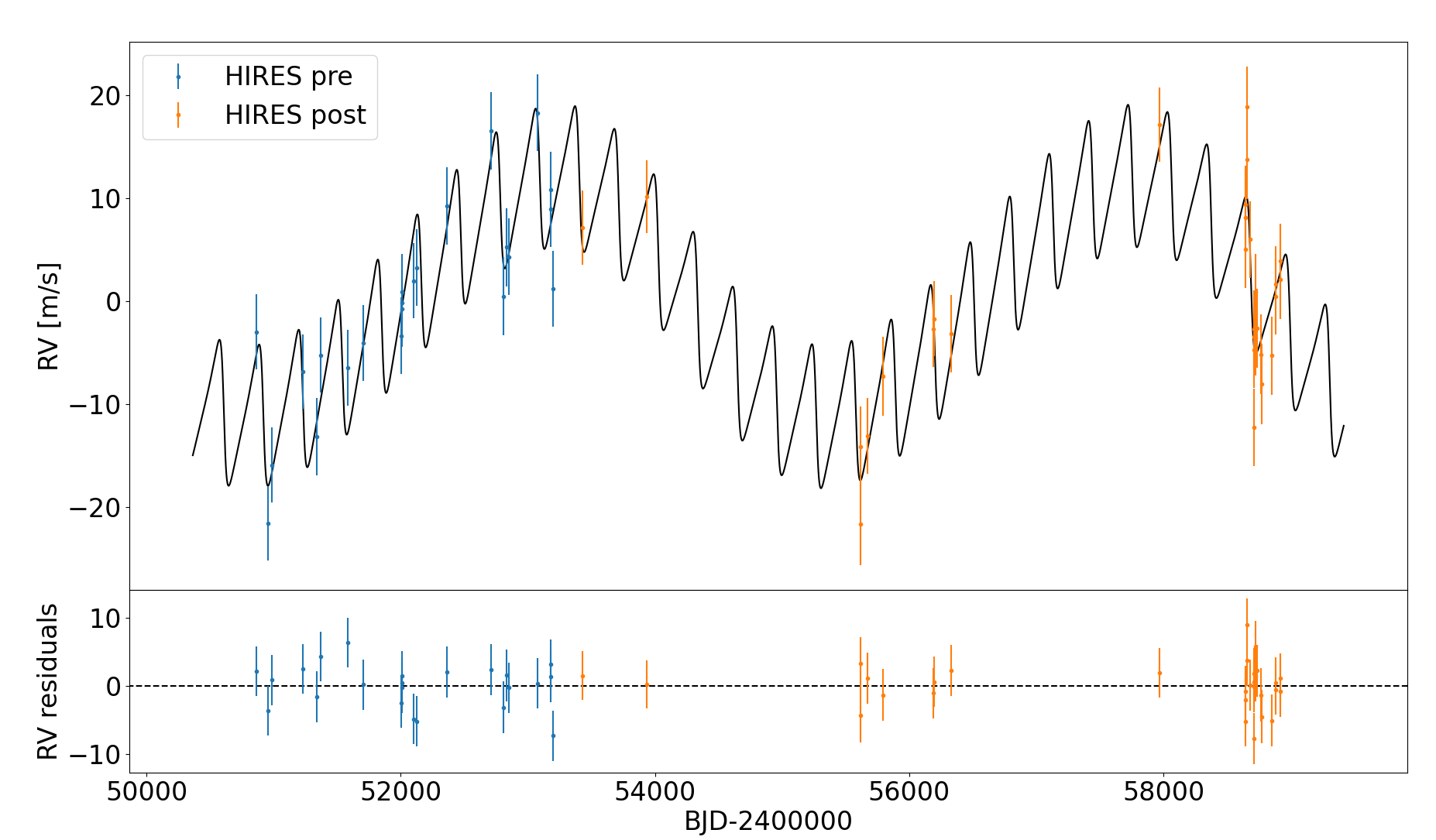}
\includegraphics[width = 0.5\textwidth]{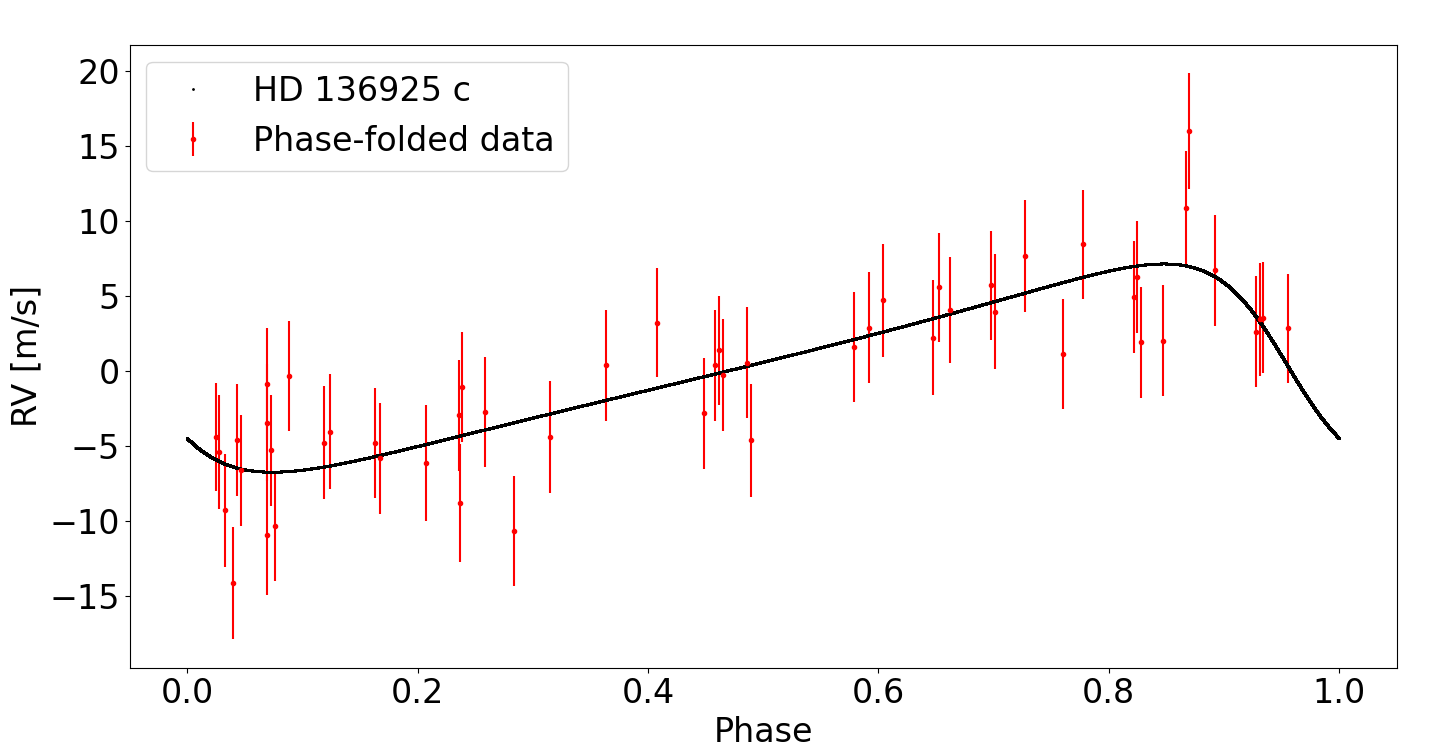}
      \caption{\textit{Top panel:} best-fit model of HD 136925. \textit{Bottom panel:} phase-folded data and Keplerian model for the new candidate HD 136925.01. 
              }
\label{fig:hd136925_phase}
\end{figure}

\noindent In M2, M3, and M4, we set a Gaussian prior on the period of the GP based on the S-index GLS peak ($=250 \pm 60$ d, wide enough to include the peak in the GLS of the RVs). In all cases, we confirm the orbital parameters for planet b. We obtain $\text{BIC(M1)} = 372.7$, $\text{BIC(M2)} = 332.3$, $\text{BIC(M3)} = 371.4$, and $\text{BIC(M4)} = 351.4$, indicating that M2 should be the better explanation of our data. In both M3 and M4, the posterior distribution of $P_{cycle}$ shows the main peak and other secondary peaks, indicating that this periodicity is not clear and dominant. On the other hand, the parameters of the second Keplerian in M2 are well determined and the error bars for the parameters of planet b are smaller than in the other cases. These results should confirm that a two-Keplerians model is the best representation of this dataset, and thus, we propose the new candidate HD 136925.01. In particular, for this new object, we find the following parameters: $K = 6.94_{-0.85}^{+0.96}$ m/s, $P = 310.76_{-0.48}^{+0.61}$ d, $e = 0.48_{-0.15}^{+0.13}$, $\omega = 64_{-72}^{+49}$, and $m\sin i = 71.4_{-8.7}^{+9.2}$ \mearth. Figure \ref{fig:hd136925_phase} shows the best-fit model and the phase-folded Keplerian of the new candidate.
\\

\subsection{Activity signatures}
\begin{figure}[htbp]
   \centering
   \includegraphics[width = 0.5\textwidth]{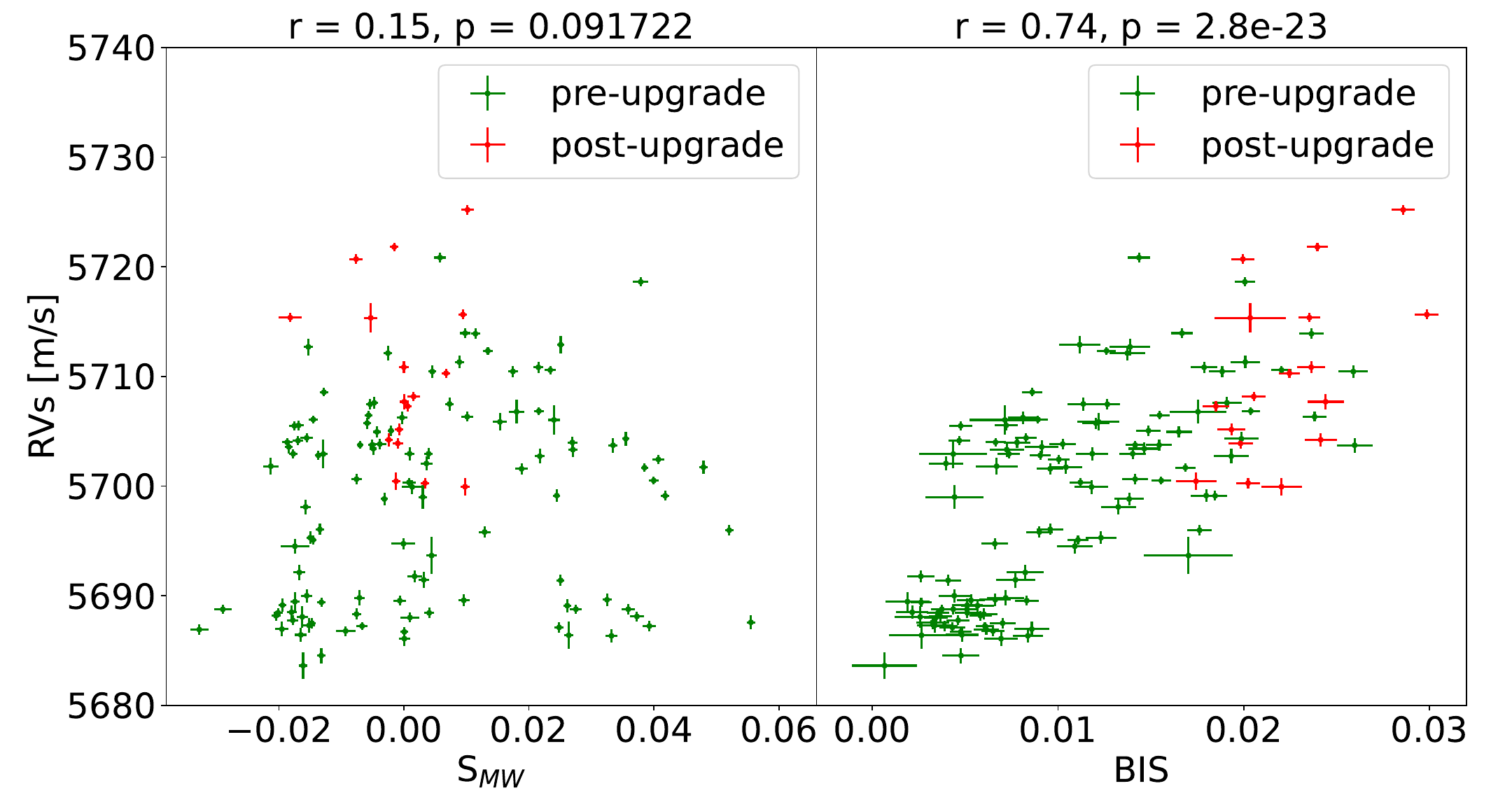}
      \caption{Correlation between RVs and two activity indicators for HD 7199, divided by pre- and post-upgrade of the instrument. }
    \label{fig:hd7199_rv_activity}
\end{figure}

\begin{figure}[htbp]
   \centering
   \includegraphics[width = 0.5\textwidth]{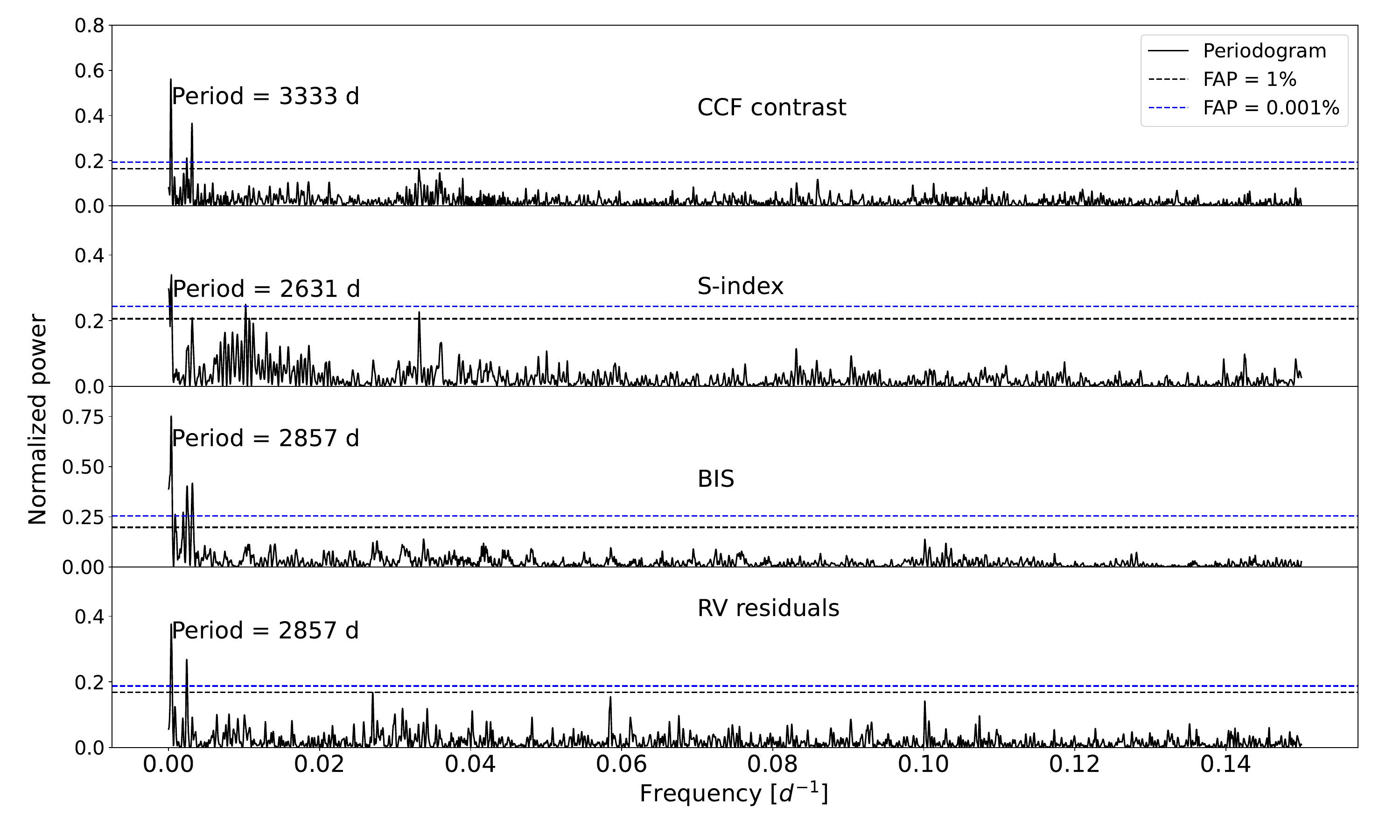}
      \caption{GLS for the CCF contrast, S-index, BIS, and RV residuals of HD 7199.}
    \label{fig:hd7199_four_gls}
\end{figure}

HD 7199 is similar to HD 204941 as they have a comparable spectral type, and the known planet has a minimum mass of 0.29 \mjup, an orbital eccentricity of 0.19, and a semi-major axis of 1.36 au ($P = 615$ d). This planet was also discovered by \cite{Dumusque2011} using 87 HARPS spectra, while now the number of available spectra is 129. As for the previous target, we studied possible correlations between RVs and activity indicators. As shown in Figure \ref{fig:hd7199_rv_activity}, we see another strong correlation between RVs and BIS, but it is not of instrumental origin this time. Even if we account for the change in offset between pre- and post-upgrade by removing the median values from the two datasets, the correlation remains with $r = 0.68$ and $p = 7.2 \times 10^{-19}$. First, we fit the RV data with a Keplerian term corresponding to planet b. Then, we searched for signals in the GLS of the RV residuals and activity indicators. As we can see in Figure \ref{fig:hd7199_four_gls}, they all display prominent peaks at similar periods, confirming a clear activity cycle affecting the total RV signal. 
Furthermore, all these periodic signals have roughly the same phase: they all peak around BJD = 54000, reach a minimum around BJD = 55500, and then peak again around BJD = 57000. For this reason, and since the shape of the signal is very similar to a sinusoidal function, we added a second Keplerian term to fit this activity cycle, yielding $P_{cycle} = 2706_{-54}^{+57}$ d and $K_{cycle} = 7.77_{-0.53}^{+0.56}$ m/s. This model with two Keplerian terms is favoured over the 1-planet model with $\Delta\text{BIC} = 117$, once again indicating that this model better fits our data.

\subsection{Stars with long-term polynomial trends}
In this section, we discuss a few objects for which there is evidence for long-term polynomial trends in addition to the known planets, indicating that a massive companion on very large orbits must be present. In almost all cases, the Keplerian origin is favoured over the possibility of the trend beign caused by an activity cycle due to the large semi-amplitudes involved \citep[$> 30$ m/s, see e.g.][]{lovis2011}. The only potentially ambiguous case is HD 106270, but no activity indicators have been published for this target so an additional analysis in this regard is currently not possible. \\

\paragraph{HD 24040}: this G0 star hosts a massive Jovian analogue with $m \sin i \sim 4$ \mjup\ and $P \sim 3600$ d discovered by \cite{wright2007} and an intermediate mass planet with $m\sin i \sim 64$ \mearth\ and $P = 500$ d, discovered by \cite{rosenthal2021}, both on very low-eccentricity orbits. Both works are based on HIRES data. In addition to the HIRES data by \cite{rosenthal2021}, we also considered 12 data points taken with ELODIE and 51 taken with SOPHIE. In addition to the two planets, \cite{rosenthal2021} also add a long-term polynomial trend to the model, and find that a quadratic trend fits the data better than a linear one. We confirm this result, finding $\Delta\text{BIC} = 17.1$ in favour of the quadratic trend compared to the linear one. (both are also strongly favoured over a model with only two planets). In particular, we find for the second order coefficient a value of $\ddot{\gamma} = (-6.1 \pm 1.3) \cdot 10^{-7}$ m s$^{-1}$ day$^{-2}$, consistent with the results of \cite{rosenthal2021}. From this, using the equations by \cite{kipping2011}, we can obtain estimates for the minimum semi-amplitude, period, and mass of the objects causing this RV trend. In particular, we derive $K > 51$ m/s, $P > 157$ yr, $a > 28$ au, and $m \sin i > 9$ \mjup, indicating that this might be a BD on a very large orbit.  \\

\paragraph{HD 163607}: this G5 star is orbited by a 0.8 \mjup\ planet at 75 d on a very eccentric orbit and a 2.2 \mjup\ planet at 1270 d on an almost circular orbit. Both were discovered by \cite{giguere2012} using HIRES data. More recently, \cite{luhn2019} announced a new candidate on a very external orbit. Since the orbit of this planet is far from closed, we fitted a model with two planets and a quadratic trend. For planets b and c, we confirm the parameters already known from the literature. As expected, the model including the quadratic trend is favoured with $\Delta\text{BIC} = 122.6$, confirming that a third very distant companion is present in the system. We find the polynomial coefficients to be $\dot{\gamma} = 0.00591_{-0.00056}^{+0.00059}$ m s$^{-1}$ d$^{-1}$ and $\ddot{\gamma} = (2.77 \pm 0.34) \cdot 10^{-6}$ m s$^{-1}$ d$^{-2}$. Using the equations by \cite{kipping2011}, we derive $K > 44$ m/s, $P > 69$ yr, and $m\sin i > 6.9$ \mjup. All these values agree with the ones reported by \cite{luhn2019}. \\

\paragraph{HD 73267}: this G5 star hosts a super-Jupiter with a minimum mass of 3 \mjup\ at about 2 au ($P \sim 1260$ d) from the star discovered by \cite{moutou2009} using 39 HARPS observations. This was later confirmed by \cite{barbato_b_2018}, \cite{feng2022}, and \cite{xiao2023}. The last two works constrained its true mass and inclination with a combination of RVs and astrometry. \cite{barbato_b_2018} also report a linear trend of $2.14_{-0.19}^{+0.20}$ m s$^{-1}$ yr$^{-1}$ that \cite{feng2022} and \cite{xiao2023} report as new candidate thanks to astrometry. Both works find a low eccentricity for the orbit of this new candidate, a period roughly between 38 and 48 yr, and true masses above 4 \mjup. First of all, we further confirm the existence of planet b with parameters compatible with the literature. Secondly, we found no correlations between RVs and activity indicators nor any significant peak in the GLS of their time series. Then, we fitted the data using both a linear and a quadratic trend in addition to planet b. Both models are strongly favoured compared to the one with only one planet, and the quadratic trend is favoured over the linear one with $\Delta\text{BIC} = 105$. In particular, we find a first-order coefficient of $\dot{\gamma} = 0.00938 \pm 0.00022$ m s$^{-1}$ d$^{-1}$ and a second-order coefficient of $\ddot{\gamma} = (2.28 \pm 0.18) \cdot 10^{-6}$ m s$^{-1}$ d$^{-2}$. Using the equations by \cite{kipping2011}, we derive $K > 72$ m/s, $P > 97$ yr, $a > 20$ au, and $m\sin i > 10.8$ \mjup. Both the period and the mass are much larger than those found by \cite{feng2022} and \cite{xiao2023}. However, this star has a RUWE of 0.978 and SNR(PMa) $= 1.25$ \citep{Kervella2022}, indicating that the astrometric signal is not very significant. In any case, since this work focuses on small planets at short periods, the exact details of the long-period modeling should have a negligible impact on the outcome of the global analysis. \\

\paragraph{HD 191806}: this K0 star hosts a super-Jupiter with $m\sin i > 8$ \mjup\ and $P \sim 1600$ d discovered by \cite{diaz2016}. The authors also find no correlation between RVs and stellar activity and point out that a long-term linear trend is present in the data, quantifying it as $\dot{\gamma} = 11.4 \pm 1.7$ m s$^{-1}$ yr$^{-1}$. They used 6 ELODIE data and 46 SOPHIE data. Now, the number of available SOPHIE data is 62, so we have 16 additional data points gathered over $\sim 4$ yr (after removing a clear outlier at BJD $= 56874$, taken on 5th August 2014, which is not present even in the time series used in the discovery paper). We fit the data with three models: one planet, one planet plus linear trend, and one planet plus quadratic trend. We find that adding a polynomial trend significantly reduces the jitter term corresponding to the SOPHIE+ part of the dataset (from 24 to 11 m/s) and the error bars of the known planet's orbital parameter, as expected. In addition, both models are favoured compared to the one with only planet b included with $\Delta\text{BIC} = 43.9$ (linear) and $\Delta\text{BIC} = 43.6$ (quadratic). However, it is not possible to distinguish between the presence of a linear or quadratic trend as they have $\Delta\text{BIC} = 0.3$ so they are virtually identical. In this case, we choose to prefer the linear term because the quadratic coefficient that results from our analysis is $\ddot{\gamma} = (3.9 \pm 1.9) \cdot 10^{-6}$ m s$^{-1}$ d$^{-2}$, which is compatible with zero within $2\sigma$ and, thus, not very significant. The linear coefficient that we obtain is $\dot{\gamma} = 9.53 \pm 1.02$ m s$^{-1}$ yr$^{-1}$ so smaller than, but still compatible with, the one found by \cite{diaz2016}, confirming their result with the addition of a few more years of data. Using the equations by \cite{kipping2011}, for the companion responsible for the linear trend we find $K > 139$ m/s, $P > 58$ yr, $a > 16$ au, and $m\sin i > 21$ \mjup. Therefore, this should be a BD on a very external orbit, and it could be an interesting target for astrometric and direct imaging studies. \\

\paragraph{HD 106270}: this G5 star hosts a very massive ($m\sin i \sim 10$ \mjup) planet at 3.3 au ($P \sim 1850$ d) with moderate eccentricity discovered by \cite{johnson2011} with HIRES data. Later, \cite{bryan2016} refined the orbital parameters and first mentioned a linear trend in the data with $\dot{\gamma} = 1.9_{-1.6}^{+1.7}$ m s$^{-1}$ yr$^{-1}$. More recently, \cite{luhn2019} used the same 29 HIRES data that we use in this work, finding $\dot{\gamma} = 3.14 \pm 0.99$ m s$^{-1}$ yr$^{-1}$. As for the previous target, we fit the data with a one-planet model, one with a linear trend, and one with a quadratic trend. In this case, we find BIC(1p) $= 251.2$, BIC(linear) $= 245.0$, and BIC(quadratic) $= 241.9$, meaning that the evidence for the long-term RV trend is strong but weaker than the previous cases as we calculate $\Delta\text{BIC} = 6.2$ in favour of the linear and $\Delta\text{BIC} = 9.3$ in favour of the quadratic model compared to the one with only planet b. In the linear case, we find $\dot{\gamma} = 3.58 \pm 1.24$ m s$^{-1}$ yr$^{-1}$, compatible with the results by \cite{luhn2019}. Using the equations by \cite{kipping2011}, we derive $K > 31$ m/s, $P > 35$ yr, $a > 12$ au, and $m \sin i > 4.4$ \mjup. In the quadratic case, we obtain $\ddot{\gamma} = 1.2 \pm 0.55$ m s$^{-1}$ yr$^{-2}$, which is a $2.2\sigma$ detection so not very significant. Using the equations by \cite{kipping2011}, we derive $K > 39$ m/s, $P > 36$ yr, $a > 12$ au, and $m \sin i > 5.5$ \mjup, so not very different from the linear case. Once again, it is not easy to distinguish between these two cases and, as mentioned before, the exact modeling of the long-term trend should have little impact on the analysis regarding close-in companions. Here, we choose the model with the quadratic trend as favoured, even if the evidence in its favour is not very strong ($\Delta\text{BIC} = 3.1$). \\

\subsection{Discrepant orbital solutions}
In this section, we mention a couple of systems for which we found best-fit solutions significantly different from the literature but without any new candidates. \\

\paragraph{HD 156098}: this F6 star has two recently discovered planets announced by \cite{feng2022}. Planet b is a Neptune-mass planet at 21 d on an almost circular orbit, while planet c has a mass of about 5 \mjup, a period of about 21 yr, and a moderately eccentric orbit. The RV dataset only includes 80 HARPS spectra that have been combined with astrometry to derive the outer planet's orbital inclination and true mass. We started fitting the data with two Keplerian terms but we found that $K_b$ tends to zero. In particular, we find $K_b = 0.32_{-0.29}^{+2.9}$ m/s, and the posterior distributions of the other parameters are completely flat. Assuming a circular orbit for planet b does not lead to different results. A model with only planet c is instead favoured with $\Delta\text{BIC} = 13.3$, indicating that the presence of the inner planet is at least dubious. Moreover, in the GLS of the residuals of the one-planet model, we see a peak at 26 d, so close to the period of the presumed planet b, but its FAP is $\sim 5.6\%$, so quite high. On top of that, we find for planet c a higher period and eccentricity compared to \cite{feng2022}, that is $e_c = 0.512 \pm 0.094$ and $P_c = 10698_{-2191}^{+4026}$ d. The large error bar is because the orbit is incomplete, i.e. our data cover only $\sim 57\%$ of the nominal period. Based on these results, we label HD 156098 b as a potential false positive and consider a one-planet model for our analysis. \\

\begin{figure}[htbp]
\centering
\includegraphics[width = 0.5\textwidth]{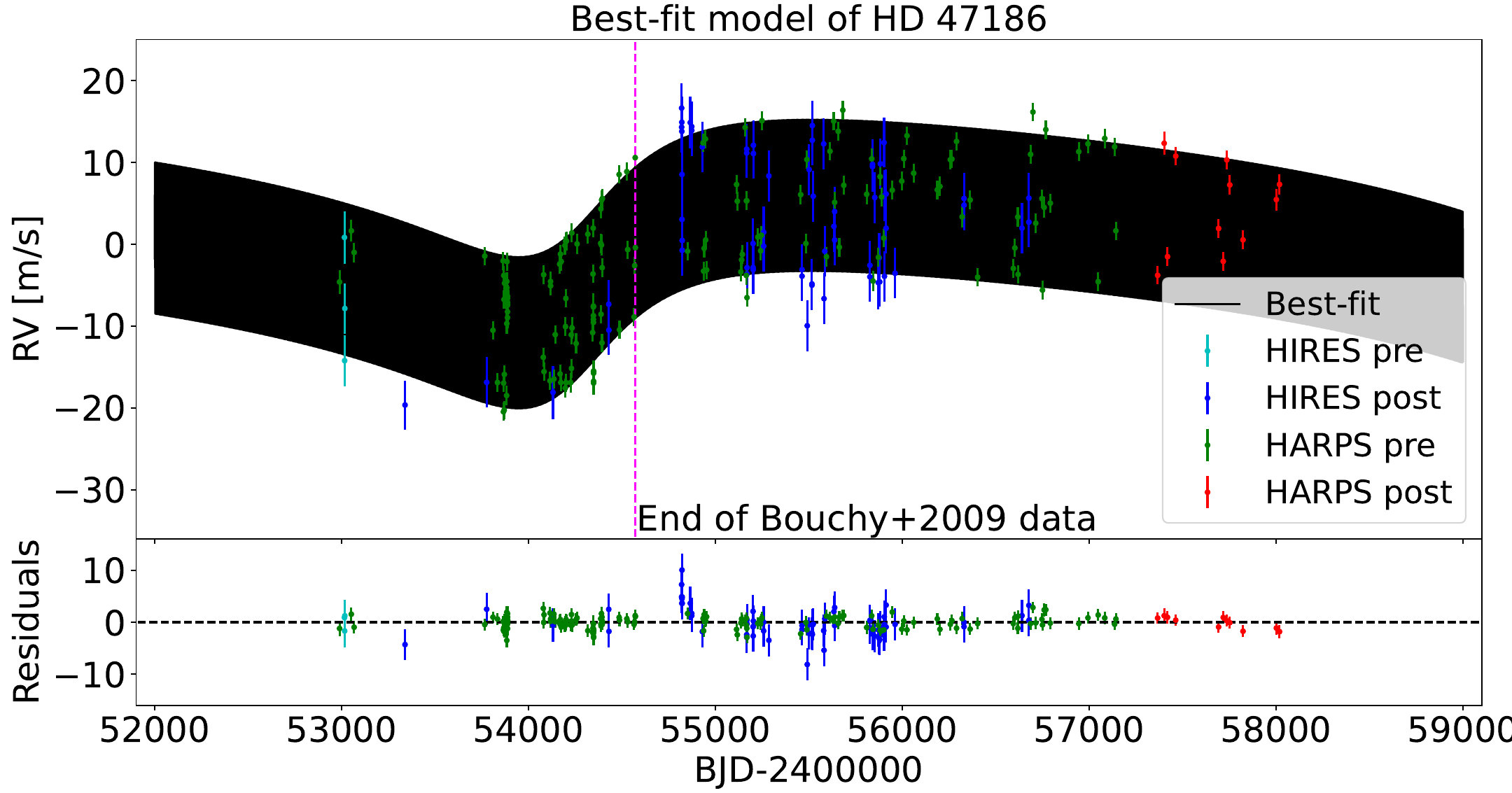}
\caption{Best-fit RV model for HD 47186. The dashed vertical line represents the end of the dataset used by \cite{bouchy2009} in their analysis (which does not include HIRES data taken before that date).}
    \label{fig:hd47186_best}
\end{figure}

\paragraph{HD 47186}: this G5 star is orbited by a 22 \mearth\ planet at 0.05 au ($P = 4$ d) and an external planet with reported minimum mass a bit larger than Saturn and 3.7 yr of period, both announced by \cite{bouchy2009} using 66 HARPS data gathered over 4.3 yr. Now, we have 67 HIRES data published by \cite{Butler2017} and 170 HARPS spectra for a total time span of 13.8 yr. We tried to reproduce the literature solution but found that this was highly inadequate to represent the likely configuration of the system with the larger dataset available today. We confirm the orbital parameters of planet b but we find significantly higher values of period, eccentricity, and mass for the external companion. In particular, we find: $K = 8.4 \pm 0.1$ m/s, $P = 5838_{-215}^{+118}$ d, $e = 0.47 \pm 0.02$, $\omega = -130.3 \pm 3.4$ degrees, and $m\sin i = 0.638 \pm 0.026$ \mjup. Such a solution has smaller jitter terms for all datasets, lower residuals, and is favoured with $\Delta\text{BIC} = 581$ compared to a solution similar to the literature one. This means that the period is not 3.7 yr but $\sim 16$ yr, the minimum mass is almost twice the value by \cite{bouchy2009}, and the orbit is rather eccentric. These results have been possible thanks to the continuous RV monitoring with high-accuracy instruments over more than a decade, once again demonstrating that this is a fundamental step for the characterization of the long-period part of the parameter space. The best-fit model is shown in Figure \ref{fig:hd47186_best}.

\newpage

\twocolumn
\section{Stellar activity and polynomial coefficients results}
\label{sec:appendix_act_pol}

\begin{table}[htbp]
\caption{Stellar activity results}
\label{tab:stellar_act_results}
%\fontsize{12}{4}\selectfont
\centering
\begin{tabular}{c c c}
\hline
\hline
\\
%Target & HIRES 1 & HIRES 2 & HARPS 1 & HARPS 2 & HARPS 3 & HARPS-N & ELODIE & SOPHIE 1 & SOPHIE 2 & HET & HJS & $\Delta t$ [yr]\\
Target & $P_{cycle}$ [d] & $K_{cycle}$ [m/s] \\
\\
\hline
\\
HD 3765 & $44.7_{-1.0}^{+1.7}$ & $2.58_{-0.23}^{+0.28}$ \\
\\
\hline
\\
HD 7199  & $2706_{-54}^{+57}$ & $7.77_{-0.53}^{+0.56}$ \\
\\
\hline
\\
HD 75898  & $781_{-57}^{+72}$ & $8.6_{-1.9}^{+2.6}$ \\
\\
\hline
\end{tabular}
\tablefoot{Targets in our sample for which we modeled stellar activity with GP regression. The second column is the period of the cycle or rotational modulation, while the last column is the contribution to the RV signal.}
\end{table}

%_{-}^{+}

\begin{table}[htbp]
\caption{Polynomial coefficients}
\label{tab:stellar_act_results}
%\fontsize{12}{4}\selectfont
\centering
\begin{tabular}{c c c}
\hline
\hline
\\
Target & $\dot{\gamma}$ [ms$^{-1}$yr$^{-1}$] & $\ddot{\gamma}$ [ms$^{-1}$yr$^{-2}$] \\
\\
\hline
\\
HD 24040  & $(5.01 \pm 0.41) \times 10^{-3}$  &  $0$  \\
\\
\hline
\\
HD 73267  &  $(9.38 \pm 0.22) \times 10^{-3}$  &  $(2.28 \pm 0.18) \times 10^{-6}$  \\
\\
\hline
\\
HD 106270  &  $(5.2 \pm 3.7) \times 10^{-3}$  &  $\left(9.0_{-4.1}^{+3.9}\right) \times 10^{-6}$  \\
\\
\hline
\\
DH 163607  &  $\left( 5.91_{-0.56}^{+0.59} \right) \times 10^{-3}$  &  $(2.77 \pm 0.34)\times 10^{-6}$  \\
\\
\hline
\\
HD 168443  &  $(-7.77 \pm 0.28) \times 10^{-3}$  &  $0$  \\
\\
\hline
\\
HD 191806  & $(2.61 \pm 0.28) \times 10^{-2}$  &  $0$  \\
\\
\hline
\end{tabular}
\tablefoot{Targets in our sample for which we added linear or quadratic polynomial trend to the RV fit. The second column is the first-order coefficient, while the second column is the second-order one.}
\end{table}

\newpage

\onecolumn

\section{Data used for systems with new candidates}

%\textcolor{red}{Table \ref{tab:candidates_rv_data} contains detailed information on the datasets used for the analyses shown in the previous section.}

\begin{table*}[htbp]
\caption{Number of spectra used for our targets discussed in the Appendix}
\label{tab:candidates_rv_data}
%\fontsize{12}{4}\selectfont
\centering
\begin{tabular}{c c c c c c c c c c c c c}
\hline
\hline
\\
%Target & HIRES 1 & HIRES 2 & HARPS 1 & HARPS 2 & HARPS 3 & HARPS-N & ELODIE & SOPHIE 1 & SOPHIE 2 & HET & HJS & $\Delta t$ [yr]\\
Target & \multicolumn{2}{c}{HIRES} & \multicolumn{3}{c}{HARPS} & HARPS-N & ELODIE & \multicolumn{2}{c}{SOPHIE} & HET & HJS & $\Delta t$ [yr]\\
 & 1 & 2 & 1 & 2 & 3 & & & 1 & 2 & & &  \\
\\
\hline
\\
HD 3765 & 29 & 261 & - & - & - & 96 & - & - & - & - & - & 26.6 \\
\\
\hline
\\
HD 204941 & - & - & 68 & 28 & 75 & - & - & - & - & - & - & 17.9 \\
\\
\hline
\\
HD 30669 & - & - & 51 & 26 & 10 & - & - & - & - & - & - & 18.9 \\
\\
\hline
\\
HD 170469 & 24 & 21 & - & - & - & - & - & - & - & - & - & 19.2 \\
\\
\hline
\\
HD 103891 & - & - & 66 & 26 & 15 & - & - & - & - & - & - & 19.0 \\
\\
\hline
\\
HD 13908 & - & - & - & - & - & - & - & 35 & 69 & - & - & 10.3 \\
\\
\hline
\\
HD 10697 & 63 & 31 & - & - & - & - & - & - & - & 34 & 27 & 22.9 \\
\\
\hline
\\
HD 136925 & 23 & 30 & - & - & - & - & - & - & - & - & - & 22.1 \\
\\
\hline
\\
HD 7199 & - & - & 112 & 17 & - & - & - & - & - & - & - & 13.9 \\
\\
\hline
\\
HD 24040 & 22 & 71 & - & - & - & - & 12 & 21 & 30 & - & - & 22.1 \\
\\
\hline
\\
HD 163607 & - & 74 & - & - & - & - & - & - & - & - & - & 13.2 \\
\\
\hline
\\
HD 73267 & - & - & 65 & 4 & 9 & - & - & - & - & - & - & 17.4 \\
\\
\hline
\\
HD 191806 & - & - & - & - & - & - & 6 & 27 & 34 & - & - & 14.6 \\
\\
\hline
\\
HD 106270 & - & 29 & - & - & - & - & - & - & - & - & - & 8.7 \\
\\
\hline
\\
HD 156098 & - & - & 51 & 22 & 7 & - & - & - & - & - & - & 16.9 \\
\\
\hline
\\
HD 47186 & 3 & 64 & 159 & 11 & - & - & - & - & - & - & - & 13.8 \\
\\
\hline
\end{tabular}
\tablefoot{Number of spectra taken with each spectrograph for each target discussed in the Appendix. HIRES 1 and 2 refer to data taken before and after the instrument upgrade described in \cite{Butler2017}. HARPS 1, 2, and 3 refer to data taken before and after the instrumental upgrade described in \cite{LoCurto2015}, and the one during the COVID lockdown on 23rd March 2020. SOPHIE 1 and 2 refer to data taken before and after the instrumental upgrade described in \cite{sophie+}. The last column displays the total time span of the observations in years.}
\end{table*}

\end{appendix}

\end{document}